\documentclass[11pt, a4paper]{article}
\pdfoutput=1
\usepackage{jheppub}
\usepackage[utf8]{inputenc}
\usepackage[T1]{fontenc}
\usepackage{amsfonts}
\usepackage[caption=false]{subfig}
\usepackage{multirow}
\usepackage{physics}
\usepackage{mathtools}
\usepackage{url}
\numberwithin{figure}{section}
\numberwithin{equation}{section}
\numberwithin{table}{section}

\hypersetup{colorlinks,linkcolor=blue,urlcolor=blue,citecolor=magenta}

\newcommand{\vt}{\vartheta}

\title{Perturbative post-quench overlaps in Quantum Field Theory}

\author[a]{Krist\'of H\'ods\'agi}
\author[a,b]{, M\'arton Kormos}
\author[a,b]{and G\'abor Tak\'acs}

\affiliation[a]{Department of Theoretical Physics, \\
Budapest University of Technology and Economics \\
1111 Budapest, Budafoki \'ut 8, Hungary}
\affiliation[b]{BME \textquotedbl{}Momentum\textquotedbl{} Statistical
Field Theory Research Group \\
1111 Budapest, Budafoki \'ut 8, Hungary}

\emailAdd{hodsagik@gmail.com}
\abstract{In analytic descriptions of quantum quenches, the overlaps between the initial pre-quench state and the eigenstates of the time evolving Hamiltonian are crucial ingredients. We construct perturbative expansions of these overlaps in quantum field theories where either the pre-quench or the post-quench Hamiltonian is integrable. Using the $E_8$ Ising field theory for concrete computations, we give explicit expressions for the overlaps up to second order in the quench size, and verify our results against numerical results obtained using the Truncated Conformal Space Approach. We demonstrate that the expansion using the post-quench basis is very effective, but find some serious limitations for the alternative approach using the pre-quench basis.}


\begin{document}

\frenchspacing

\maketitle
\flushbottom

\section{Introduction}
\label{sec:intro}

The current interest in the dynamics of quantum many-body systems out of equilibrium drives intensive research, both on the experimental and the theoretical side. The former have witnessed a rapid expansion due to the realisation of isolated quantum many-body dynamics in cold atom systems \cite{2006Natur.440..900K,2007Natur.449..324H,2011Natur.472..307S,2012NatPh...8..325T,2012Natur.481..484C,2012Sci...337.1318G,2013NatPh...9..640L,2013Natur.502...76F,2013PhRvL.111e3003M,2015Sci...348..207L,2016Sci...353..794K}. These experiments shed light on fundamental issues regarding closed quantum systems out of equilibrium while providing fresh viewpoint on well-known theoretical models of quantum many body systems \cite{2011RvMP...83..863P,2016RPPh...79e6001G,2016JSMTE..06.4001C}.

A paradigmatic protocol to realise out-of-equilibrium dynamics is the quantum quench \cite{2006PhRvL..96m6801C,2007JSMTE..06....8C} which has become the cornerstone of recent research in the field. The quantum quench is a sudden change in the parameters of the Hamiltonian operator determining the dynamics of the system. Here we focus on so-called global quenches when both the pre-quench and the post-quench systems are translationally invariant. On the basis of the Eigenstate Thermalisation Hypothesis \cite{1991PhRvA..43.2046D,1994PhRvE..50..888S}, it is expected that generic systems show relaxation to a thermal equilibrium \cite{2008Natur.452..854R}. Thermalisation is absent for integrable systems and they are expected to reach a steady state described by a Generalised Gibbs Ensemble (GGE) \cite{2007PhRvL..98e0405R}. However, it was discovered by studying quenches in the XXZ spin chain that  the identification of the set of conserved charges necessary to build GGE is a non-trivial task \cite{2014PhRvL.113k7202W,2014PhRvL.113k7203P,2014PhRvA..90d3625G,2014JSMTE..09..026P} requiring the construction of new `quasi-local' conserved quantities \cite{2014NuPhB.886.1177P,2015PhRvL.115o7201I,2015arXiv151204454I,2016JSMTE..06.4008I}, with similar issues found in quantum field theories \cite{2015PhRvA..91e1602E}.

Besides the eventual asymptotic steady states, the details of the relaxation process are also of considerable interest both from theoretical and experimental point of view. However, a solution for the whole time evolution after the quench is a very challenging task.  A complete exact description of the post-quench dynamics is so far only available for models that can be mapped to free particles \cite{2006PhRvL..97o6403C,2008PhRvL.101l0603S,2009EL.....8720002S,2010NJPh...12e5015F,2011PhRvL.106o6406D,2011PhRvL.106v7203C,2012JSMTE..07..016C,2012JSMTE..07..022C,2012PhRvL.109x7206E,2013PhRvL.110x5301C,2013PhRvL.110m5704H,2014JPhA...47q5002B,2014PhRvA..89a3609K,2014JSMTE..07..024S,2016PhRvA..94c1605S,2018PhRvA..97c3609C} and in conformal field theory \cite{2006PhRvL..96m6801C,2007JSMTE..06....8C}.

An important input for many approaches to the post-quench dynamics are the overlaps of the initial state with the post-quench eigenstates, including  the Quench Action approach \cite{2013PhRvL.110y7203C, 2015JPhA...48QFT01D} or form factor expansions \cite{2012JSMTE..04..017S,2014JSMTE..10..035B}. For quenches in free theories these are easily obtained using a Bogoliubov transformation, and they are also known for a number of quenches in interacting Bose gases and integrable spin chains \cite{2012JSMTE..05..021K,2014PhRvA..89c3601D,2014PhRvL.113k7202W,2014JSMTE..12..009B,2014JPhA...47n5003B,2014JSMTE..06..011P,2014JSMTE..05..006B,2014JPhA...47H5003B,2016PhRvL.116g0408P,2017JSMTE..08.3103M,2018JSMTE..05.3103P}, including results in the context of the AdS/CFT correspondence \cite{2015JHEP...08..098D,2015arXiv151202533F,2016JHEP...02..052B,2016PhLB..763..197D}.


In all of these cases multi-particle overlaps are factorised in terms of zero-momentum pairs, and recently the notion of integrable quenches was introduced to classify those quenches which admit such a structure \cite{2017arXiv170904796P,2018arXiv181211094P}. In the case of quenches in integrable quantum field theories, this factorisation would correspond to an initial state which can be written as a squeezed state in terms of the post-quench eigenstates. In certain models there are analytic methods to determine these overlaps approximately \cite{2014PhLB..734...52S,2016NuPhB.902..508H,2018JHEP...08..170H}, or alternatively, they can be obtained numerically using truncated Hamiltonian methods \cite{2017PhLB..771..539H,2018ScPP....5...27H}. When the post-quench dynamics is integrable, the knowledge of the overlaps opens the way for an analytic treatment of the dynamics \cite{2010NJPh...12e5015F,2012JSMTE..02..017S,2013PhRvL.111j0401M,2014JSMTE..10..035B,2014PhLB..734...52S,2015JSMTE..11..004S,2016NuPhB.902..508H,2016JSMTE..03.3115C,2016JSMTE..06.3102B,2016JSMTE..08.3107C,2017JSMTE..01.3103E}.

Unfortunately, analytic knowledge of these overlaps is rather restricted. However, sufficiently ``small'' quenches can be addressed using  perturbation theory which also has the promise to be able to study the effects of integrability breaking, as suggested recently \cite{2014JPhA...47N2001D,2017JPhA...50h4004D}. In this work we follow this route and work out two different perturbative expansions of post-quench overlaps. In our recent work \cite{2018ScPP....5...27H}, comparison with truncated Hamiltonian methods suggested that such an approach is indeed promising, but it also revealed some limitations, which we hope to clarify and address by  a more thorough investigation.

The overlaps are defined as the scalar product of the initial state with post-quench eigenstates which eventually suggests two different approaches depending on whether one uses perturbation expansion in the post-quench or pre-quench basis. In principle, the latter approach does not require integrability of the time evolving Hamiltonian and thus has the potential to capture overlaps for quenches breaking integrability. To benchmark these methods, we choose a specific interacting integrable model, the scaling Ising Field Theory in a magnetic field. The spectrum of the model consists of eight particle species \cite{1989IJMPA...4.4235Z} so it is sufficiently rich to test our approach in detail. It has the additional advantage that its quench dynamics can be accurately captured using truncated Hamiltonian methods \cite{2016NuPhB.911..805R,2018ScPP....5...27H} that provide a numerical verification of our analytic results.

The paper is organised in the following way. Sec. \ref{sec:pert_exps} defines the overlap functions and constructs them using perturbation theory on the post-quench basis. Sec. \ref{sec:IFT} provides a short description of the Ising Field Theory in a magnetic field, describing its spectrum and form factors. Regarding the latter, we go beyond existing results and construct a few more form factors that are necessary for our subsequent calculations. Sec. \ref{sec:TCSA} briefly summarises the Hamiltonian truncation method and compares the perturbatively constructed overlaps to the numerical results. Sec. \ref{ssec:pre_exp} discusses the perturbative expansion in the pre-quench basis and compares the results to numerics. Our conclusions and a brief outlook are presented in Sec. \ref{sec:conc}. Some technical details regarding the form factor bootstrap, the perturbative expansions, and the numerics can be found in the Appendices.

\section{Perturbation theory for the overlaps}
\label{sec:pert_exps}

\subsection{Quench overlaps}
Quantum quenches correspond to a sudden change of parameters in the Hamiltonian of a quantum system at a given time instant $t=0$ changing it from  the pre-quench ($t<0$) to the post-quench ( $t>0$) Hamiltonian. Here we are interested in global quantum quenches from the ground state, in which case both systems are translationally invariant, and the initial state at $t=0$ is the ground state of the pre-quench Hamiltonian. The initial state is an excited state of the post-quench Hamiltonian of finite energy density containing particle excitations. This can be formalised by expanding the pre-quench vacuum $\ket*{\Omega}$ on the basis of asymptotic multi-particle states of the post-quench Hamiltonian:
\begin{equation}
\label{eq:overlaps_def}
\ket*{\Omega}=\mathcal{N}\left\{\ket*{0}+\sum\limits_{N=1}^\infty \sum_{a_1,\dots a_N}\left(\prod_{a=1}^{s}\frac{1}{k^{(N)}_a!}\right)\left(\prod\limits_{i=1}^N \int\frac{\dd\vt_i}{2\pi} \right)
{K}_{a_1\dots a_N}({\vt_1,\dots ,\vt_N}) \ket*{\vt_1 \dots \vt_N}_{a_1\dots a_N}
\right\}\,,
\end{equation}
where $s$ denotes the number of particle species of the post-quench system, $k^{(N)}_a$ is the number of particles of species $a$ in the state consisting of $N$ particles, and the particle momenta are parameterised by the usual relativistic rapidity $\vartheta.$ In terms of the rapidity, the energy and momentum of a multi-particle state is given by
\begin{align}
\hat H_\text{post} \ket*{\vt_1 \dots \vt_N}_{a_1\dots a_N} &= \sum_{i=1}^N m_{a_i} \cosh(\vartheta_i)\ket*{\vt_1 \dots \vt_N}_{a_1\dots a_N} \,,\\
\hat P \ket*{\vt_1 \dots \vt_N}_{a_1\dots a_N} &= \sum_{i=1}^N m_{a_i} \sinh(\vartheta_i)\ket*{\vt_1 \dots \vt_N}_{a_1\dots a_N} \,.
\end{align}
The $K_{a_1\dots a_N}({\vt_1,\dots ,\vt_N})$ are the overlap functions, i.e. the scalar products of the post-quench eigenstates with the initial state. Our goal is to give a perturbative expansion for these functions for small quenches.

It is also possible to express the state using the cumulants ${\bar K}$ of the overlap functions:
\begin{eqnarray}
\label{eq:cumulants_overlaps_def}
\ket*{\Omega}=\mathcal{N}\exp\Bigg\{&&\sum\limits_{N=1}^\infty \sum_{a_1,\dots a_N}\left(\prod_{a=1}^{s}\frac{1}{k^{(N)}_a!}\right)\left(\prod\limits_{i=1}^N \int\frac{\dd\vt_i}{2\pi} \right)
{\bar K}_{a_1\dots a_N}({\vt_1,\dots ,\vt_N}) \nonumber\\
&& {A_{a_1}^\dagger(\vt_1) \dots A_{a_N}^\dagger(\vt_N)}
\Bigg\}\ket*{0}
\,,
\end{eqnarray}
where the $A^\dagger_a(\vartheta)$ are the asymptotic particle creation operators. When the post-quench dynamics is integrable, there exists a notion of an integrable initial state \cite{2017arXiv170904796P} which parallels the concept of integrable boundary states introduced by Ghoshal and Zamolodchikov \cite{1994IJMPA...9.3841G}. For such a state all cumulants $N>2$ vanish, and the two-particle cumulant is diagonal in the mass
\begin{equation}
{\bar K}_{a_1a_2}(\vt_1,\vt_2)=0\quad\textrm{ if }\quad m_{a_1}\neq m_{a_2}\,.
\label{eq:integrable_two_particle_overlap}\end{equation}
In this case the quench is a source of independently created particle pairs (and zero-momentum particles, whenever 
the one-particle overlap is finite), i.e. multi-particle overlaps factorise in terms of the one- and the two-particle overlap functions. This structure is important since the one- and two-particle overlaps completely characterise the finite density state resulting after the quench.

For quenches with small enough post-quench density, i.e. when the average density of particles is smaller than the inverse of the interaction range, one expects on physical grounds that a similar factorisation holds with a very good approximation. 
Such a factorisation is important since the form factor based approaches to time evolution developed in \cite{2012JSMTE..04..017S,2014JSMTE..10..035B,2017JSMTE..10.3106C} use a resummation technique to obtain predictions for relaxation times which uses the vanishing of cumulants with $N>2$. Note that since the form factor approach itself assumes a suitably small post-quench density, it is self-consistent to assume factorisation and neglect the $N>2$ cumulants even when the more stringent condition of integrability does not hold for the initial state. Therefore despite the fact that the post-quench state itself has a finite density, small quenches can be described in terms of one- and two-particle overlaps.

\subsection{Perturbation theory for the overlaps}

The overlaps can be computed by expanding the pre-quench vacuum $\ket*{\Omega}$ in the basis of post-quench eigenstates. Assuming that the pre- and post-quench Hamiltonians are related by
\begin{equation}
H_\mathrm{pre}=H_\mathrm{post}+\lambda\int \dd{x} \phi(x)
\label{eq:quench}
\end{equation}
where $\phi$ is a local field, one can use ordinary Rayleigh--Schr\"odinger perturbation theory to express the overlaps in terms of matrix elements of $\phi$:
\begin{equation}
\begin{split}
\ket*{\Omega}&=\ket*{0}-\lambda\sum_{N=1}^{\infty}\sum_{a_1,\dots a_N}\frac{2\pi}{k^{(N)}_a!}\left(\prod_{i=1}^{N}\int_{-\infty}^{\infty}\frac{\dd{\vt_i}}{2\pi} \right) \delta\left(\sum_{i=1}^{N}p_{a,i}\right)\times\\
&\times\frac{F^{\phi*}_{a_1,\dots a_N}(\vt_1,\dots ,\vt_N)}{\sum_{i=1}^{N}E_{p_{a,i}}}\ket*{\vt_1,\dots , \vt_N}_{a_1,\dots a_N}+O(\lambda^2)\,,
\end{split}
\end{equation}
where $p_{a,i}=m_{a_i}\sinh\vartheta_i$ and $E_{p_{a,i}}=m_{a_i}\cosh\vartheta_i,$ and $m_{a_i}$ is the mass of particle $i$ that is of type $a_i.$ The first order correction includes the following matrix element:
\begin{equation}
\label{eq:pert_ff}
\bra*{k}\int\dd{x}\phi(x)\ket*{0}=2\pi F^{\phi*}_{a_1,\dots a_N}(\vt_1,\dots ,\vt_N)\delta\left(\sum_{i=1}^{N}p_{a,i}\right)\,,
\end{equation}
where $\ket*{k} = \ket*{\vt_1,\dots , \vt_N}_{a_1,\dots a_N}$  and 
\begin{equation}
\label{eq:FFdef}
F_{a_1,\dots a_N}^\phi(\vt_1,\dots ,\vt_N)=\bra*{0}\phi(0)\ket*{\vt_1,\dots ,\vt_N}_{a_1,\dots a_N}\,
\end{equation}
is the $N$-particle form factor of the operator $\phi$, while the Dirac-delta corresponds to momentum conservation.

The second order contribution can be written, using Eq. \eqref{eq:pert_es} of Appendix \ref{app:pert}, as
\begin{eqnarray}
&&\ket*{\Omega^{(2)}}=\sum_{N,M=1}^{\infty}\sum_{\substack{a_1,\dots a_N \\ b_1,\dots b_M}}
\frac{(2\pi)^2}{k^{(N)}_a! k^{(M)}_b!} \left(\prod_{i=1}^{N}\int_{-\infty}^{\infty}\frac{\dd{\vt_i}}{2\pi}\right)\left(\prod_{j=1}^{M}\int_{-\infty}^{\infty}\frac{\dd{\vt'_j}}{2\pi}\right) \delta\left(\sum_{j=1}^{M}p'_{b,j}\right) \times\nonumber\\ 
&&\times\delta\left(\sum_{i=1}^{N}p_{a,i}-\sum_{j=1}^{M}p'_{b,j}\right)\frac{F^{\phi}_{\{a_i\},\{b_j\}}(\{\vt_i-\imath\pi\},\{\vt'_j\})F^{\phi*}_{\{b_j\}}(\{\vt'_j\})}{\sum_{i=1}^{n}E_{p_{a,i}}\sum_{j=1}^{m}E_{p'_{b,j}}}\ket*{\vt_1,\dots ,\vt_N}_{\{a_i\}}-\\
&&-\sum_{N=1}^{\infty}\sum_{a_1,\dots a_N}\frac{2\pi}{k^{(N)}_a!}\left(\prod_{i=1}^{N}\int_{-\infty}^{\infty}\frac{\dd{\vt_i}}{2\pi}\right)2\pi\delta(0) \delta\left(\sum_{i=1}^{N}p_{a,i}\right)\frac{\expval{\phi}F^{\phi*}_{\{a_i\}}(\vt_1,\dots ,\vt_N)}{\left(\sum_{i=1}^{N}E_{p_{a,i}}\right)^2}\ket*{\vt_1,\dots ,\vt_N}_{\{a_i\}}\,,\nonumber
\end{eqnarray}
where $a$ and $b$ indices refer to particle species, while $\{a_i\}$ and $\{\vt_i\}$ denote a set of species indices and rapidity variables, respectively. Note that this expression contains an explicit divergence $\delta(0)$ and the first form factor in the first term also has poles whenever $\vartheta'_j=\vartheta_i$ for some $i$ and $j$ with $a_i=b_j$. The perturbed state is not normalised in this convention so all quantities have to be divided by $\mathcal{N}=1+\mathcal{O}(\lambda^2)$. However, since we perform overlap calculations up to $\mathcal{O}(\lambda^2)$, the normalisation can be neglected as the leading order contribution to any overlap is of $\mathcal{O}(\lambda)$. Consequently, the divergences can not be removed by dividing with $\mathcal{N}$. Instead, as shown in the following sections, they can be regularised by switching to finite volume.

\subsection{Finite volume regularisation}
\label{ssec:finvolreg}
Divergences involving $\delta(0)$ originate from the volume integral in Eq. \eqref{eq:quench} and can be handled by switching to finite volume $L$:
\begin{equation}
\label{eq:finvol}
\int_{-\infty}^{\infty}\dd{x}\phi(x)\longrightarrow \int_0^L\dd{x}\phi(x)\,,
\end{equation}
where for simplicity we impose periodic boundary conditions. Using this prescription \eqref{eq:pert_ff} is modified as
\begin{equation}
\label{eq:finvolmel}
\bra*{k}\int\dd{x}\phi(x)\ket*{0}=L \bra*{k}\phi(0)\ket*{0}_L{|}_{p_k=0} \,,
\end{equation}
where the $L$ subscript signals that the matrix element is understood in finite volume and $p_k=0$ makes it explicit that the eigenstate $\bra*{k}$ has zero overall momentum.

When $L$ is sufficiently large, the finite volume eigenstates can still be described as multi-particle states with rapidities $\{\vt_i\}$ that are quantised according to the Bethe--Yang equations:
\begin{equation}
\label{eq:BY}
Q_i=mL\sinh\vt_i+ \sum_{j\neq i} \delta(\vt_i-\vt_j) =2\pi I_i\,,
\end{equation}
where the quantum numbers $I_i$ are integers and
\begin{equation}
\delta(\vartheta)=-\imath\log S(\vartheta)
\end{equation}
is the two-particle scattering phase-shift. As a result, the momentum integrals are replaced by discrete sums running over the quantum numbers or, equivalently, the rapidities labelling different states in finite volume. Matrix elements can be expressed by the finite volume form factor formula \cite{2008NuPhB.788..167P}:
\begin{equation}\label{eq:finvolFF}
\bra*{\{\vt_i\}}\phi\ket*{\{\vt'_j\}}_L=\frac{F^{\phi}_{\{a_i\},\{a_j\}}(\{\vt_i-\imath\pi\},\{\vt'_j\})}{\sqrt{\rho_{\{a_i\}}(\{\vt_i\})}\sqrt{\rho_{\{a_j\}}(\{\vt'_j\})}}\,,
\end{equation}
where the $\rho$ density factors are defined by the following determinant:
\begin{equation}\label{eq:rhodef}
\rho_{a_1,\dots a_N}(\vt_1, \dots \vt_N)=\det(\frac{\partial Q_k}{\partial\vt_l}),\qquad k,l=1,\dots N\,.
\end{equation}
Note that the N-particle density factor scales as $\mathcal{O}(L^N)$ with respect to the volume. Putting everything together one obtains for the perturbative expansion up to second order in finite volume
%
\begin{equation}
\begin{split}
\ket*{\Omega}_L&=\ket*{0}_L-\lambda L\sum_{N=1}^{\infty}\left(\prod_{a=1}^{{s}}\frac{1}{k^{(N)}_a!}\right)\sum_{\{\vt_1,\dots \vt_N\}} \delta_{\sum_{i=1}^{N}m_{a_i}\sinh\vt_i,0}\times\\
& \times\frac{F_{a_1,\dots a_N}^{\phi*}(\{\vt_1,\dots \vt_N\})}{\sqrt{\rho_{a_1,\dots a_N}(\vt_1,\dots \vt_N)}\sum_{i=1}^{N}m_{a_i}\cosh\vt_i}\ket*{\{\vt_1,\dots \vt_N\}}_{\{a_i\},L}+\\
& +\lambda^2\ket*{\Omega^{(2)}}_L+O(\lambda^3)\,.\label{eq:finvolvac}
\end{split}
\end{equation}
%
Although the first order correction seems to be proportional to $L,$ the volume factor is cancelled by the density factor of the finite volume states so the correction to the overlaps is finite after taking the $L\rightarrow\infty$ limit. The second order correction has the explicit form
%
\begin{equation}
\label{eq:fvsec}
\begin{split}
&\ket*{\Omega^{(2)}}_L=L^2\sum_{N,M=1}^{\infty}\sum_{\substack{a_1,\dots a_N \\ b_1,\dots b_M}}\left(\prod_{a=1}^{{s}}\frac{1}{k^{(N)}_a! k^{(M)}_b!}\right) \sum_{\{\vt_1,\dots \vt_N\}}\sum_{\{\vt'_1,\dots \vt'_M\}}\delta_{\sum_{i=1}^{M}m_{b_j}\sinh\vt'_j,0}\times\\
&\times\delta_{\sum_{i=1}^{N}m_{a_i}\sinh\vt_i, \sum_{j=1}^{M}m_{b_j}\sinh\vt'_j} 
\frac{\bra*{\{\vt'_j\}}\phi\ket*{0}_L\bra*{\{\vt_i\}}\phi\ket*{\{\vt'_j\}}_L}{\sum_{i=1}^{N}m_{a_i}\cosh\vt_i\sum_{j=1}^{M}m_{b_j}\cosh\vt'_j}\ket*{\{\vt_i\}}_{\{a_i\},L}-\\
&-L^2\sum_{N=1}^{\infty}\sum_{a_1,\dots a_N}\left(\prod_{a=1}^{{s}}\frac{1}{k^{(N)}_a!}\right) \sum_{\{\vt_1,\dots \vt_N\}} \delta_{\sum_{i=1}^{N}m_{a_i}\sinh\vt_i,0}\frac{\expval{\phi}_L\bra*{\{\vt_i\}}\phi\ket*{0}_L}{\left(\sum_{i=1}^{N}m_{a_i}\cosh\vt_i\right)^2}\ket*{\{\vt_i\}}_{\{a_i\},L}\,.
\end{split}
\end{equation}
%
The divergence appears in the last term, since the density factor coming from the finite volume matrix element is insufficient to suppress the $L^2$ factor. However, the divergence is cancelled by the disconnected part of the finite volume form factor, as shown below.

\subsection{Results for one- and two-particle overlaps}
\label{ssec:post_exp}

\subsubsection{One-particle overlaps}
One-particle overlaps are obtained by taking the scalar product of the pre-quench vacuum with a post-quench eigenstate containing a single stationary particle:
\begin{equation}
\frac{g_i}{2}=\mathbin{^i\braket*{\{0\}}{\Omega}}\,,
\end{equation}
where $\mathbin{^i\bra*{\{0\}}}$ denotes a single-particle state of species $i$ with zero rapidity. From Eq. \eqref{eq:overlaps_def} the single-particle overlap can be expressed as
\begin{equation}
\frac{g_i}{2}=K_i(0)\,.
\end{equation}
For simplicity we suppress the species index here, the generalisation to the case of multiple species is presented later. The amplitude $g$ is connected to the finite volume scalar product as \cite{2010JHEP...04..112K}:
\begin{equation}
\label{eq:onepnorm}
\frac{g}{2}=\frac{\braket*{\{0\}}{\Omega}_L}{\sqrt{mL}}\,,
\end{equation}
where the factor of $1/\sqrt{mL}$ results from the relation between the normalisation of the infinite and finite volume states. Using  \eqref{eq:finvolvac} we can read off the result of perturbation theory up to the second order:
\begin{align}
\label{eq:onep01}
\frac{g}{2}=&\frac{1}{\sqrt{mL}}\left[-\lambda L\frac{F_1^\phi}{m\sqrt{mL}}+\lambda^2L^2\left(-\frac{\expval{\phi} F_1^\phi}{m^2\sqrt{mL}}+\right.\right.\\\nonumber&
\left.\left.+\frac{F_1^\phi F_2^\phi(\imath\pi,0)}{m^2\sqrt{mL}\sqrt{mL}\sqrt{mL}}+\frac{\expval{\phi} F_1^\phi}{m^2\sqrt{mL}}+
\dots \right)+ O\left(\lambda^3\right)\right]\,,
\end{align}
where the ellipsis denote the contribution of higher multi-particle form factors and we used the results for finite volume form factors derived in 
\cite{2008NuPhB.788..167P,2008NuPhB.788..209P}:
\begin{equation}
\label{eq:onepnorm01}
\braket*{\{0\}|\phi}{0}_L=\frac{F_1^\phi}{\sqrt{mL}}\,,\qquad \braket*{\{0\}|\phi}{\{0\}}_L=\frac{F_2^\phi(\imath\pi,0)}{\sqrt{mL}\sqrt{mL}}+\expval{\phi}\,.
\end{equation}
Note that the diagonal form factor includes a disconnected contribution, which exactly cancels the divergent term appearing in the second order of perturbation theory. Eq. \eqref{eq:onep01} can be simplified in the form:
\begin{equation}
\label{eq:onep02}
\frac{g}{2}=-\lambda \frac{F_1^\phi}{m^2}+\lambda^2\left(\frac{F_1^\phi F_2^\phi(\imath\pi,0)}{m^4}+\dots \right)+ O\left(\lambda^3\right)\,,
\end{equation}
where the ellipses denote contributions from higher form factors, which can also be evaluated. The first such term corresponds to the $\{\vt'_j \}=\{-\vt_1,\vt_1\}$ term in Eq. \eqref{eq:fvsec} and reads
\begin{equation}
\frac{\lambda^2L^2}{2}\sum_{\vt_1}\frac{F_3^\phi(\imath\pi,-\vt_1,\vt_1) F_2^{\phi*}(-\vt_1,\vt_1)}{mL\rho_2(\vt_1,-\vt_1)2m^2\cosh\vt_1}\,,
\end{equation}
where $\rho_2$ is the density factor defined in \eqref{eq:rhodef} and overall momentum conservation was used to eliminate one of the rapidity summations. In the infinite volume limit the summation is transformed into an integral 
\begin{equation}
\sum_{\vt_1}\rightarrow\int\frac{\dd\vt}{2\pi}\tilde{\rho}(\vt)\,,
\end{equation}
where $\tilde{\rho}(\vt)$ is the density of zero-momentum states, which can be obtained by enforcing zero overall momentum on the Bethe--Yang equations \eqref{eq:BY}:
\begin{equation}
\label{eq:rho1}
\tilde{\rho}_1(\vt_1)=\left. \frac{\partial Q_1(\vt_1,\vt_2)}{\partial \vt_1}\right|_{m\sinh\vt_1+m\sinh\vt_2=0}\,.
\end{equation}
The quotient of these two density factors yields
\begin{equation}
\frac{\tilde{\rho}(\vt)}{\rho_2(\vt,-\vt)}=\frac{1}{mL\cosh\vt}\,.
\end{equation}
and so the powers of $L$ cancel, leading to the infinite volume limit
\begin{equation}
\label{eq:onepFFcont}
\frac{\lambda^2}{2}\int\frac{\dd\vt}{2\pi}\frac{F_3^\phi(\imath\pi,-\vt,\vt) F_2^{\phi*}(-\vt,\vt)}{2m^4\cosh^2\vt}\,.
\end{equation}
This expression contains another possible source of divergence due to kinematic poles of the form factors \eqref{eq:FFkpol} since $F_3(\imath\pi,-\vt,\vt)$ has a simple pole for $\vt\rightarrow0$. However, $F_2\propto\vt$ for $\vt\ll 1$ due to the two-particle $S$-matrix satisfying $S(0)=-1$ (this reflects an effective exclusion statistics satisfied by these particles, see Eq. \eqref{eq:ffeq_exchange} below). So the product is regular and the integral is well-defined. Adding this term to Eq. \eqref{eq:onep02} results in:
\begin{equation}
\label{eq:onep03}
\frac{g}{2}=-\lambda \frac{F_1^\phi}{m^2}+\lambda^2\left(\frac{F_1^\phi F_2^\phi(\imath\pi,0)}{m^4}+\frac{1}{2}\int\frac{\dd\vt}{2\pi}\frac{F_3^\phi(\imath\pi,-\vt,\vt) F_2^{\phi*}(-\vt,\vt)}{2m^4\cosh^2\vt}+\dots \right)+ O\left(\lambda^3\right)\,.
\end{equation}
The above considerations can easily be generalised to a theory with multiple particle species, resulting in the following expression for the overlap of a particle of species $a$
\begin{equation}
\begin{split}
\frac{g_a}{2}&= -\lambda \frac{F_{a}^\phi}{m_a^2}+\lambda^2\Bigg(\sum_{b=1}^{s}\frac{F_b^\phi F_{ab}^\phi(\imath\pi,0)}{m_a^2 m_b^2}+\\
&\sum_{b\leq c}\frac{1}{(2\delta_{bc})!}\int\frac{\dd\vt}{2\pi}\ \frac{F_{abc}^\phi(\imath\pi,\vt,\vt_{bc}) F_{bc}^{\phi*}(\vt,\vt_{bc})}{m_a^2\left(m_b\cosh(\vt)+\sqrt{m_c^2+(m_b\sinh\vt)^2}\right)\sqrt{m_c^2+(m_b\sinh\vt)^2}}\\
&+\text{ conributions from higher form factors}\Bigg)+ O\left(\lambda^3\right)\,,
\label{eq:onepmultspec}
\end{split}
\end{equation}
with
\begin{equation}
\label{eq:vtbc}
\vt_{bc}=-\text{arcsinh}\left(\frac{m_b\sinh\vt}{m_c}\right)\,,
\end{equation}
which is a straightforward generalisation of Eq. \eqref{eq:onep03}. Note that this expression is regular, since $F_{aac}$ does not have a kinematic pole in the case $a\neq c$ for $\vt\rightarrow0$ due  to $S_{ac}(0)=+1$ for two different species $a$ and $c$.

\subsubsection{Two-particle overlaps}

The next term in the expansion \eqref{eq:overlaps_def} corresponds to two-particle states with zero total momentum. In the case of a single particle species, their contribution is characterised by a single rapidity-dependent overlap function $K(-\vt,\vt)$:
\begin{equation}
K(-\vt,\vt)=\braket*{\{\vt,-\vt\}}{\Omega}\,,
\end{equation}
which is related to the corresponding finite volume inner product by the relation \cite{2010JHEP...04..112K}:
\begin{equation}
\label{eq:twopnorm}
K(-\vt,\vt)=\frac{\tilde{\rho}(\vt)\braket*{\{\vt,-\vt\}}{\Omega}_L}{\sqrt{\rho_2(\vt,-\vt)}}\,.
\end{equation}
The perturbative contributions to $K(-\vt,\vt)$ can be easily read off from the $n=2$ terms in Eq. \eqref{eq:finvolvac}:
\begin{equation}
\label{eq:twop01}
\begin{split}
K(-\vt,\vt)&=\frac{\tilde{\rho}(\vt)}{\sqrt{\rho_2(\vt,-\vt)}} \Bigg[   -\lambda L\frac{F_2^{\phi*}(-\vt,\vt)}{2m\cosh\vt\sqrt{\rho_2(\vt,-\vt)}}+ \\
& +\lambda^2L^2\left(\frac{-\expval{\phi} F_2^{\phi*}(-\vt,\vt)} {(2m\cosh\vt)^2\sqrt{\rho_2(\vt,-\vt)}}+\frac{F_1^\phi F_3^{\phi*}(\imath\pi,-\vt,\vt)}{2m^2\cosh\vt\sqrt{mL}\sqrt{mL}\sqrt{\rho_2(\vt,-\vt)}}+\dots \right)\\
& + O\left(\lambda^3\right)
\Bigg]
\,,
\end{split}
\end{equation}
where the ellipses again correspond to higher multi-particle form factor contributions. Using 
\begin{equation}
\label{eq:densnorms}
\frac{\tilde{\rho}(\vt)}{\rho_2(\vt,-\vt)}=\frac{1}{mL\cosh\vt}\,
\end{equation}
we obtain
\begin{equation}
\label{eq:twop02}
\begin{aligned}
K(-\vt,\vt)= -\lambda \frac{F_2^{\phi*}(-\vt,\vt)}{2m^2\cosh^2\vt}+
\lambda^2L\left( -\frac{\expval{\phi} F_2^{\phi*}(-\vt,\vt)} {4(m\cosh\vt)^3}+\frac{F_1^\phi F_3^{\phi*}(\imath\pi,-\vt,\vt)}{2m^3\cosh^2\vt mL}+\dots \right)+ O\left(\lambda^3\right)\,,
\end{aligned}
\end{equation}
The above expression contains an apparent infinite-volume divergence; similarly to  the one-particle case, it is expected to cancel with the disconnected piece of the next term in the form factor expansion. To verify this we consider the next term in the form factor expansion:
\begin{equation}
\label{eq:FFtwop}
\sum_{\vt'>0}\frac{\braket*{\{\vt,-\vt\}}{\phi|\{-\vt',\vt'\}}_L \braket*{\{\vt',-\vt'\}}{\phi|0}_L}{4m^2\cosh\vt\cosh\vt'}\,,
\end{equation}
where a disconnected term appears for $\vt=\vt'$. In this case the relation between the finite volume matrix element and the infinite volume form factors reads as \cite{2008NuPhB.788..209P}:
\begin{align}
\label{eq:disctwop}
\braket*{\{\vt,-\vt\}}{\phi|\{-\vt,\vt\}}_L=&\frac{1}{\rho_2(\vt,-\vt)} \left(F_4^{\phi,s}(\imath\pi+\vt,\imath\pi-\vt,-\vt,\vt)+\right.\\\nonumber&\left.+ 2mL\cosh\vt F_2^\phi(\imath\pi,0)+ \rho_2(\vt,-\vt)\expval{\phi}\right)\,,
\end{align}
where the superscript $s$ denotes that the form factor is evaluated symmetrically at $\vt=\vt'$, i.e.
\begin{equation}
\label{eq:FFsym}
F_4^{\phi,s}(\imath\pi+\vt,\imath\pi-\vt,-\vt,\vt)=\lim_{\epsilon\rightarrow0}F_4^{\phi}(\imath\pi+\vt+\epsilon,\imath\pi-\vt+\epsilon,-\vt,\vt)\,,
\end{equation}
which is a regular expression \cite{1996NuPhB.473..469D,2008NuPhB.788..209P}. Note that the last term exactly cancels term proportional to $L$ in \eqref{eq:twop02}. The final result for the two-particle overlap is
\begin{equation}
\label{eq:twop03}
\begin{aligned}
K(-\vt,\vt)=
    -\lambda \frac{F_2^{\phi*}(\vt,-\vt)}{2m^2\cosh^2\vt}+
\lambda^2\left(\frac{F_1^\phi F_3^{\phi*}(\imath\pi,-\vt,\vt)}{2m^4\cosh^2\vt}+ \frac{F_2^{\phi*}(-\vt,\vt)F_2^\phi(\imath\pi, 0)}{2m^4\cosh^4\vt}+\right.\\\left.
+\frac{1}{2}\int_{-\infty}^{\infty}\frac{\dd{\vt'}}{2\pi} \frac{F_4^{\phi,s}(\imath\pi+\vt,\imath\pi-\vt,-\vt',\vt')F_2^{\phi*}(-\vt',\vt')}{4m^4\cosh^2\vt\cosh^2\vt'} + \dots \right)+ O\left(\lambda^3\right)\,,
\end{aligned}
\end{equation}
The generalisation to multiple particle species is not as straightforward as in the case of one-particle overlaps. The new feature is that in infinite volume divergent disconnected pieces appear even when there is only one particle that appears with the same rapidity on both sides. However, this cannot happen in finite volume, due to the quantisation of rapidities according to Eq. \eqref{eq:BY} \cite{2008NuPhB.788..209P}. Consequently, it is necessary to be more careful when taking the limit $L\rightarrow\infty$, cf.  Appendix \ref{app:twopcalcs}.

The final result is	
%
\begin{multline}
\label{eq:Kaafin}
K_{aa}(-\vt,\vt)=
 -\lambda \frac{F_{aa}^{\phi*}(-\vt,\vt)}{2m_a^2\cosh^2\vt}+
\lambda^2\Bigg(\sum_{b=1}^{N_{\text{spec}}}\frac{F_b^\phi F_{baa}^{\phi*}(\imath\pi,-\vt,\vt)}{2m_a^2\cosh^2\vt m_b^2} + \\
+\frac{F_{aa}^{\phi*}(-\vt,\vt)F_{aa}^\phi(\imath\pi, 0)}{2m_a^4\cosh^4\vt}+\sum_{b=1}^{N_{\text{spec}}}D_{ab}(\vt,-\vt) + \\
+\sum_{(c,d)\neq(a,b)}\frac{1}{(2\delta_{cd})!}\int_{-\infty}^{\infty}\frac{\dd{\vt'}}{2\pi} \frac{F_{aacd}^{\phi,s}(\imath\pi+\vt,\imath\pi-\vt,\vt',\vt'_{cd})F_{cd}^{\phi*}(\vt',\vt'_{cd})}{2m_a^2\cosh^2\vt(m_c\cosh\vt'+m_d\cosh\vt'_{cd})m_d{\cosh\vt'_{cd}}} + \dots \Bigg)\\
 + O\left(\lambda^3\right)\,,
\end{multline}
for a pair composed of two particles in the same species $a$, and
%
\begin{multline}
K_{ab}(\vt,\vt_{ab})=
-\lambda \frac{F_{ab}^{\phi*}(\vt,\vt_{ab})}{C_{ab}(\vt,\vt_{ab})}
+\lambda^2 \Bigg[
\sum_{c=1}^{N_{\text{spec}}}\frac{F_c^\phi F_{cab}^{\phi*}(\imath\pi,\vt,\vt_{ab})}{m_c^2C_{ab}(\vt,\vt_{ab})}+\\
+\frac{F_{ab}^{\phi*}(\vt,\vt_{ab})}{C_{ab}(\vt,\vt_{ab})(m_a\cosh\vt+m_b\cosh\vt_{ab})}\left(\frac{F_{aa}^\phi(\imath\pi, 0)}{m_b\cosh\vt_{ab}}+\frac{F_{bb}^\phi(\imath\pi, 0)}{m_a\cosh\vt}\right)+\\[1em]
+G_{ab}^a(\vt,\vt_{ab})+G_{ab}^b(\vt,\vt_{ab})+\\[0.5em]
+\sum_{\substack{c\neq a,b \\ d\neq a,b}} \frac{1}{(2\delta_{cd})!}\int_{-\infty}^{\infty}\frac{\dd{\vt'}}{2\pi}  \frac{F_{abcd}^{\phi,s}(\imath\pi+\vt,\imath\pi+\vt_{ab},\vt',\vt'_{cd})F_{cd}^{\phi*}(\vt',\vt'_{cd})}{C_{ab}(\vt,\vt_{ab})(m_c\cosh\vt'+m_d\cosh\vt'_{cd})m_d\cosh\vt'_{cd}}+ \dots 
\Bigg]
\\
+ O\left(\lambda^3\right)\,,
\label{eq:Kabfin}
\end{multline}
for a pair composed of two particles in different species $a\neq b$, 
with the functions $C$, $D$, $G$ defined in Appendix  \ref{app:twopcalcs}.

\section{Testing ground: the $E_8$ Ising field theory}
\label{sec:IFT}

Now we turn to putting our approach to quench overlaps to the test, for which we need a model which satisfies the following important criteria. First, it must be rich enough to test all aspects of our results; second, there must be enough information about its spectrum and form factors for the evaluation of the analytic expressions for the overlaps and third, it must be amenable to an effective alternative treatment.	

\subsection{Action and particle spectrum}
The model we consider is the scaling limit of critical Ising quantum spin chain in an external magnetic field, which can also be obtained as a relevant perturbation of the $c=1/2$ Conformal Field Theory a.k.a. the massless free Majorana fermion:
\begin{equation}
\label{eq:A_op}
\mathcal{A}=\mathcal{A}_{\text{CFT},\: c=1/2}+h\int \sigma(x)d^2x\,,
\end{equation}
where $\sigma$ is the magnetisation field, which is a primary field of conformal weight $(1/16,1/16)$. This model is integrable and its spectrum consists of eight stable particles \cite{1989IJMPA...4.4235Z} (labelled as $A_i,\:\: i=1,..,8$), with the following masses \cite{1994PhLB..324...45F}
\begin{align}
\label{eq:e8spectrum}
m_2 =  2m_1\cos\frac{\pi}{5}\:,\quad m_3 =  2&m_1\cos\frac{\pi}{30}\:,\quad m_4 =  2m_2\cos\frac{7\pi}{30}\:,\quad m_5 =  2m_2\cos\frac{2\pi}{15}\:,\nonumber\\
m_6 =  2m_2\cos\frac{\pi}{30}\:,\quad m_7 &=  4m_2\cos\frac{\pi}{5}\cos\frac{7\pi}{30}\:,\quad m_8 =  4m_2\cos\frac{\pi}{5}\cos\frac{2\pi}{15}\:,
\end{align}
where all masses are expressed in terms of the mass gap $m_1$ (the mass of the lightest particle $A_1$) which is related to the coupling $h$ as
\begin{equation}
\label{eq:e8gap}
m_1=(4.40490857\dots )|h|^{8/15}.
\end{equation}
The fundamental two-particle $S$-matrix describing the scattering process $A_1A_1\rightarrow A_1A_1$ is given by
\begin{equation}
S_{11}(\vartheta)=\left(\frac{2}{3}\right)_\vartheta \left(\frac{2}{5}\right)_\vartheta \left(\frac{1}{15}\right)_\vartheta\qquad \left(x\right)_\vartheta=\frac{\sinh\vartheta+i\sin\pi x}{\sinh\vartheta-i\sin\pi x}\,,
\label{eq:ordSmat}
\end{equation}
while those involving higher ones can be computed using the $S$-matrix bootstrap  \cite{1990IJMPA...5.1025F,1996NuPhB.473..469D}. The above spectrum is rich enough to test our results in detail, thus satisfying our first condition stated above.

Another consequence of integrability is that the elementary form factors 
\begin{equation}
\label{eq:FFdefmp}
F_{i_1\dots i_n}^\mathcal{O}(\vt_1,\dots ,\vt_n)=\bra*{0}\mathcal{O}(0,0)\ket*{\vt_1,\dots ,\vt_n}_{i_1\dots i_n}\,,
\end{equation}
can be determined using the form factor bootstrap, for a review of which the interested reader is referred to \cite{1992ASMP...14.....S}. Matrix elements of the local operator between two general multi-particle states can be obtained by crossing symmetry from the elementary ones. This provides enough information to evaluate our expressions for the overlaps, satisfying our second condition.

The third condition is fulfilled by the fact that the quench dynamics of the model is captured to a high precision by the Truncated Conformal Space Approach \cite{2016NuPhB.911..805R,2018ScPP....5...27H}. 		

\subsection{Digression: form factor bootstrap in the $E_8$ field theory}
\label{sec:FFbs}

Due to integrability, the form factors of local operators of the $E_8$ field theory satisfy the so-called form factor bootstrap equations \cite{1992ASMP...14.....S} which use the exact $S$-matrix as an input. For completeness we give a list of these equations for the case of a non-degenerate mass spectrum (corresponding to a scattering theory diagonal in species space):
\begin{enumerate}
\item \emph{Lorentz invariance}
\begin{equation}
F_{n}^{\phi}(\vt_1+\lambda,\dots \vt_n+\lambda)=e^{s_\phi\lambda}F_{n}^{\phi}(\vt_1,\dots \vt_n)\,,
\label{eq:ffeq_lorentz}\end{equation}
where $s_\phi$ is the Lorentz spin of the local field $\phi$, which is zero for Lorentz scalars.
\item \emph{Exchange property}
\begin{equation}
F_{n}^{\phi}(\vt_1,\dots \vt_j,\vt_{j+1},\dots \vt_n)=S(\vt_j-\vt_{j+1})F_{n}^{\phi}(\vt_1,\dots \vt_{j+1},\vt_j,\dots \vt_n)\,.
\label{eq:ffeq_exchange}\end{equation}
\item \emph{Cyclic property}
\begin{equation}
F_{n}^{\phi}(\vt_1,\dots \vt_j,\vt_{n-1},\vt_n +2\pi i)=F_{n}^{\phi}(\vt_n,\vt_1,\dots \vt_{n-1})\,.
\label{eq:ffeq_cyclic}\end{equation}
\item \emph{Kinematical singularities}
\begin{equation}
\label{eq:FFkpol}
-\imath \lim_{\tilde{\vt}\rightarrow\vt}(\tilde{\vt}-\vt) F_{n+2}^{\phi}(\tilde{\vt}+\imath\pi,\vt,\vt_1,\vt_2,\dots \vt_n)=\left(1-\prod_{i=1}^{n}S(\vt-\vt_i)\right) F_{n}^{\phi}(\vt_1,\vt_2,\dots \vt_n)\,.
\end{equation}
\item \emph{Bound state singularities}
\begin{equation}
\label{eq:FFboundpol}
-\imath \lim_{\vt_{ab}\rightarrow\imath u_{ab}^c}(\vt_{ab}-\imath u_{ab}^c) F_{n+2}^{\phi}(\vt_a,\vt_b,\vt_1,\vt_2,\dots \vt_n)=\Gamma_{ab}^c F_{n+1}^{\phi}(\vt_c,\vt_1,\vt_2,\dots \vt_n)\,,
\end{equation}
with $\vt_{ab}=\vt_a-\vt_b$ and $\vt_c=\vt_a-\imath(\pi-u_{bc}^a)=\vt_b+\imath(\pi-u_{ac}^b)$, where $u_{ab}^c$ is the position of the bound state pole corresponding to the occurrence of particle $A_c$ in the scattering of $A_a$ and $A_b$:
\begin{equation}
\label{eq:Smatbpole}
S_{ab}(\vt\sim\imath u_{ab}^c)\sim \frac{\imath (\Gamma_{ab}^c)^2}{\vt-\imath u_{ab}^c}\,.
\end{equation}
\end{enumerate}	
In the above equations we suppressed the species indices which can easily be restored when needed.

For the form factors containing only the lightest species $A_1$, the first three conditions can be satisfied by considering an Ansatz 
\begin{equation}
\label{eq:FFAnsmain}
F_{n}^{\phi}(\vt_1,\vt_2,\dots \vt_n)=H_n\frac{\Lambda_n(x_1,\dots ,x_n)}{(x_1\dots x_n)^n} \prod_{i<j}^{n}\frac{F_{11}^{\text{min}}(\vt_i-\vt_j)}{D_{11}(\vt_i-\vt_j)(x_i+x_j)}
\end{equation}
with $x_i\equiv e^\vt_j$. $H_n$ is a constant factor, $\Lambda_n$ is a symmetric polynomial, the $D_{11}$ factors ensure the correct positions of the bound state poles, while kinematic poles are included in $(x_i+x_j)$ factors. The function $F^\text{min}_{11}$ is the so-called minimal form factor that have no poles in the strip $\Im\vt\in[0,2\pi]$ and satisfies
\begin{equation}
 F_{11}^{\text{min}}(\imath\pi-\vt)=F_{11}^{\text{min}}(\imath\pi+\vt)\qquad
 F_{11}^{\text{min}}(\vt)=S_{11}(\vt)=F_{11}^{\text{min}}(-\vt)\,.
\end{equation}
The kinematical and bound state equations then yield recurrence relations for the polynomials $\Lambda_n$ which can be solved inductively in particle number $n$. For the $E_8$ model, solutions of these equations were previously constructed in \cite{1996NuPhB.473..469D} and later in \cite{2006NuPhB.737..291D}; the latter article is accompanied with an explicit set of form factors at \cite{IsingFF}. The available functions have been used in a number of works \cite{2008NuPhB.788..167P,2018ScPP....5...27H}, providing strong evidence that they are correct. Form factors involving higher species can be constructed by repeated use of the bound state singularity equation.

However, for our purposes we need to construct more form factors than available in the above sources. Unfortunately, the bound state recursive equation appearing in \cite{1996NuPhB.473..469D} has some misprints. Therefore for the sake of clarity we present a detailed re-derivation of the bound state recursion formula in Appendix  \ref{app:FF1brec} using the Ansatz \eqref{eq:FFAnsmain}. The result is
\begin{equation}
\label{eq:bFFres}
\frac{\Lambda_{n+2}(x e^{\imath\pi/3},x e^{-\imath\pi/3},x_1,\dots ,x_n)}{x^4\prod_{i=1}^{n}(x-e^{-11\imath\pi/15}x_j)(x-e^{11\imath\pi/15}x_j)(x+x_j)}=(-1)^n\Lambda_{n+1}(x,x_1,\dots ,x_n)
\end{equation}
with
\begin{equation}
\label{eq:Hres}
\frac{H_{n+2}}{H_{n+1}}=\frac{\Gamma_{11}^1}{2\cos[2](\pi/3)\cos[2](\pi/5)\cos[2](\pi/30)G_{11}(2\pi\imath/3)}\left[\frac{\sin[2](11\pi/30)\gamma}{4\cos[2](\pi/3)\cos[2](\pi/5)\cos[2](\pi/30)}\right]^n\,.
\end{equation}
Once the constants $H_n$ are fixed, a similar equation can be derived from the kinematic singularity equation \eqref{eq:FFkpol}. The final expression is \cite{1996NuPhB.473..469D}
\begin{equation}
\label{eq:kFFrec}
(-1)^n \Lambda_{n+2}(-x,x,x_1,\dots ,x_n)=\mathcal{A}_n U(x,x_1,\dots ,x_n)\Lambda_n(x_1,\dots ,x_n)
\end{equation}
with
\begin{align}
U(x,x_1,\dots ,x_n)=\frac{1}{2}x^5\sum_{k_1,k_2,\dots ,k_6=0}^{n}(-1)^{k_1+k_3+k_5} x^{6n-(k_1+\dots +k_6)}\\
\sin(\frac{\pi}{15}(10(k_1-k_2)+6(k_3-k_4)+(k_5-k_6)))\omega_{k_1}\dots \omega_{k_6}\,,
\end{align}
and
\begin{equation}
\mathcal{A}_n=\frac{4\gamma\sin ^2\left(\frac{11 \pi }{30}\right) \left(\cos \left(\frac{\pi }{3}\right) \cos \left(\frac{\pi }{5}\right) \cos \left(\frac{\pi }{30}\right)\right)^2 \left(G_{11}\left(\frac{2 \pi  \imath}{3}\right)\right)^2}{\left(\Gamma_{11}^1 \sin \left(\frac{2 \pi }{15}\right) \sin \left(\frac{11 \pi }{30}\right) \sin \left(\frac{8 \pi }{15}\right) \sin \left(\frac{3 \pi }{10}\right)\right)^2} \left(\frac{\sin \left(\frac{2 \pi }{3}\right) \sin \left(\frac{2 \pi }{5}\right) \sin \left(\frac{\pi }{15}\right)}{8 \sin ^4\left(\frac{11 \pi }{30}\right) G_{11}(0)\gamma^2}\right)^n\,.
\end{equation}
Following the procedure outlined in \cite{1996NuPhB.473..469D}, these two equations can be used to obtain many-particle form factors of the lightest particle, from which form factors containing heavier species can be constructed. An example calculation is presented in Appendix  \ref{app:FFbdown}; in fact, the computation process can be automated using a software capable of symbolic processing such as e.g. \texttt{Wolfram Mathematica}. A useful shortcut is provided by including bound state singularities involving two particles of either species $A_2$ or $A_3$. Using this approach we constructed several new exact form factors which can be found in the following \texttt{Mathematica} file;\footnote{See the Supplementary Material; and also the ancillary file attached to the arXiv preprint.} it is an extension of the results available in \cite{IsingFF}.

\section{Comparison with TCSA}

In this section we compare the  results of the  perturbation theory calculation performed in the post-quench basis with the numerical overlaps extracted from the TCSA simulations.

\label{sec:TCSA}

\subsection{Method and notations}
The quenches considered in this section are governed by the following action:
\begin{equation}
\label{eq:A_quench}
\mathcal{A}=\mathcal{A}_{CFT,\: c=1/2}-h_i\int \sigma(x)d^2x-(h_f-h_i)\int\sigma(x)\Theta(t)d^2x\,,
\end{equation}
which corresponds to a sudden change of $h$ from the initial $h_i$ to the final $h_f$ at $t=0$, expressed by the Heaviside function $\Theta(t)$. For $t\leq 0$ the system is in the ground state of the pre-quench Hamiltonian, which is the initial state of the out-of-equilibrium time evolution which happens for $t>0$.

To obtain a non-perturbative description of the quantum quench, we use the Truncated Conformal Space Approach (TCSA) developed by Yurov and Zamolodchikov \cite{1990IJMPA...5.3221Y,1991IJMPA...6.4557Y}, which was shown to be an effective tool to describe the overlaps \cite{2017PhLB..771..539H} and the time evolution \cite{2016NuPhB.911..805R,2018ScPP....5...27H} after a quantum quench in perturbed conformal (or free) field theories, including the Ising Field Theory considered in the present work.

The method is based on the numerical diagonalisation of the finite volume Hamiltonian matrix in the unperturbed basis, in our case in the conformal basis of the Ising CFT. In the TCSA the finite volume matrix elements of the perturbing operator can be calculated exactly using the conformal Ward identities \cite{2007NuPhB.772..227K}, after mapping the space-time cylinder to the conformal plane. For a detailed review of truncated space methods the interested reader is referred to Ref. \cite{2018RPPh...81d6002J}. The Hamiltonian matrix corresponding to the action Eq. \eqref{eq:A_op} for $t>0$ in finite volume $R$ with periodic boundary conditions can be written on the plane in the following dimensionless form 
\begin{equation}
H/m_1=(H_\text{CFT}+H_\sigma)/m_1=\frac{2\pi}{r}\left(L_0+\bar{L}_0-c/12+ \tilde h \frac{r^{2-x_\sigma}}{(2\pi)^{1-x_\sigma}}M_\sigma\right)\,.
\label{eq:H_mat}
\end{equation}
Here $r=m_1R$ is the dimensionless volume and $\tilde h = h m_1^{x_\sigma-2}$ is the dimensionless magnetic field [cf.  Eq. \eqref{eq:e8gap}]. This means that all quantities are measured in the units of the gap $m_1$ of the above Hamiltonian. $x_\sigma=1/8$ is the scaling dimension of the field $\sigma$, $L_0$ and $ {\bar L}_0$ are the standard Virasoro generators, $c=1/2$ is the central charge of the Ising CFT,  and $M_\sigma$ is the matrix of the perturbing operator. In finite volume the spectrum is discrete, and the Hilbert space is truncated such that only states having energy lower than a given cut-off $E_\text{cut}$ are kept. The truncation is carried out on the level of the conformal field theory spectrum where it is parameterised by the maximal conformal level, 
\begin{equation}
N_\text{cut}=\frac{R}{2\pi} E_\text{cut}\,.
\end{equation}
Overlaps are computed by first determining the ground state of the pre-quench Hamiltonian with $\tilde h=\tilde h_i$ in \eqref{eq:H_mat}, and then taking its scalar product with the eigenvectors of the post-quench Hamiltonian with $\tilde h_f$. 
This corresponds to the quench protocol described by Eq. \eqref{eq:A_quench} which can be characterised by the dimensionless quench magnitude
\begin{equation}
\label{eq:xidef}
\xi\equiv\frac{h_f-h_i}{h_f}\,.
\end{equation}
We measure everything in appropriate powers of the post-quench mass gap $m_1,$ so the perturbative parameter $\lambda$ of Eqs. \eqref{eq:onepmultspec},\eqref{eq:Kaafin},\eqref{eq:Kabfin} is obtained by multiplying $\xi$ with the post-quench parameter $h_f$.
All the TCSA calculations are performed in finite volume, so in order to compare the numerical results to the perturbative predictions it is convenient use the finite volume normalisation of Eqs. \eqref{eq:onepnorm} and \eqref{eq:twopnorm}.

The accuracy of the results obtained from TCSA can be improved by extrapolating in the cut-off.  The results have a power-law dependence in $1/N_\text{cut}$ with the leading and subleading exponents determined by the conformal weights and the operator product algebra of the perturbing operator(s) \cite{2016NuPhB.911..805R,2018ScPP....5...27H,2015JHEP...09..146L,2019JHEP...01..177L}. For the $E_8$ field theory, the leading dependence of overlaps on the cut-off level can be expressed as
\begin{equation}
\label{eq:epol}
\braket*{n}{\Omega}_\text{TCSA}=\braket*{n}{\Omega}+A_n N_\text{cut}^{-7/4}+ B_n N_\text{cut}^{-11/4}+\dots\, .
\end{equation}
The extrapolation was performed by first identifying the eigenstates at various cut-off levels by comparing their energy and matrix elements obtained from TCSA with results following from the Bethe--Yang equations \eqref{eq:BY}, and the finite volume form factor formalism \cite{2008NuPhB.788..167P,2008NuPhB.788..209P}. For a state with some fixed multi-particle content and quantum numbers, the dependence of its overlap with the pre-quench ground state was then fitted the with function in Eq. \eqref{eq:epol}, and the extrapolated result was identified as the constant term of the best fitting function. This procedure worked remarkably well, however in some cases it was not possible to eliminate the cut-off dependence of the overlaps, especially for two-particle states as discussed later.

Let us remark that the overlap functions are defined up to a phase factor, since we can freely choose the phase of any quantum state. The TCSA uses a basis in which all vectors are real, consequently the overlaps obtained from this approach are also real. Thus the comparison is performed such that we take the absolute value of Eqs. \eqref{eq:onepmultspec},\eqref{eq:Kaafin},\eqref{eq:Kabfin}.

Before turning to the discussion of the comparison with TCSA calculations, let us comment on the numerical evaluation of the perturbative formulae. We observe that the contribution from the terms involving an integral over the momentum of a pair state is very small in most of the cases. Consequently, we argue that the error we make by truncating the form factor expansion at two-particle intermediate states is orders of magnitude smaller than the main contributions in second order. This argument is supported by Appendix  \ref{app:tables}.

\subsection{Perturbative expansion against TCSA overlaps}

Now we turn to the actual comparison of the perturbative results with the TCSA data for the one-particle and two-particle overlaps.

\subsubsection{One-particle overlaps}

\begin{figure}[t!]
\begin{tabular}{cc}
\begin{subfloat}
{\includegraphics[width=0.45\textwidth]{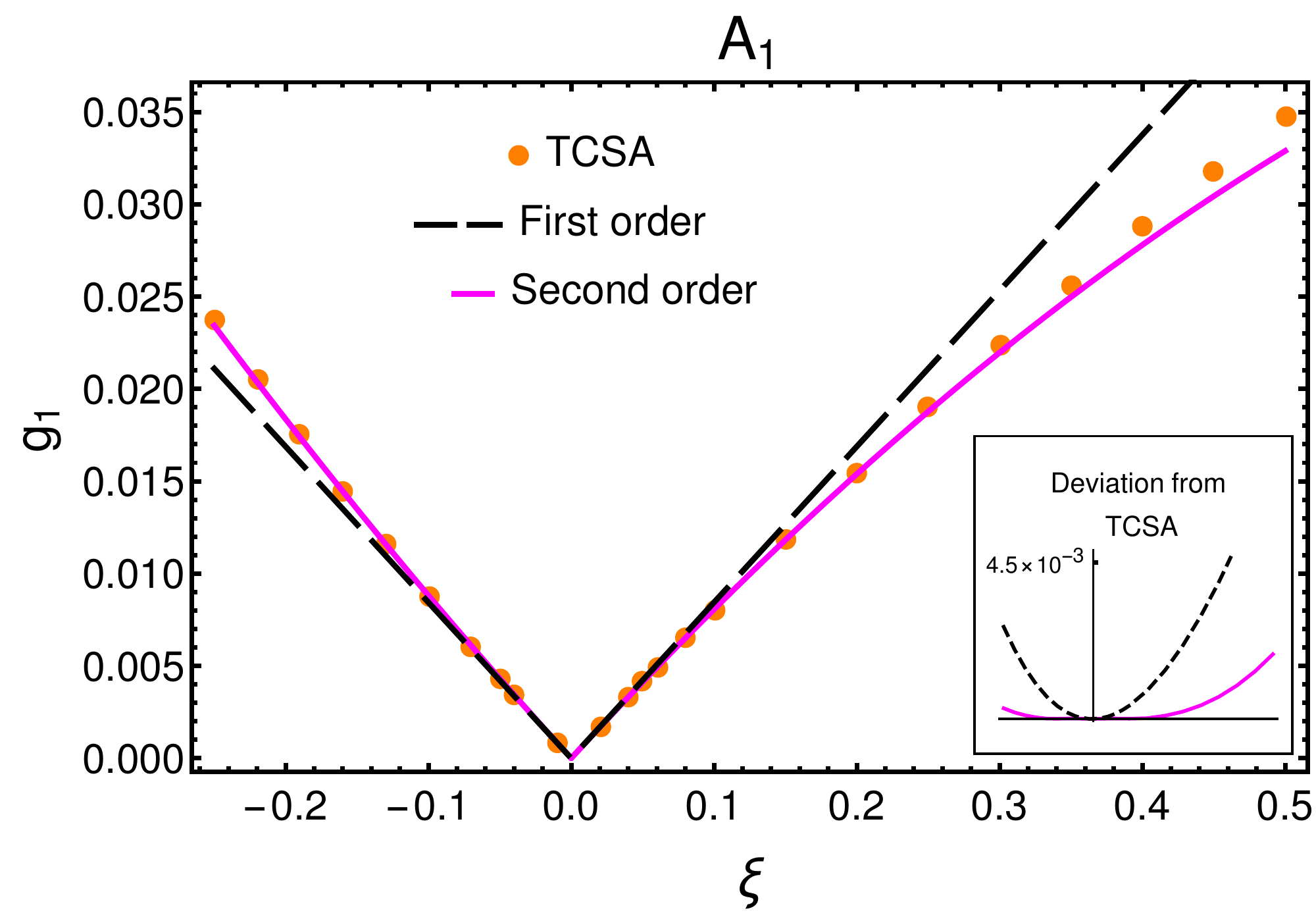}
\label{subfig:g1post}}
\end{subfloat} &
\begin{subfloat}
{\includegraphics[width=0.45\textwidth]{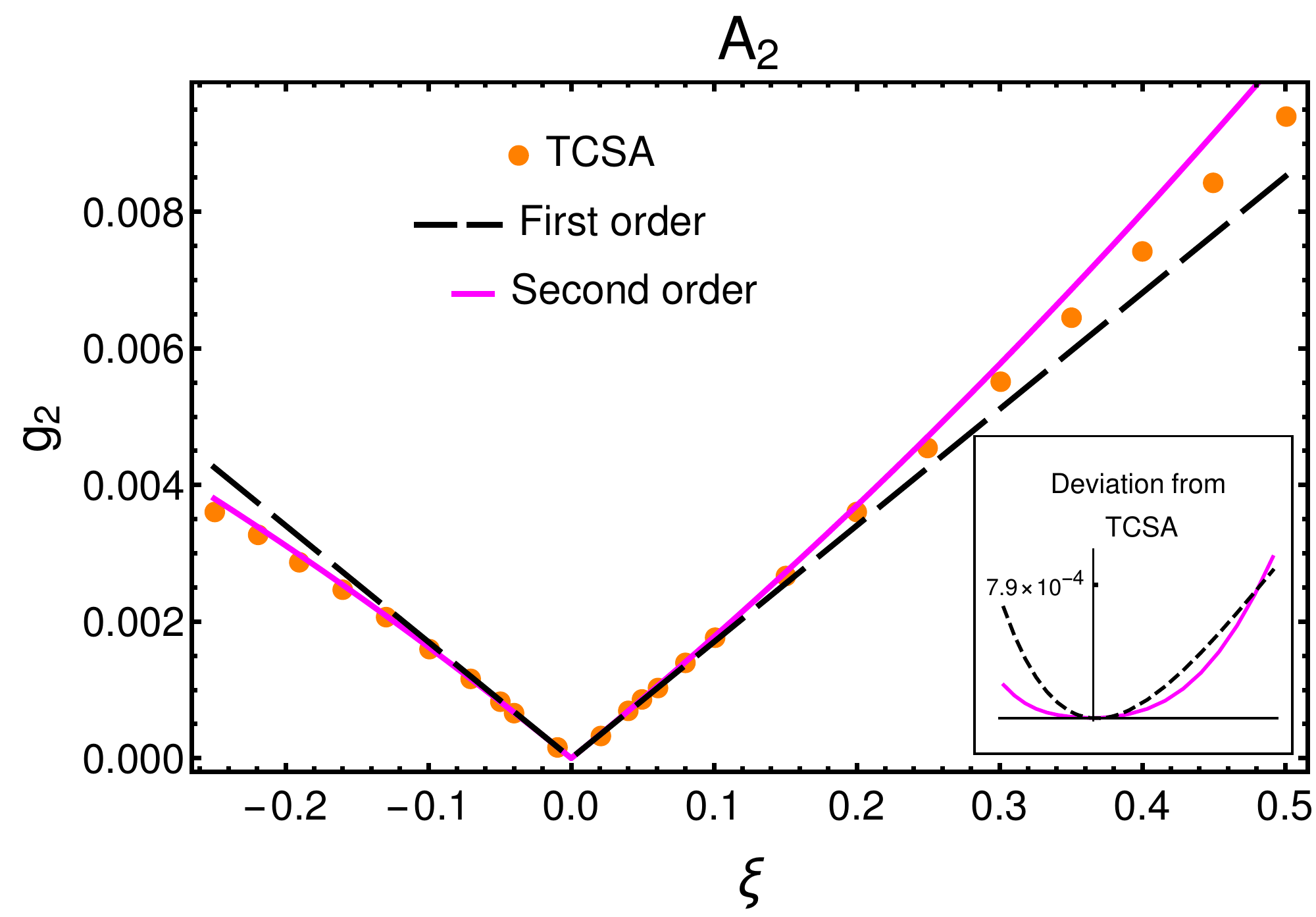}
\label{subfig:g2post}}
\end{subfloat} \\
\begin{subfloat}
{\includegraphics[width=0.45\textwidth]{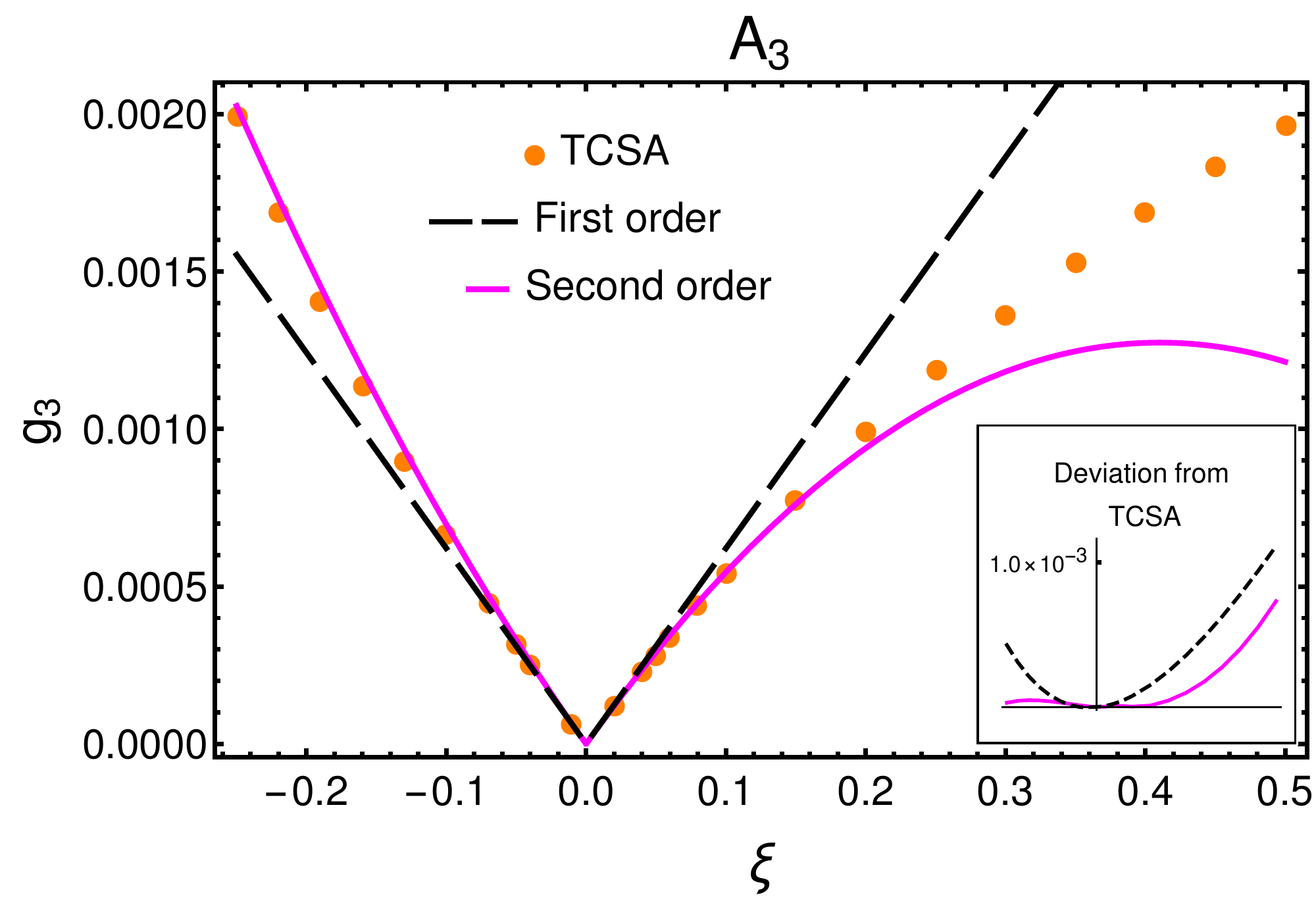}
\label{subfig:g3post}}
\end{subfloat} &
\begin{subfloat}
{\includegraphics[width=0.45\textwidth]{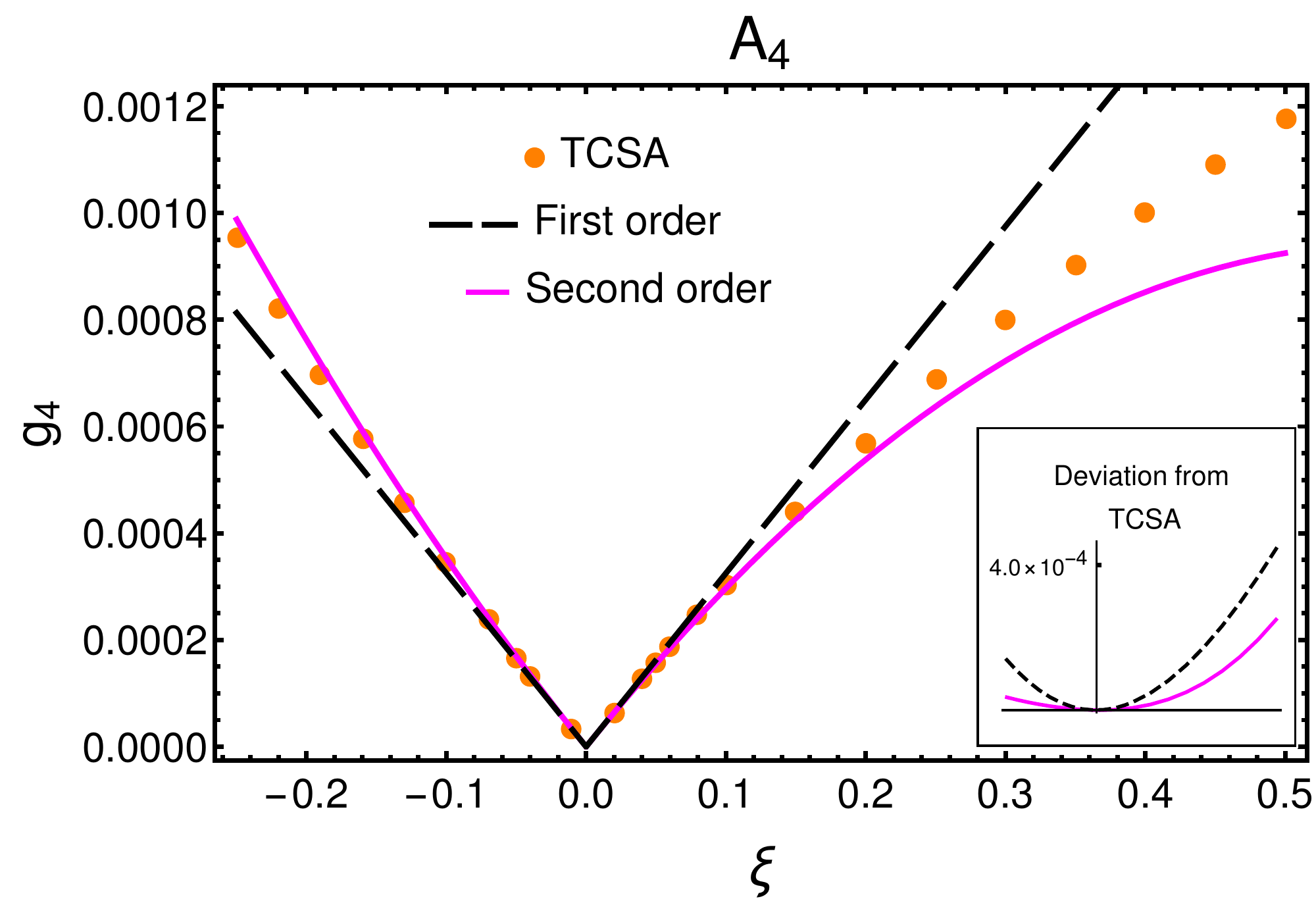}
\label{subfig:g4post}}
\end{subfloat}
\end{tabular}
\caption{Comparison between TCSA and perturbative overlaps of the lowest-lying four one-particle states as functions of the quench magnitude $\xi=(h_f-h_i)/h_f$ for quenches along the $E_8$ axis in volume $m_1R=40$. Dashed lines indicate the first-order predictions, and continuous lines depict the sum of first two orders. TCSA data is shown by dots. The insets show the deviations of the first and second order results from the numerical data.}
\label{fig:onep_comp}
\end{figure}

Let us  start with the overlaps of the one-particle eigenstates.
The result of the comparison is presented in Fig. \ref{fig:onep_comp}, where the numerical TCSA results are shown in dots and the first and second order perturbative results are plotted in dashed and solid lines, respectively.
It is clear that the perturbative expression describes the overlaps very well for a quite wide range of quench magnitudes. Note that Eq. \eqref{eq:onepmultspec} involves a sum over three-particle form factors which can only be evaluated in a truncated manner since the list of available form factors is incomplete. The largest number of these form factors are accessible for the case of the lightest particle $A_1$, therefore the agreement is the best for this case and the domain of validity almost covers the whole region of the plot. For heavier particles it is expected that extending the set of available form factors would result in a better agreement with TCSA data, although the domain of validity presumably remains smaller than for $A_1$ (see Table \ref{tab:onep}). Note also that including the second order leads to a major improvement of the agreement between the perturbative and TCSA results in almost all the parameter region presented here.

\subsubsection{Two-particle overlap functions}
The multiple particle species present in the Ising Field Theory provide an opportunity to observe both kinds of $K$ functions calculated in Section \ref{sec:pert_exps}. In this case we used the data of quenches at a few different values of $\xi$ and plotted the overlaps as functions of the momentum parameterising the particle pair. 
Unfortunately, for most two-particle states it was not possible to fit the overlaps obtained from TCSA with a cut-off dependence resembling Eq. \eqref{eq:epol}. As expected, for states with energy close to the cut-off and for overlaps obtained from TCSA in larger volume tended to produce less accurate fits than the others. Consequently, we decided to present the two-particle data at the highest available cut-off without extrapolation. Further comments and illustrations using some examples are presented in Appendix  \ref{app:epol}.
\begin{figure}[t!]
\centering
\includegraphics[width=0.6\textwidth]{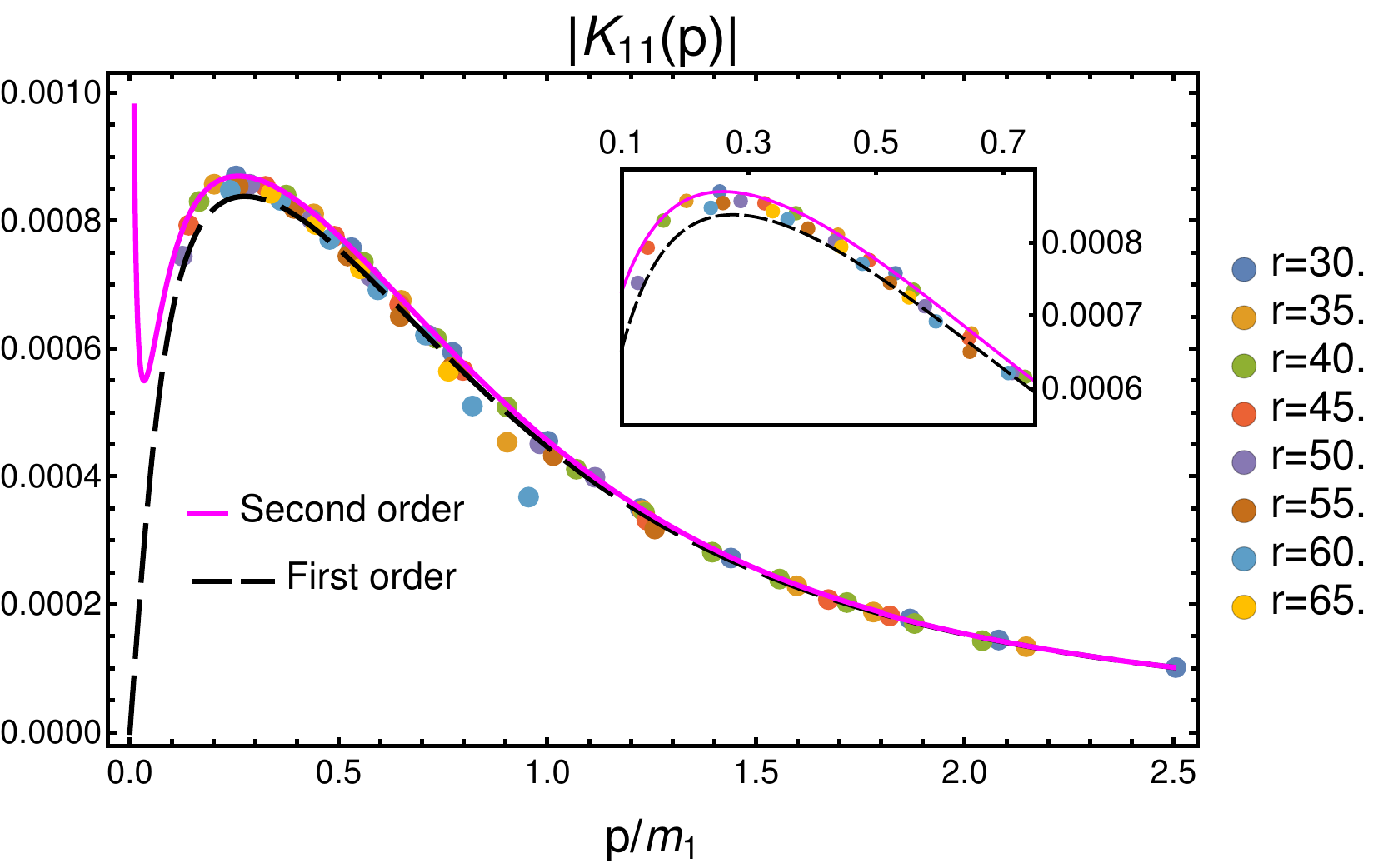}
\caption{Perturbative predictions against the TCSA data for the two-particle overlap $K_{11}$ as a function of the dimensionless momentum $p_1/m$ after a quench of size $\xi=(h_f-h_i)/h_f=0.05.$ The overlap was determined by TCSA using volumes $m_1R=30\dots65,$ as the colours indicate. Dashed lines correspond to the first-order predictions, and continuous lines depict the sum of first two orders. The inset shows that adding the $\mathcal{O}(\lambda^2)$ term improves the agreement considerably even for this small quench.}
\label{fig:K11p05post}
\end{figure}

The comparison for the overlap function of a pair of the lightest particle is presented in Fig. \ref{fig:K11p05post}. which shows that the perturbative expansion performs well in matching the numerical results of TCSA also for pair overlaps for a small quench with $\xi=0.05$. The change from first to second order is less spectacular as it was in the one-particle case, but it still significantly improves the agreement. In addition, the second order correction dramatically alters the qualitative behaviour of the overlap since it introduces a pole for zero momentum. This is consistent with the results of the work \cite{2018JHEP...08..170H} where it was demonstrated that pair overlaps $K_{aa}(p)$ have a pole whenever the one-particle overlap $g_a$ is non-vanishing, and Eq. \eqref{eq:Kaafin} correctly accounts for the expected residue of the pole in the leading order of perturbation theory.

\begin{figure}[t!]
\begin{tabular}{cc}
\begin{subfloat}[$\xi=0.1$]
{\includegraphics[width=0.45\textwidth]{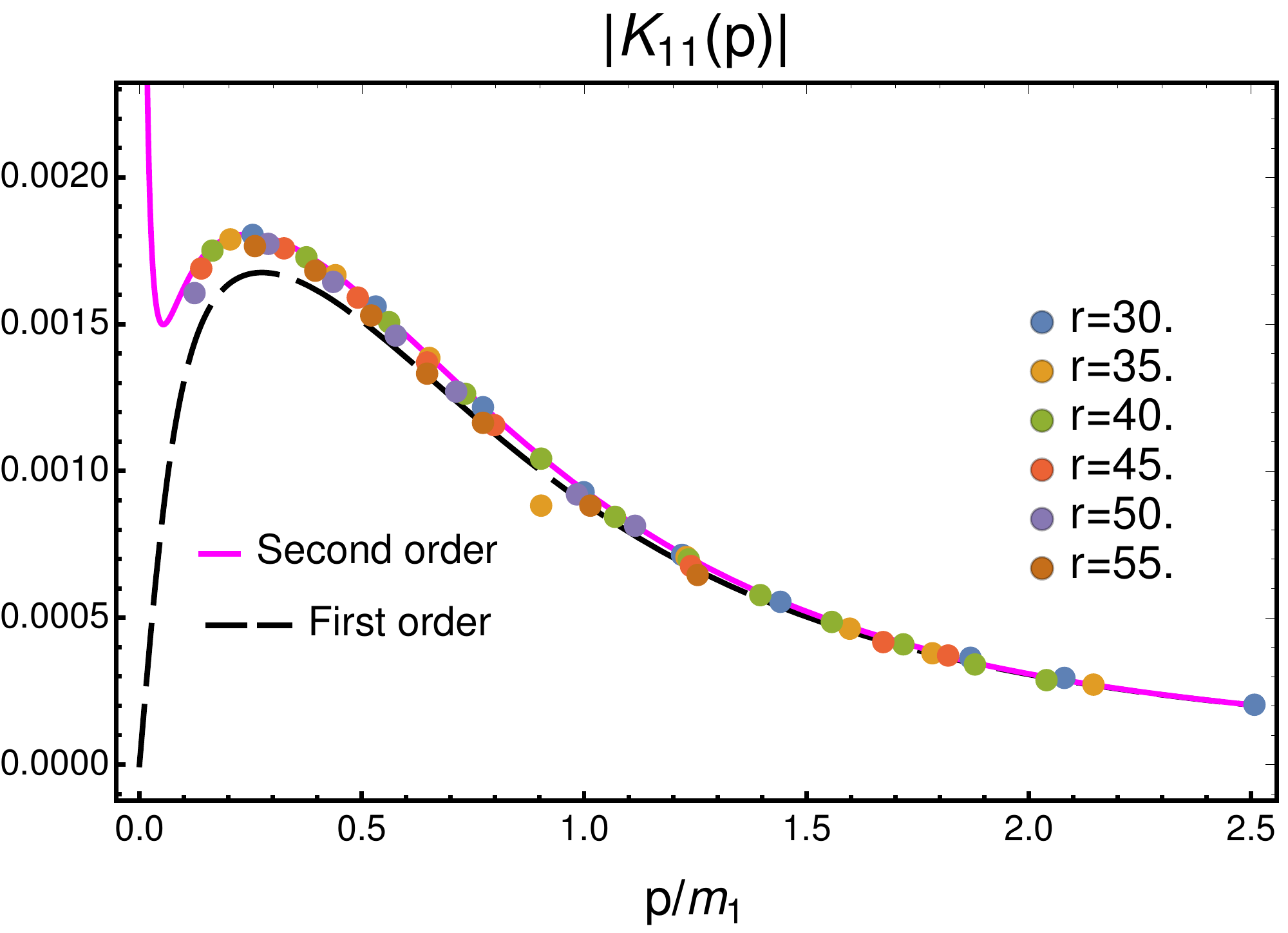}
\label{subfig:K11p1}}
\end{subfloat} &
\begin{subfloat}[$\xi=0.2$]
{\includegraphics[width=0.45\textwidth]{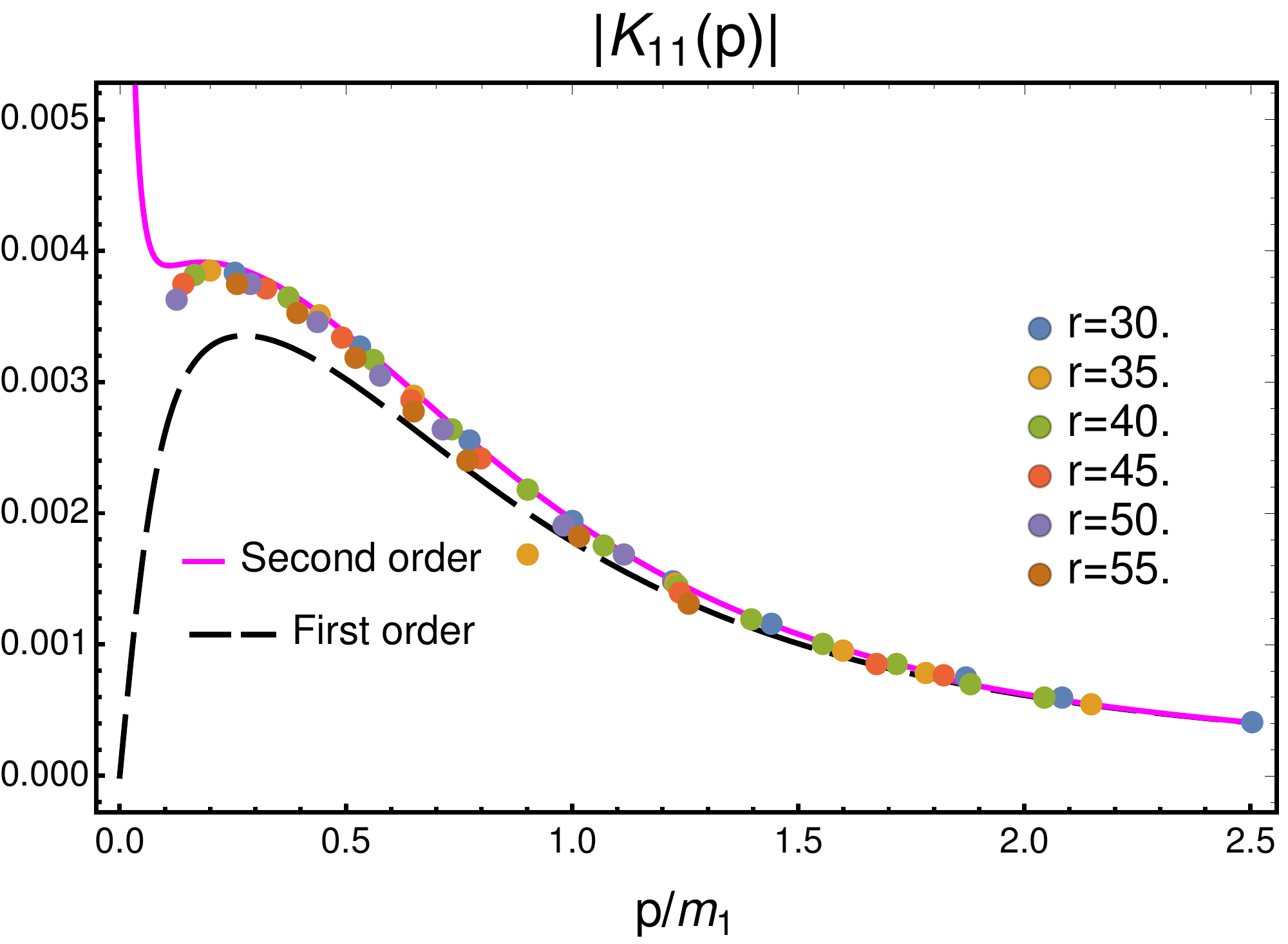}
\label{subfig:K11p2}}
\end{subfloat}
\end{tabular}
\caption{Numerical and analytical two-particle overlaps after quenches of size (a) $\xi=0.1$ and (b) $\xi=0.2$ along the $E_8$ axis. The overlap was determined by TCSA using volumes in the range $m_1R=30\dots55.$ Conventions are as in Fig. \ref{fig:K11p05post}. While at high energies TCSA results are predicted correctly, the low-energy points tend to deviate further from the perturbative prediction upon increasing the quench amplitude.}
\label{fig:K11plarge}
\end{figure}

We can also explore the limitations of the perturbative description by increasing the quench amplitude $\xi$, as illustrated in Fig. \ref{fig:K11plarge}. Note that while the high-energy behaviour is still captured correctly by the post-quench expansion, there is some observable deviation at low momenta. Unfortunately, it was not possible to obtain TCSA data for low enough momenta to investigate the presence of the pole, as that would require large volume where TCSA becomes less accurate. Even so, the agreement shown in Fig. \ref{fig:K11plarge} is still quite convincing.

\begin{figure}[t!]
\begin{tabular}{cc}
\begin{subfloat}[$K_{12}(p)$]
{\includegraphics[width=0.45\textwidth]{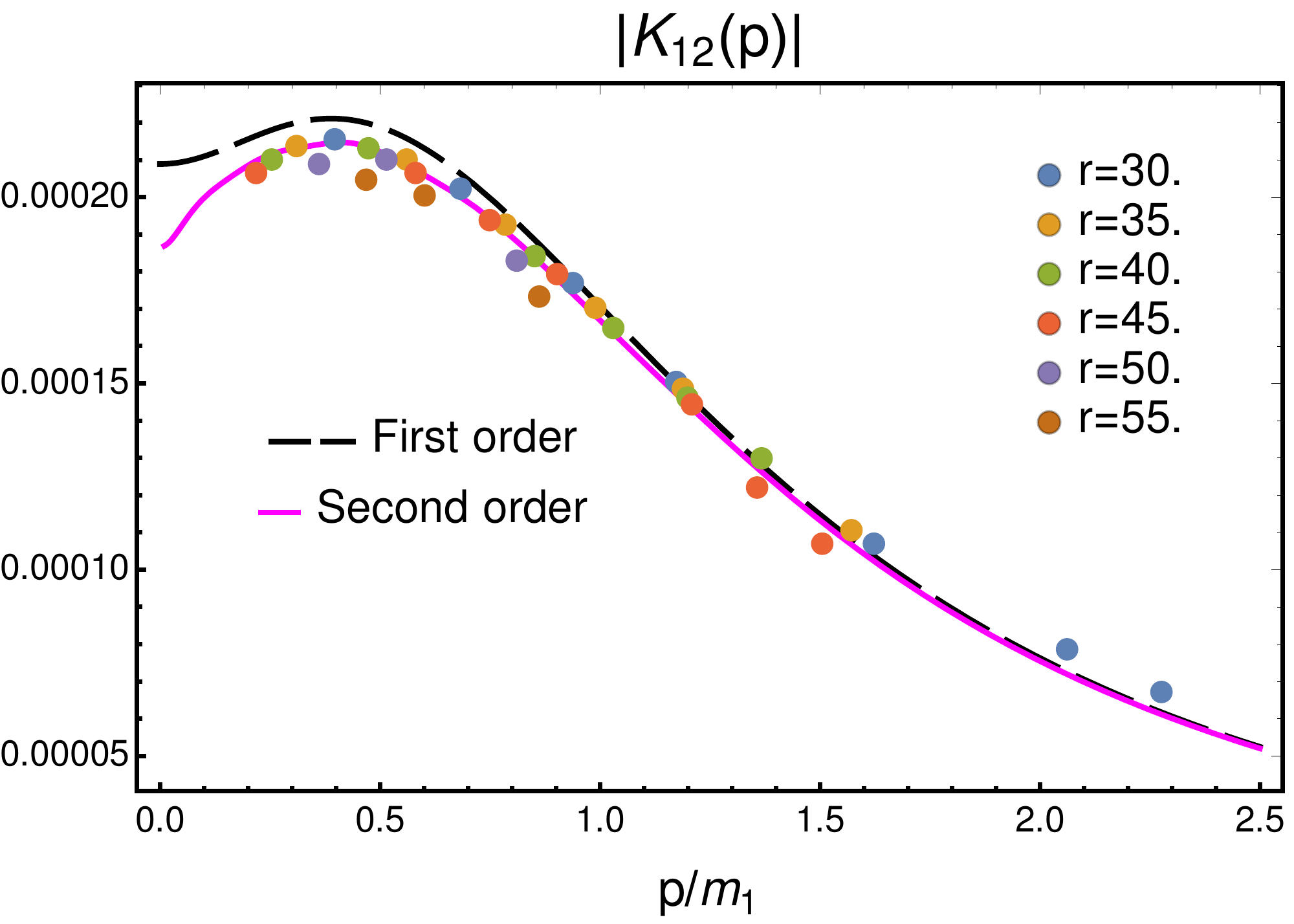}
\label{subfig:K12post05}}
\end{subfloat} &
\begin{subfloat}[$K_{22}(p)$]
{\includegraphics[width=0.45\textwidth]{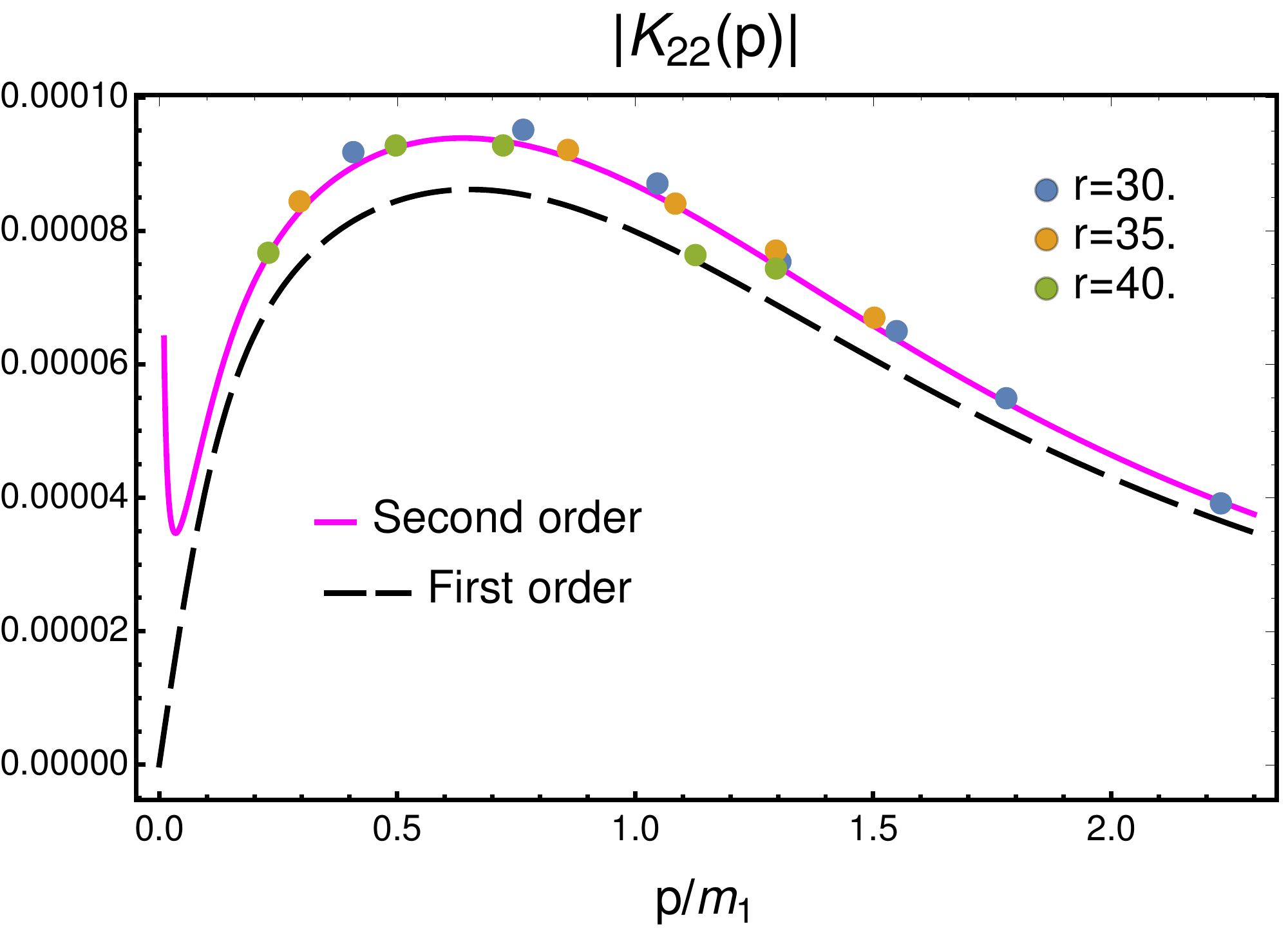}
\label{subfig:K22p05}}
\end{subfloat}
\end{tabular}
\caption{Overlap functions of pairs of heavier particles after a quench of size $\xi=0.05$ along the $E_8$ axis. Conventions are as in Fig. \ref{fig:K11p05post}. }
\label{fig:K12K22}
\end{figure}

It is also possible to compare the predictions for overlap functions of pair states of heavier particles as well as that of pairs composed of different species. Fig. \ref{fig:K12K22} illustrates that, similar to the case of $K_{11}$, the analytic prediction still agrees very well with the TCSA numerics, with the difference that the second order contribution plays a larger role compared to that case shown in Fig. \ref{fig:K11p05post}. 

\section{Perturbative expansion on the pre-quench basis}
\label{ssec:pre_exp}
Instead of using the post-quench state as in Section \ref{ssec:post_exp}, it is also possible to obtain a perturbative expansion of the overlap functions by expanding the post-quench eigenstates on the pre-quench basis. In contrast to the previous calculations where the task was to obtain the correction to the pre-quench vacuum order by order, here it is necessary the compute part of the perturbative correction to each eigenstate that is proportional to the pre-quench ground state $\ket*{\Omega}$. However, the steps of the calculations are rather similar to the preceding ones so instead of a detailed calculation we focus on the results while commenting on the differences.

\subsection{One-particle overlaps}
For the $g$ amplitudes the major changes compared to Eq. \eqref{eq:onepmultspec} are in the denominators: the energy differences appearing in the perturbative expansion are now with respect to the one-particle mass instead of the vacuum. Cancellation of divergent parts is again due to disconnected pieces. Apart from the energy denominators, the only difference appears in the ordering of rapidities in the form factors. The final result is given by:
\begin{align}
\label{eq:onepmultpre}
\frac{g^{(0)}_a}{2}=& -\lambda \frac{F_{a}^\phi}{m_a^2}-\lambda^2\left(\sum_{b=1,b\neq a}^{N_{\text{spec}}}\frac{F_b^\phi F_{ab}^\phi(\imath\pi,0)}{m_a^2 m_b(m_b-m_a)}+\frac{F^\phi_{aa}(\imath\pi,0)F_a^\phi}{m_a^4}+\sum_{b\leq c}\frac{1}{(2\delta_{bc})!}\times\right.\\
&\left.\times\int\frac{\dd\vt}{2\pi}\right.\left. \frac{F_{bca}^\phi(\imath\pi+\vt,\imath\pi+\vt_{bc},0) F_{bc}^{\phi}(\vt,\vt_{bc})}{m_a^2\left(m_b\cosh(\vt)+\sqrt{m_c^2+(m_b\sinh\vt)^2}-m_a\right)\sqrt{m_c^2+(m_b\sinh\vt)^2}}+\dots \right)\nonumber\\&+ O\left(\lambda^3\right)\nonumber\,,
\end{align}
where now all masses and form factors are those of the pre-quench theory. The dots in the parenthesis indicate contributions of intermediate states with more than two particles. Note that if $m_a>m_b+m_c$ then the denominator of integrand has a zero and the integrand has a pole. This will occur for all one-particle states with mass $m_a>2m_1$ in one or more such integral terms.
This pole is the consequence of a disappearing energy difference between a one-particle state and the two-particle continuum in infinite volume. We postpone the discussion of this singularity after presenting the two-particle overlaps. 

\subsection{Two-particle overlaps}
Let us start with the discussion of the $K_{aa}$ function. The first order contribution is simply
\begin{equation}
-\lambda\frac{F_{aa}^\phi(-\vt,\vt)}{2m_a^2\cosh[2](\vt)}\,.
\end{equation}
The second order contribution is given as a sum over eigenstates [c.f. the third term of Eq. \eqref{eq:pert_es}]. Inserting the vacuum yields a divergent term which is cancelled by the disconnected piece of the fourth term, analogously to the argument in Section \ref{ssec:post_exp}. The connected part of that term disappears in the infinite volume limit due to the corresponding  density factor and therefore the only remaining term resulting from inserting vacuum is
\begin{equation}
-\frac{\lambda^2}{2}\frac{F_{aa}^\phi(\imath\pi,0)F_{aa}^{\phi}(-\vt,\vt)}{m_a^4\cosh[4](\vt)}\,.
\end{equation}
Moving forward and inserting the one-particle states yields
\begin{equation}
\frac{\lambda^2}{2}\sum_{b=1}^{{s}}\frac{F_b^\phi F_{baa}^\pi(\imath\pi,-\vt,\vt)}{m_a^2m_b\cosh[2](\vt)(2m_a\cosh(\vt)-m_b)}\,,
\end{equation}
where the aforementioned pole manifests itself as a divergence of the pair overlap function $K_{aa}(\vt)$ whenever there is a particle with $m_b>2m_a$. 

Proceeding to the insertion of two-particle states, we can consider inserting a pair of particles $A_b$ with $b\neq a$, in which case the form factor has no pole. In finite volume, the corresponding contribution reads
\begin{equation}
\label{eq:pre_2p_sum1}
\frac{\lambda^2 L^2}{2} \sum_{\vt'} \frac{F_{bb}^\phi(-\vt',\vt')F_{bbaa}^\phi(\imath\pi+\vt',\imath\pi-\vt',-\vt,\vt)}{\rho_{bb}(\vt',-\vt')2m_a\cosh\vt(2m_a\cosh\vt-2m_b\cosh\vt')m_a L\cosh\vt}\,.
\end{equation}
Note that the pole is only present in infinite volume since for any finite $L$ there are no exact degeneracies in the spectrum due to the Bethe-Yang equations \eqref{eq:BY}. Hence one might expect that the finite volume regularisation technique detailed in Section \ref{ssec:finvolreg} is able to treat its effect properly. In the limit  $L\rightarrow\infty$ limit the energy difference can be zero at
\begin{equation}
\vt_*=\text{arccosh}\left(\frac{m_a\cosh\vt}{m_b}\right)\,.
\end{equation}
Note that $\vt_*$ is imaginary if $\vt<\text{arccosh}(m_b/m_a)$. However, for larger $\vt$ the pole is on the real axis. Eq. \eqref{eq:pre_2p_sum1} can be rewritten as
\begin{equation}
\label{eq:pre_2p_sum2}
\frac{\lambda^2}{8}\sum_{\vt'}\frac{F_{bb}^\phi(-\vt',\vt')F_{bbaa}^\phi(\imath\pi+\vt',\imath\pi-\vt',-\vt,\vt)}{\tilde{\rho}_{b}(\vt')m_a^2m_b\cosh\vt'\cosh^2\vt(m_a\cosh\vt-m_b\cosh\vt')}\,.
\end{equation}
The sum can be represented as a sum of contour integrals using (cf. Appendix \ref{app:twopcalcs}) 
\begin{equation}
\sum_{\vt'}\frac{f(\vt')}{\tilde{\rho}_b(\vt')}=\sum_{\vt'}\oint_{\vt'}\frac{\dd\vt}{2\pi}\frac{-f(\vt)}{1+e^{\imath\tilde{Q}_b(\vt)}}\,,
\end{equation}
where $f(\vt)$ is assumed to be regular at $\vt'$, the contours encircle the $\vt'$ values on the real axis that are given by the quantisation condition
\begin{equation}
\label{eq:pre_2p_quant}
\tilde{Q}_b(\vt')=m_bL\sinh\vt'+\delta_{bb}(2\vt')=2\pi J\,,
\end{equation}
where $J$ is a half-odd integer. The contours can be joined into two infinite lines below and above the real axis. On the first the integrand vanishes in the infinite volume limit while the second one yields
\begin{equation}
\frac{\lambda^2}{8}\int_{-\infty+\imath\epsilon}^{\infty+\imath\epsilon} \frac{\dd\vt'}{2\pi}\frac{F_{bb}^\phi(-\vt',\vt')F_{bbaa}^\phi(\imath\pi+\vt',\imath\pi-\vt',-\vt,\vt)}{m_a^2m_b\cosh^2\vt\cosh\vt'(m_a\cosh\vt-m_b\cosh\vt')}\,.
\end{equation}
When joining the contours, it is necessary to subtract the residue of the pole at $\vt=\vt^*,$
\begin{equation}
\frac{\lambda^2}{8}\frac{\imath F_{bb}^\phi(-\vt_*,\vt_*)F_{bbaa}^\phi(\imath\pi+\vt_*,\imath\pi-\vt_*,-\vt,\vt)}{m_a^2m_b\cosh^2\vt\cosh\vt' m_b\sinh\vt_*(1+e^{\imath\tilde{Q}_b(\vt_*)})}\,.
\end{equation}
(for a detailed calculation, see Appendix  \ref{app:twopcalcs}.) Even though the result is finite, it does not have a $L\rightarrow\infty$ limit due the factor $e^{\imath\tilde Q_b(\vt^*)}\sim e^{\imath m_b L \sinh\vt^*}$ Consequently, the sum in $\eqref{eq:pre_2p_sum2}$ still fails to have a well-defined infinite volume limit.

Therefore the singularities corresponding to vanishing energy denominators are intractable by the method of finite volume regularisation. One may try other ways to circumvent this problem and arrive at a regular expression well defined in the $L\rightarrow\infty$ limit, however we failed in all our attempts so far. So the proper treatment of these singularities remains an interesting open question.

Nevertheless, there exists particular quenches which are free of the complications discussed above. If the matrix elements of the operator $\phi$ are proportional to the energy of the involved states, the divergence is cancelled and the sum can be readily transformed to an integral. This is precisely what happens for quenches presented in Section \ref{sec:TCSA} where the perturbing operator is $\sigma(x)$ which is proportional to the trace of the energy-momentum tensor \cite{1996NuPhB.473..469D}, and as a consequence all of its form factors are proportional to the total energy of the appropriate states. 

This can also be intuitively understood by noting that in general the pre-quench basis is not an optimal choice to express the time evolution of the post-quench Hamiltonian. For example, heavy particles whose kinematically allowed decays in the pre-quench system are only prohibited by integrability become unstable. 
These particles are expected to acquire a finite lifetime which is reflected by the divergent terms of the perturbative series and therefore requires a resummation which is expected to shift the singularity away from the real axis. Similarly, two-particle states acquire a finite lifetime due to inelastic processes.  The situation is radically different for quenches in the direction of the original $\sigma$ perturbation. Such a protocol is simply equivalent to a rescaling of parameters describing the spectrum, but its structure remains intact. Accordingly, one does not expect divergences in perturbation theory and in fact, they are absent -- apart from the ones tractable with the method of finite volume regularisation.

Assuming such a quench protocol one obtains the following results for overlaps with pair of two particles of the same species:
\begin{align}
\label{eq:Kaapre}
&K^{(0)}_{aa}(-\vt,\vt)= -\lambda \frac{F_{aa}^{\phi}(-\vt,\vt)}{2m_a^2\cosh^2\vt}+
\lambda^2\Bigg(\sum_{b=1}^{N_{\text{spec}}}\frac{F_b^\phi F_{baa}^{\phi}(\imath\pi,-\vt,\vt)}{2m_a^2\cosh^2\vt m_b(2m_a\cosh\vt-m_b)}+\\
&+\frac{F_{aa}^{\phi}(-\vt,\vt)F_{aa}^\phi(\imath\pi, 0)}{2m_a^4\cosh^4\vt}+\sum_{b=1}^{N_{\text{spec}}}D^{(0)}_{ab}(\vt,-\vt)+\nonumber\\
&+\sum_{(c,d)\neq(a,b)}\frac{1}{(2\delta_{cd})!}\int_{-\infty}^{\infty}\frac{\dd{\vt'}}{2\pi} \frac{F_{aacd}^{\phi,s*}(\imath\pi+\vt,\imath\pi-\vt,\vt',\vt'_{cd})F_{cd}^{\phi}(\vt',\vt'_{cd})}{2m_a^2\cosh^2\vt(2m_a\cosh\vt-E^{(0)}_{cd}(\vt'))m_d{\cosh\vt'_{cd}}} + \dots \Bigg)+ O\left(\lambda^3\right)\,,\nonumber
\end{align}
while for a pair composed of particles of different species the result is
\begin{align}
\label{eq:Kabpre}
&K^{(0)}_{ab}(\vt,\vt_{ab})=
-\lambda \frac{F_{ab}^{\phi}(\vt,\vt_{ab})}{C^{(0)}_{ab}(\vt,\vt_{ab})}
+\lambda^2\Bigg(\sum_{c=1}^{N_{\text{spec}}}\frac{F_c^\phi F_{cab}^{\phi}(\imath\pi,\vt,\vt_{ab})}{m_c(E^{(0)}_{ab}(\vt)-m_c)C^{(0)}_{ab}(\vt,\vt_{ab})}-
\\
&-\frac{F_{ab}^{\phi}(\vt,\vt_{ab})}{C^{(0)}_{ab}(\vt,\vt_{ab})E^{(0)}_{ab}(\vt)}\left(\frac{F_{aa}^\phi(\imath\pi, 0)}{m_b\cosh\vt_{ab}}+\frac{F_{bb}^\phi(\imath\pi, 0)}{m_a\cosh\vt}\right)+G_{ab}^{b(0)}(\vt,\vt_{ab})+G_{ab}^{a(0)}(\vt,\vt_{ab})+\nonumber\\[.5em]
&
+\sum_{\substack{c\neq a,b \\ d\neq a,b}} \frac{1}{(2\delta_{cd})!}\int_{-\infty}^{\infty}\frac{\dd{\vt'}}{2\pi} \frac{F_{abcd}^{\phi,s*}(\imath\pi+\vt,\imath\pi+\vt_{ab},\vt',\vt'_{cd})F_{cd}^{\phi}(\vt',\vt'_{cd})}{C^{(0)}_{ab}(\vt,\vt_{ab})(E^{(0)}_{ab}(\vt)-E^{(0)}_{cd}(\vt'))m_d\cosh\vt'_{cd}}+ \dots \Bigg)+ O\left(\lambda^3\right)\,,\nonumber
\end{align}
where $E^{(0)}_{ab}(\vt)=m_a\cosh\vt+m_b\cosh\vt_{ab}$ is the (pre-quench) energy of an $A_a-A_b$ particle pair with zero overall momentum. The $C$, $D$, and $G$ functions can be simply transformed to the pre-quench basis by replacing the corresponding quantities in the definitions of Appendix  \ref{app:twopcalcs}. 

\subsection{Comparison with TCSA}
Similarly to the pre-quench expansion, the results of the pre-quench expansion can be compared to TCSA numerics. There are two main points of interest. First, for the quenches considered in Section \ref{sec:TCSA} one can compare both the pre- and post-quench expansions to TCSA at the same time. Second, the pre-quench calculation only requires the knowledge of the pre-quench spectrum and form factors and so it can also be used for the case when the post-quench dynamics is governed by a non-integrable Hamiltonian. In the latter case, the comparison is more limited since to avoid the unresolved singularities the states for which the overlap is considered must be below the continuum, i.e. the two-particle threshold. Nevertheless, it is still worthwhile to perform such an examination.

\begin{figure}[t!]
\begin{tabular}{cc}
\begin{subfloat}
{\includegraphics[width=0.45\textwidth]{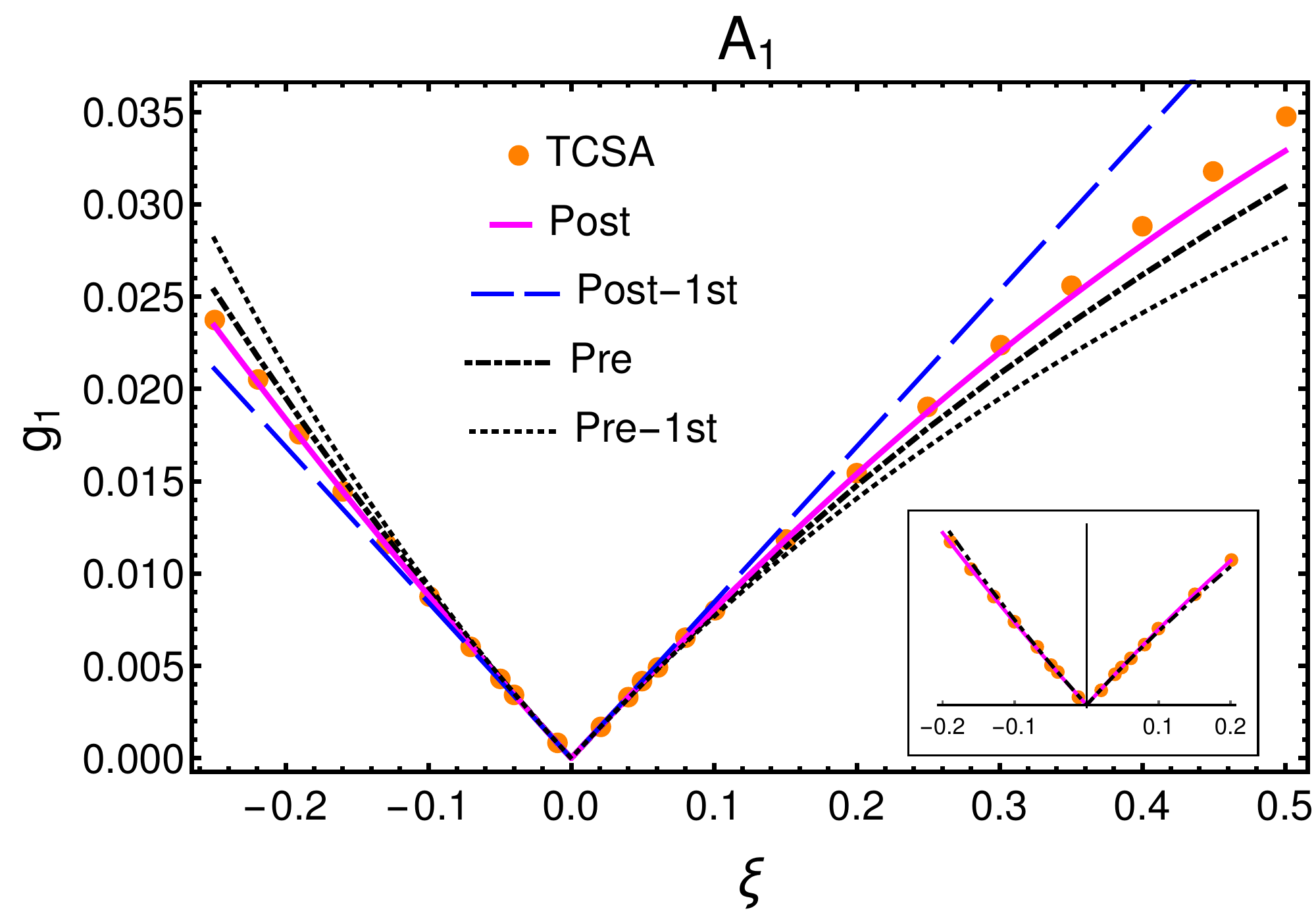}
\label{subfig:g1pp}}
\end{subfloat} &
\begin{subfloat}
{\includegraphics[width=0.45\textwidth]{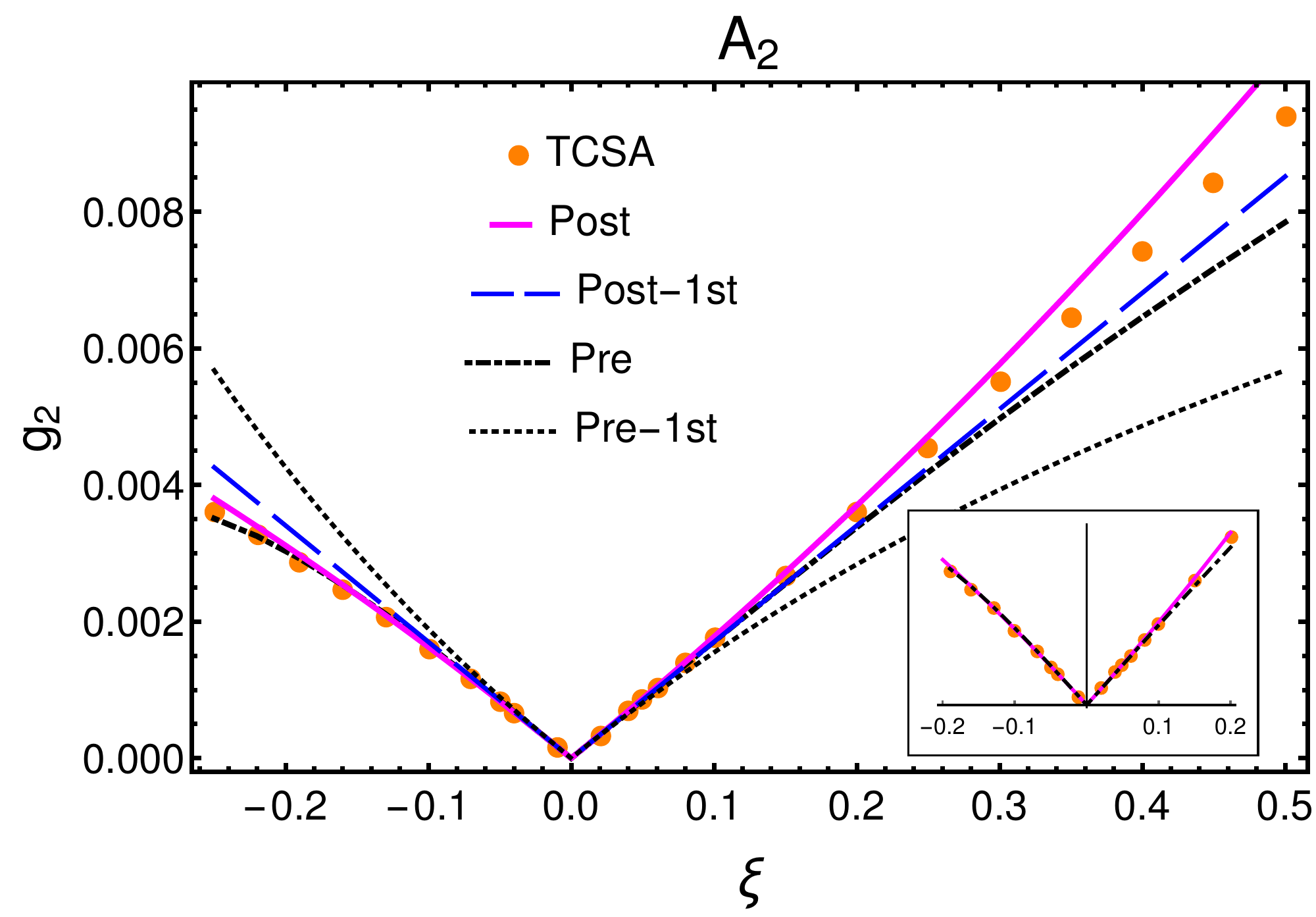}
\label{subfig:g2pp}}
\end{subfloat} \\
\begin{subfloat}
{\includegraphics[width=0.45\textwidth]{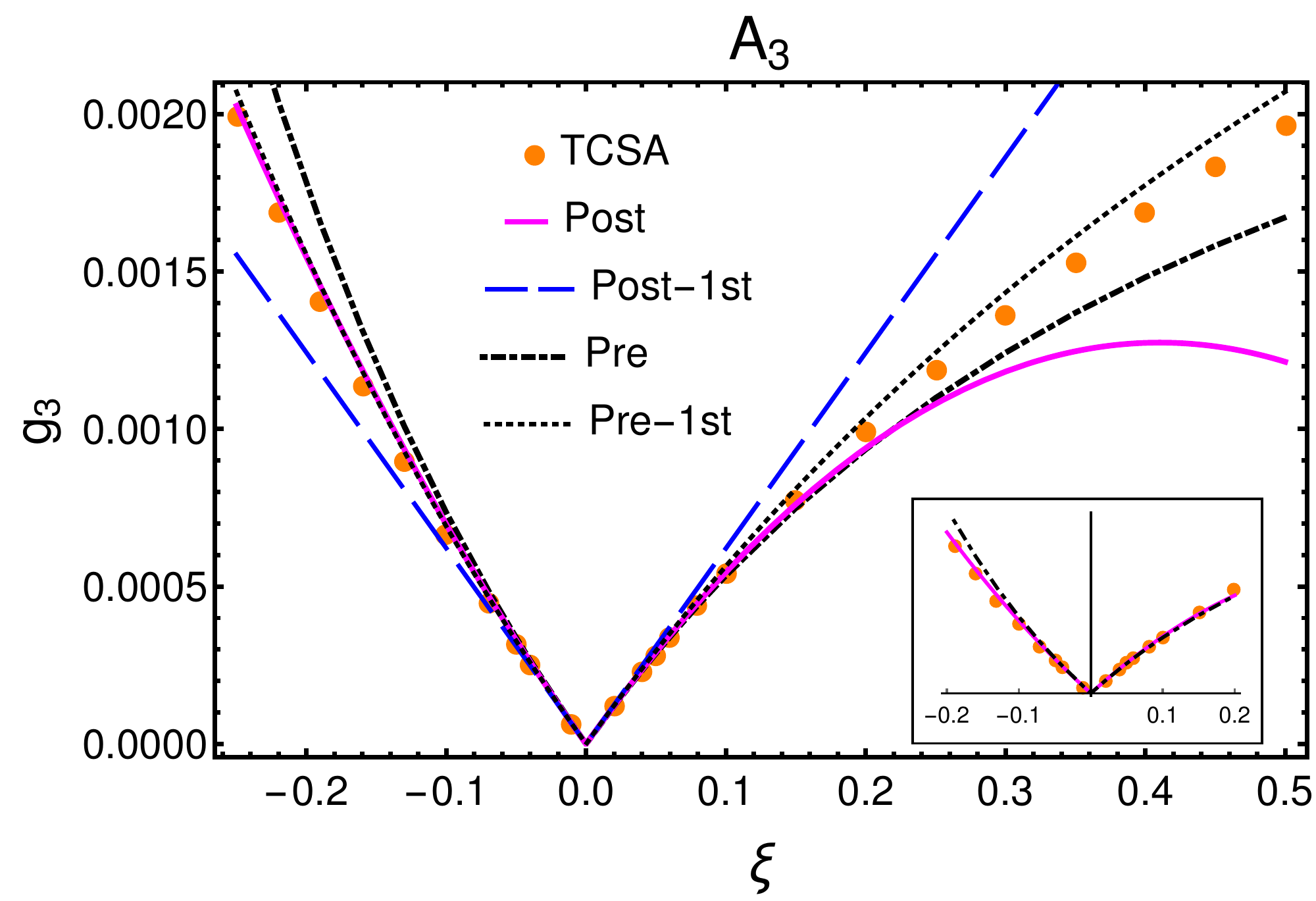}
\label{subfig:g3pp}}
\end{subfloat} &
\begin{subfloat}
{\includegraphics[width=0.45\textwidth]{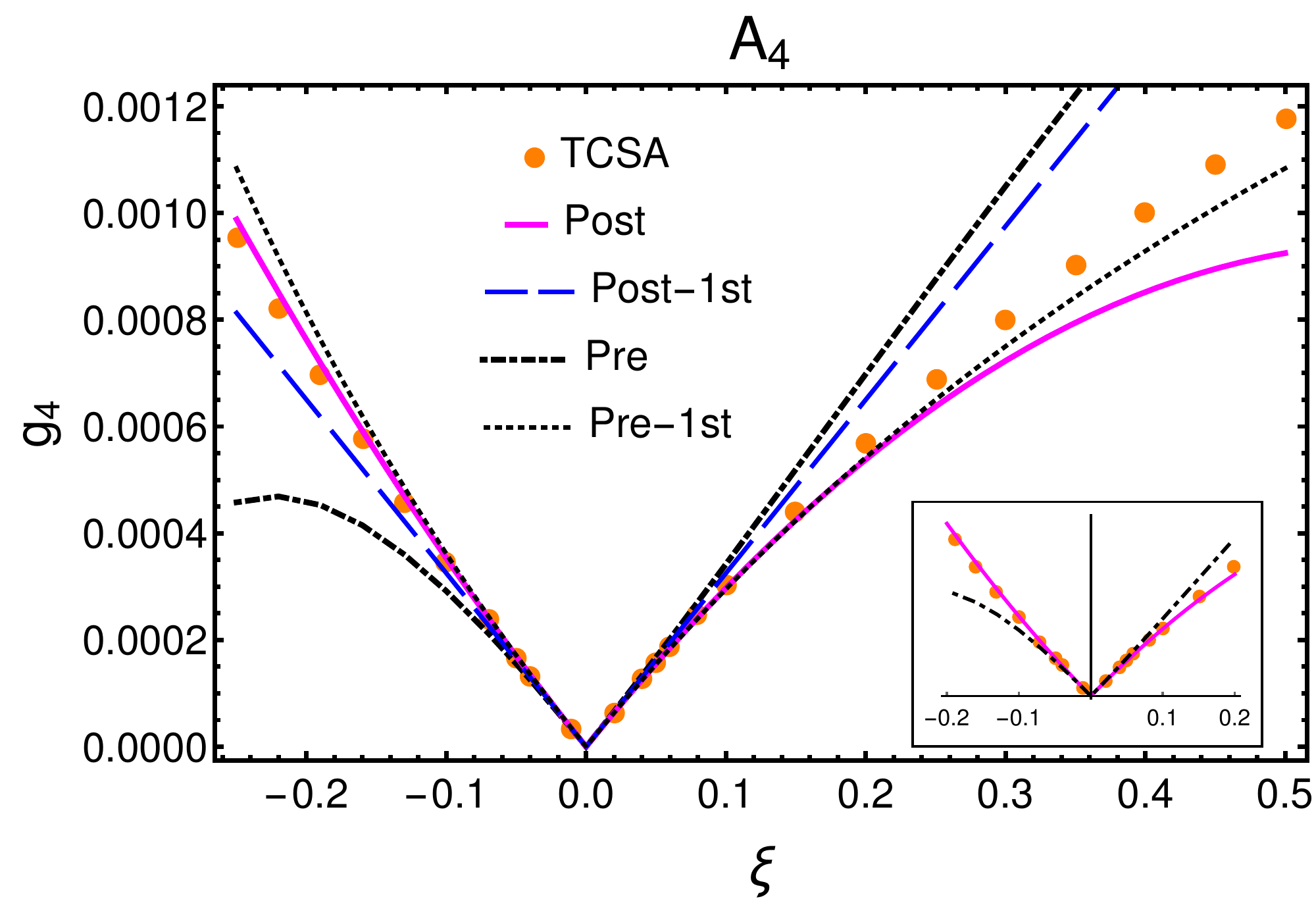}
\label{subfig:g4pp}}
\end{subfloat}
\end{tabular}
\caption{Comparison between TCSA overlaps of the lightest four one-particle states and two different perturbative expansions as functions of the quench magnitude $\xi$ for quenches along the $E_8$ axis in volume $m_1R=40$. Dashed lines indicate the first-order predictions of the post-quench expansion  and continuous lines depict the sum of the first two orders. The pre-quench result up to the first and second order is shown in dotted and dot-dashed lines, respectively. TCSA data is shown by dots. Inset: second order results vs. TCSA for $|\xi|<0.2$.}
\label{fig:onep_pp}
\end{figure}

\subsubsection{Integrable post-quench dynamics}

We begin with the presentation of results for the case of integrable post-quench dynamics. Overlaps for the first four one-particle states are presented in Fig. \ref{fig:onep_pp}. 
The behaviour of the pre-quench expansion is very similar the post-quench expansion with the exception of a slightly narrower domain of validity. The insets show that for $|\xi|<0.2$ the post-quench result is very accurate, while for the heavier particle overlaps the pre-quench perturbative expansion becomes less accurate in this region.

\begin{figure}[t!]
\begin{tabular}{cc}
\begin{subfloat}[$A_1-A_1$ pair]
{\includegraphics[width=0.45\textwidth]{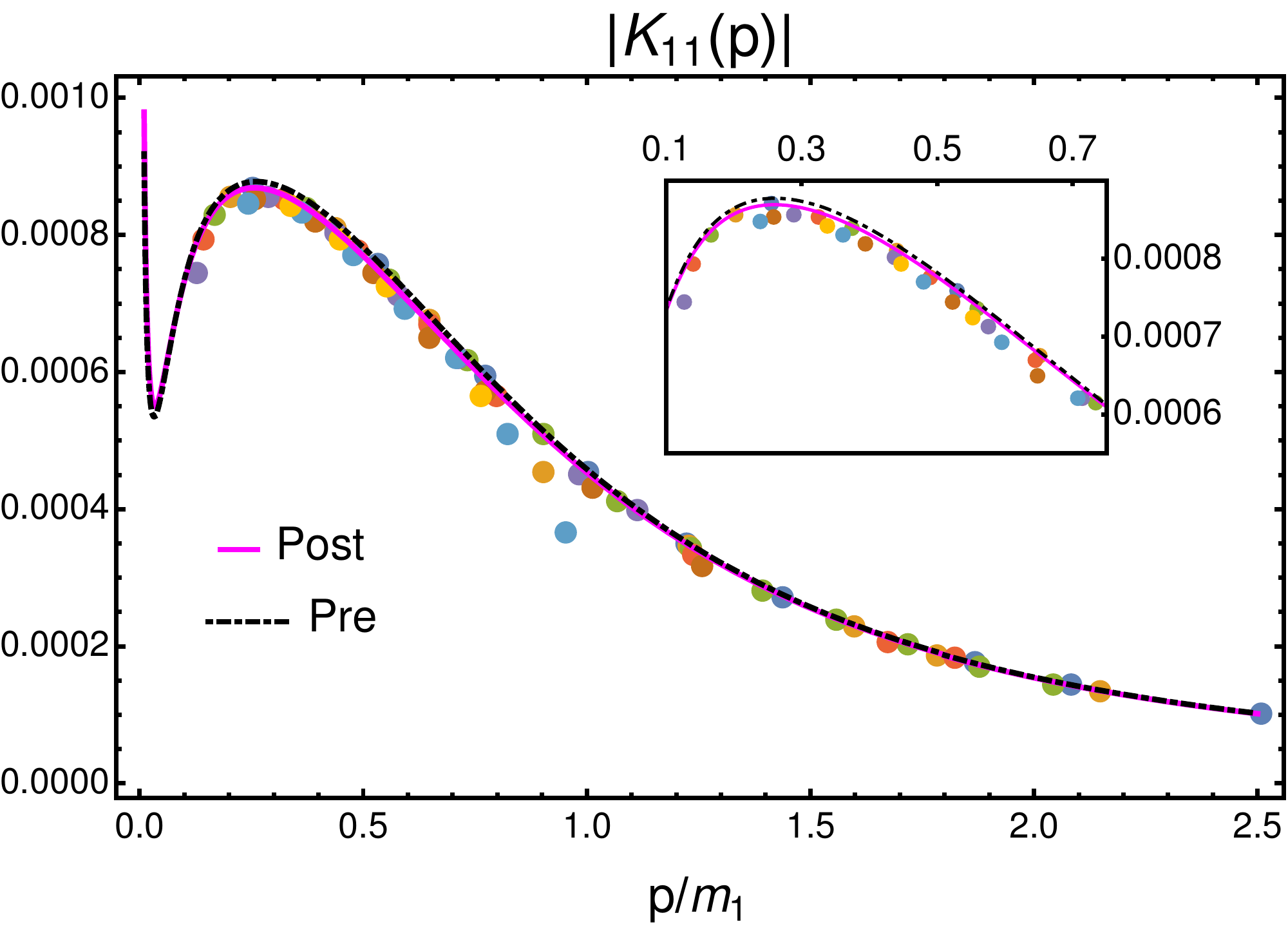}
\label{subfig:K11p05pp}}
\end{subfloat} &
\begin{subfloat}[$A_1-A_2$ pair]
{\includegraphics[width=0.45\textwidth]{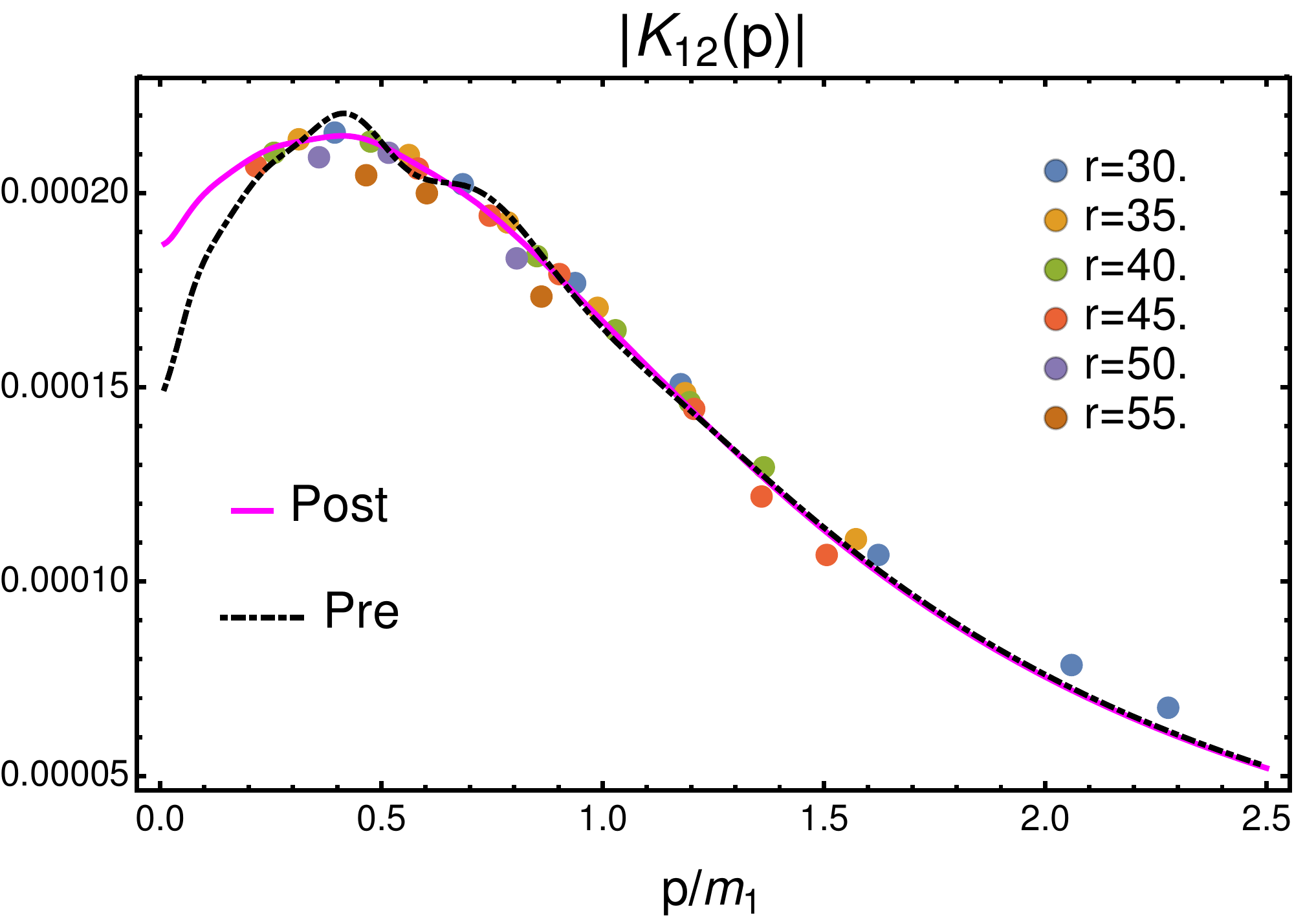}
\label{subfig:K12p05pp}}
\end{subfloat}
\end{tabular}
\caption{Predictions of the two perturbative expansions against the TCSA data for the two-particle overlaps $K_{11}$ and $K_{12}$ as functions of the dimensionless momentum $p/m_1$ after a quench of size $\xi=0.05$ in volume (a) $m_1R=30\dots65$ and (b) $m_1R=30\dots55.$ The two expansions almost completely agree. The inset illustrates that the post-quench expansion is slightly closer to TCSA data.}
\label{fig:Kpp}
\end{figure}

Fig. \ref{fig:Kpp} shows the results for two-particle overlaps. Within the perturbative region both expansions closely resemble each other and both provide an accurate description of the TCSA data. 

\subsubsection{Non-integrable post-quench dynamics}
Finally, let us turn to the discussion of quenches from an integrable pre-quench Hamiltonian to a non-integrable post-quench one. They were implemented by adding the other Ising primary $\epsilon(x)$ field with scaling dimension $x_\epsilon=1$ to the action \eqref{eq:A_quench}.
\begin{equation}
\label{eq:A_oquench}
\mathcal{A}=\mathcal{A}_{CFT,\: c=1/2}-h\int \sigma(x)d^2x-\frac{M}{2\pi}\int\epsilon(x)\Theta(t)d^2x\,,
\end{equation}
For $h=0$ this action describes a free massive fermion of mass $|M|$ \cite{1961AnPhy..16..407L,1970AnPhy..57...79P}. Positive $M$ corresponds to the ferromagnetic phase, while $M<0$ is the paramagnetic direction. In the vicinity of the $h=0$ axis in the ferromagnetic phase $M>0$ the spectrum is shaped by the weak confinement of the free fermionic excitations as described by the McCoy--Wu scenario \cite{1978PhRvD..18.1259M}. The interpolation between the $E_8$ spectrum corresponding to $M=0,h>0$ and the free fermionic excitations for $h=0$  in both phases was the subject of a thorough quantitative analysis in Refs. \cite{2006hep.th...12304F,2013arXiv1310.4821Z}.

The position in the $M-h$ parameter space can be characterised with the dimensionless combination
\begin{equation}\label{off_eta}
\eta=\frac{M}{|h|^{8/15}}\,,
\end{equation}
which can be taken as the parameter describing the magnitude of the quench. In the following we work in units of $m_1$ given by Eq. \eqref{eq:e8gap} using the coupling $h$ in the integrable pre-quench model, i.e. the mass gap of the pre-quench system. There is an additional modification to the earlier comparison: when we calculate the finite volume normalisation of TCSA data given by Eqs. \eqref{eq:onepnorm} we have to use the post-quench masses. These can be obtained from TCSA using a cut-off extrapolation scheme similar to the case of the overlaps (see Appendix  \ref{app:epol}).
\begin{figure}[t!]
\begin{tabular}{cc}
\begin{subfloat}
{\includegraphics[width=0.45\textwidth]{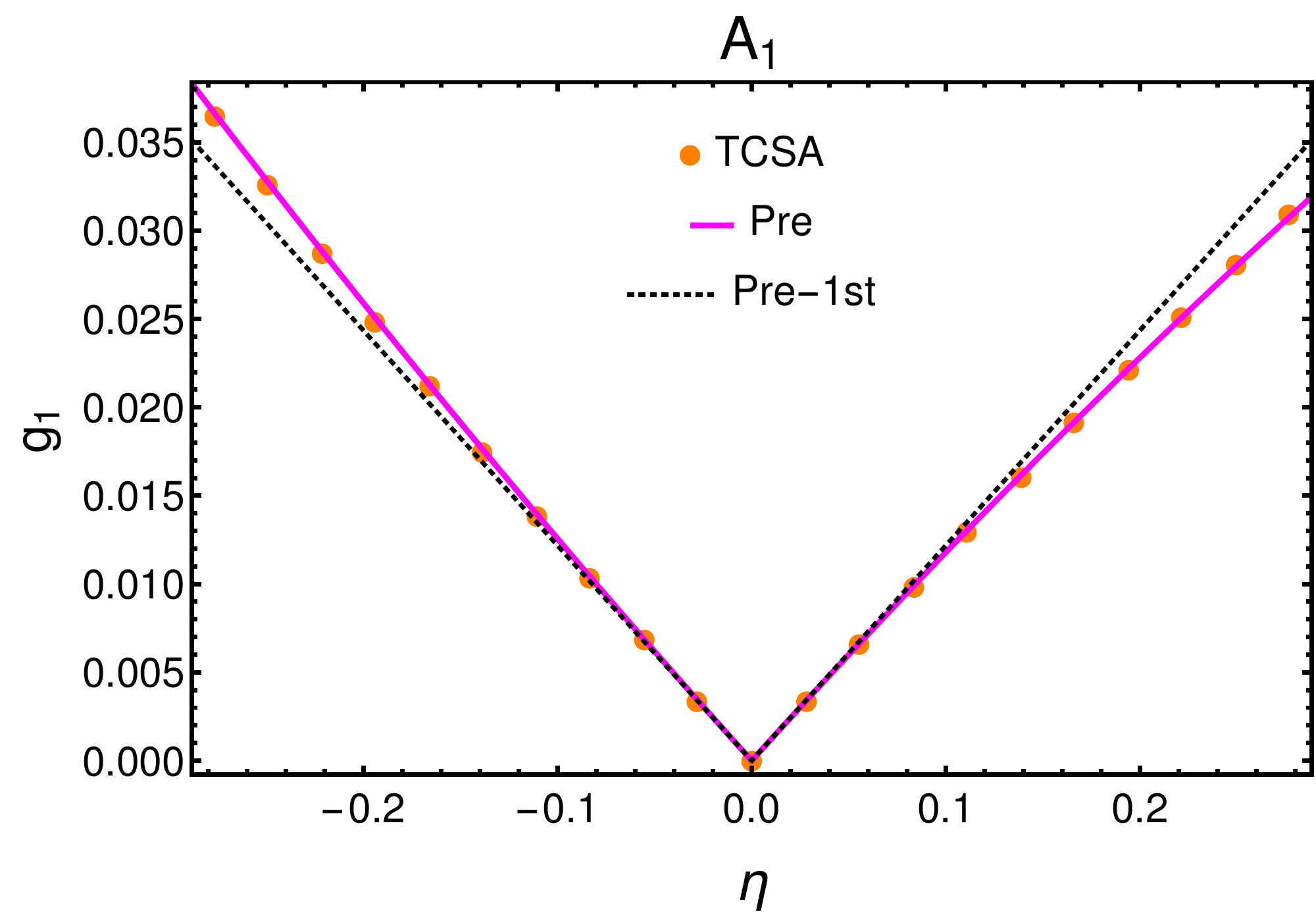}
\label{subfig:goff1}}
\end{subfloat} &
\begin{subfloat}
{\includegraphics[width=0.45\textwidth]{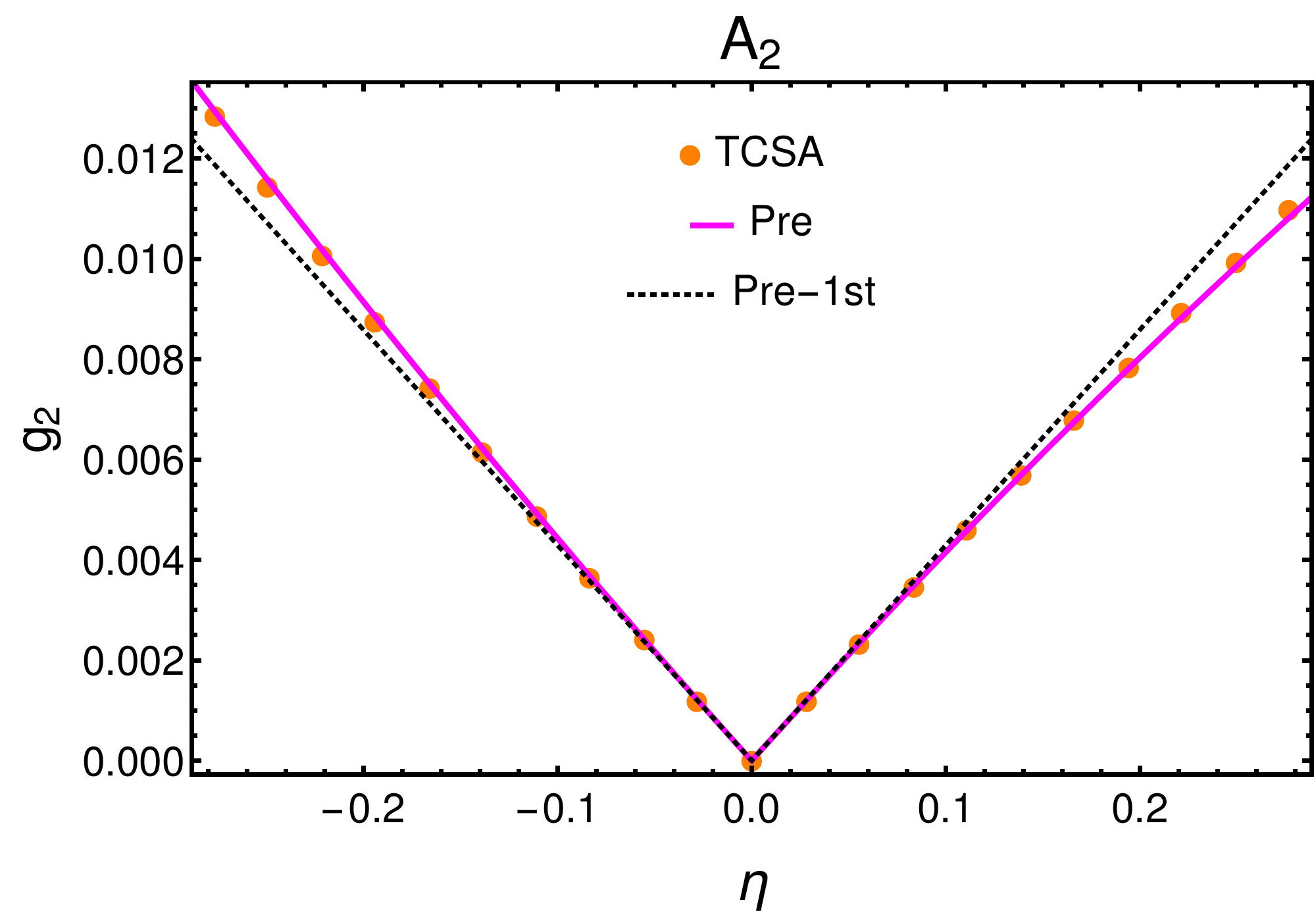}
\label{subfig:goff2}}
\end{subfloat} \\
\begin{subfloat}
{\includegraphics[width=0.45\textwidth]{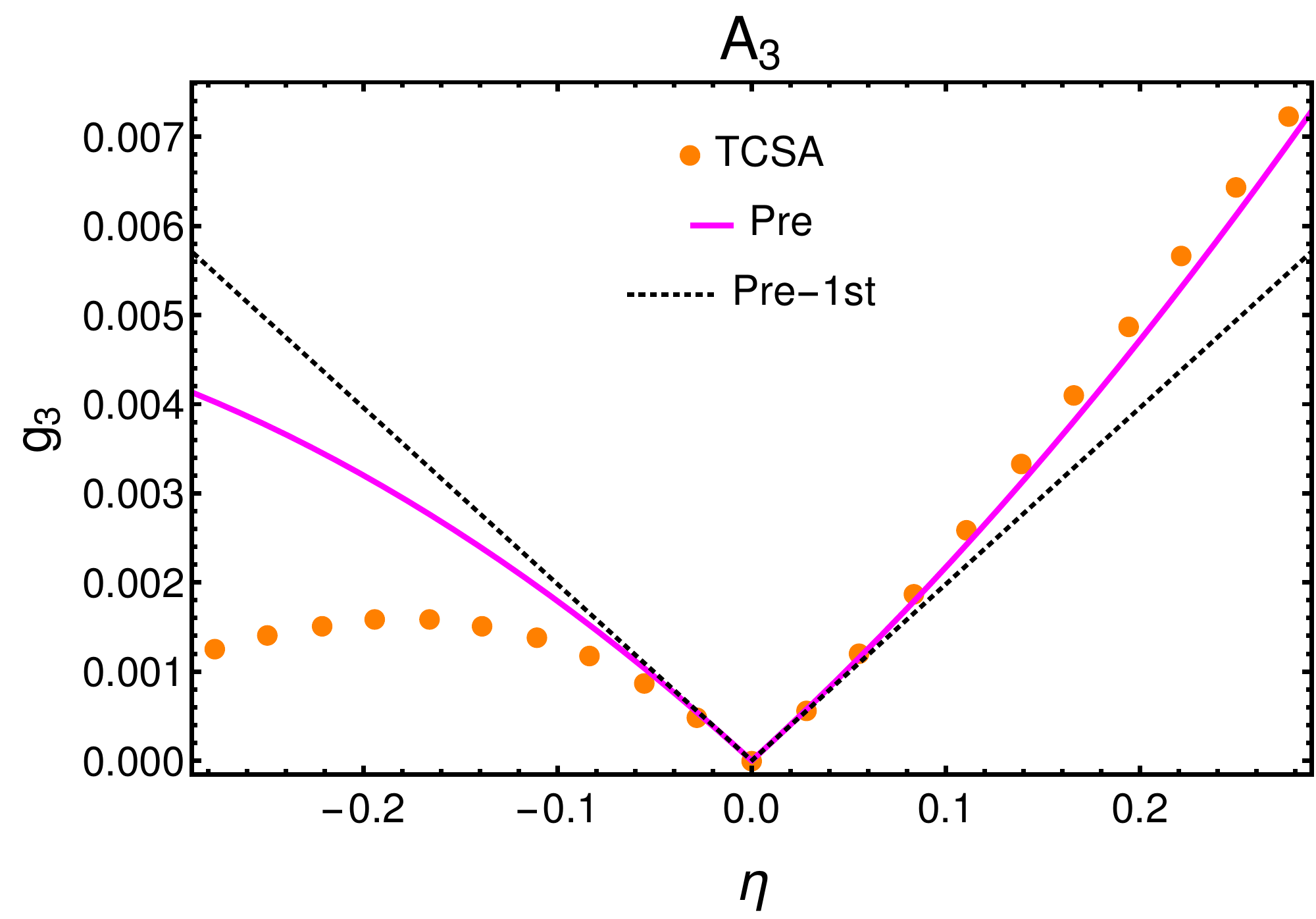}
\label{subfig:goff3}}
\end{subfloat} &
\begin{subfloat}
{\includegraphics[width=0.45\textwidth]{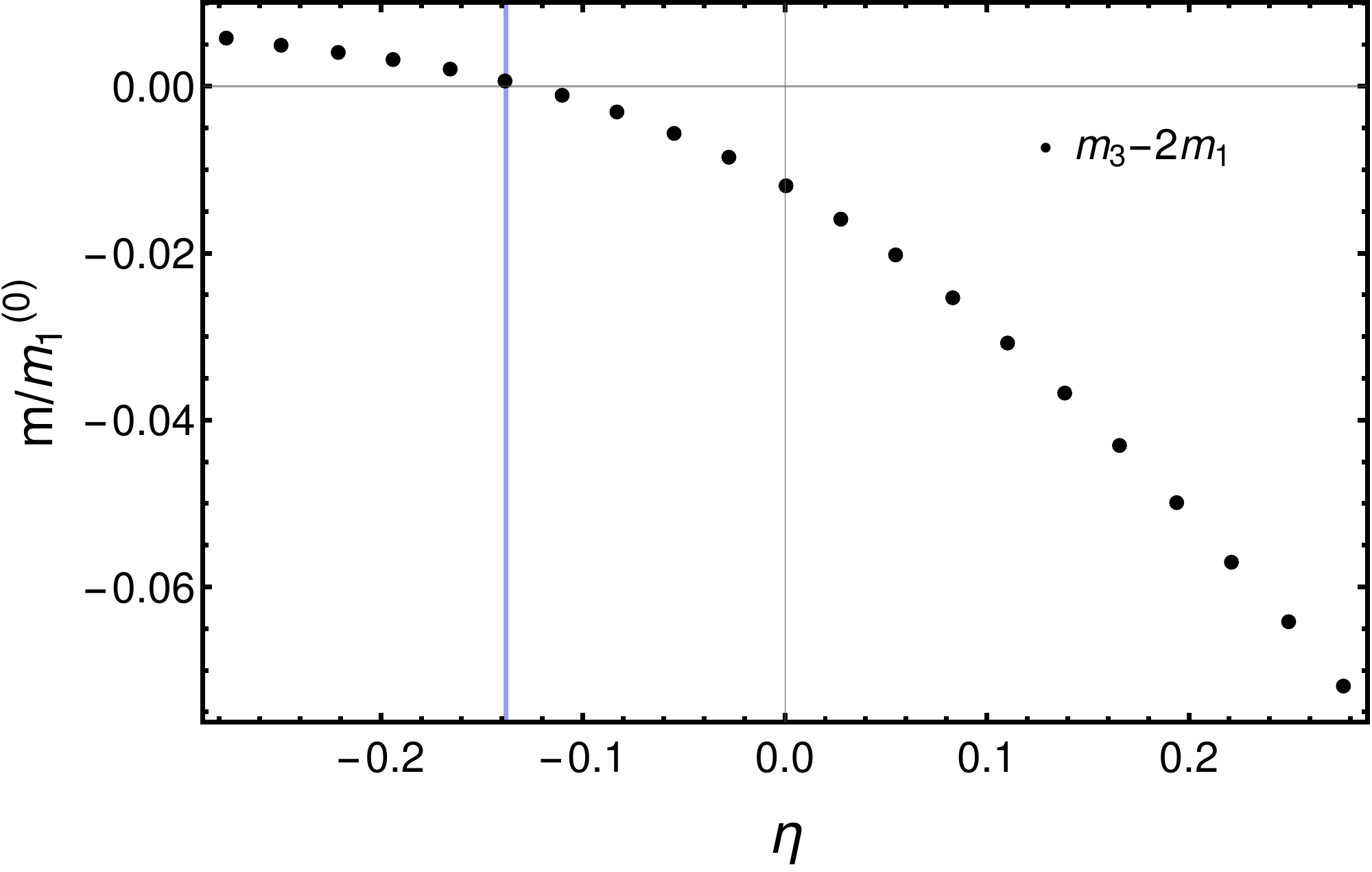}
\label{subfig:m3}}
\end{subfloat}
\end{tabular}
\caption{The overlaps of the three lowest-lying one-particle states for non-integrable quenches as functions of $\eta$ \eqref{off_eta} compared with TCSA data for $m_1^\text{pre}R=40$. Adding the second order yields better agreement in all cases. We observe sizeable deviation only in the case of the third particle in the paramagnetic direction due to the state becoming unstable, i.e. $m_3>2m_1,$ as it is shown in the bottom right panel. The stability threshold $\eta_3=-0.138$ is indicated by a blue grid line.}
\label{fig:onep_off}
\end{figure}

The overlaps of the first two one-particle states are excellently captured by the perturbative expansion. The third particle shows an asymmetry in terms of agreement with TCSA data: we observe a small deviation from the perturbative result in the ferromagnetic direction, but a sizeable difference in the paramagnetic region. This can be understood in the light of the instability of the third particle in the paramagnetic regime. As $\eta$ decreases, it crosses a threshold value $\eta_3=-0.138$ where $m_3$ becomes smaller than $2m_1$, so the corresponding particle state becomes unstable for $\eta<\eta_3$ \cite{2013arXiv1310.4821Z}. This is a non-perturbative phenomenon which is reflected by the deviation of TCSA data from the perturbative prediction, observable even for small negative $\eta$.

\section{Conclusions}
\label{sec:conc}

In this work we used perturbation theory to construct overlaps of post-quench eigenstates with the ground state of the pre-quench Hamiltonian for quenches between integrable and close to integrable Hamiltonians. As a testing ground for our approach, we used quenches in the scaling Ising Field Theory (IFT).

The first approach we developed is based on a perturbative expansion of the initial state in the basis of post-quench eigenstates. This approach is effective when the post-quench dynamics is integrable, since then the required matrix elements can be expressed in terms of form factors which can often be constructed exactly using the form factor bootstrap. Naive application of ordinary non-degenerate perturbation theory leads to divergences which can be handled by putting the system in finite volume using the formalism developed in \cite{2008NuPhB.788..167P,2008NuPhB.788..209P}. Separating the contributions of connected and disconnected parts yields a finite result for the overlap functions. In practice, we performed this procedure for states consisting of at most two particles up to second order in perturbation theory.  The second order contribution is eventually expressed as an infinite sum of form factors which can however be truncated due to its excellent convergence properties. To reduce the residual error from truncating the sum over multi-particle intermediate states, we extended the construction of  form factors in the IFT beyond the results derived in Refs. \cite{1996NuPhB.473..469D,2006NuPhB.737..291D}. The numerical evaluation of the perturbative formulae established that the second order contribution is nearly complete using the available form factors (see Appendix \ref{app:tables}).

The validity of our perturbative calculation can be checked by comparison with numerical simulations of quenches in the Ising Field Theory obtained by the Truncated Conformal Space Approach. This method provides an accurate description of the dynamics for various quenches \cite{2016NuPhB.911..805R,2018ScPP....5...27H}. The post-quench perturbative expansion was shown to agree with the TCSA results to a very high precision over a wide range of the parameter describing the magnitude of the quench. 
This was also the case for overlaps with two-particle states composed of particles of different species which violate the property \eqref{eq:integrable_two_particle_overlap} of integrable initial states. This provides further support for the claim that the quench is not necessarily integrable even if both the pre-quench and post-quench dynamics are \cite{2014JPhA...47N2001D,2017JPhA...50h4004D}.
The results for the pair overlap function also  showed the appearance of the zero-momentum pole for $K_{aa}(p)$, which was proved to hold generally if there is non-zero overlap with one-particle states \cite{2018JHEP...08..170H}.

The second approach we examined used perturbation theory in the pre-quench basis. This approach is effective when the pre-quench dynamics is integrable, and is directly related to a recently suggested perturbative approach to describe the post-quench dynamics \cite{2014JPhA...47N2001D,2017JPhA...50h4004D}. In principle, this allows for the post-quench system to be non-integrable and the study of the effects of integrability breaking on the non-equilibrium dynamics. However, it turned out that the approach ran into difficulties for overlaps of states lying in the multi-particle continuum due to the presence of energy poles in the perturbative expansion, precluding a well-defined infinite volume limit. This difficulty is only absent for quenches with integrable post-quench dynamics, where the pre-quench expansion yields results very similar to the post-quench one. However, for quenches to a non-integrable post-quench Hamiltonian, only the overlaps for the lowest-lying states can be constructed using this approach. For such overlaps we found good agreement with the numerical TCSA data, where the main limitation is posed by the instability of the third lightest particle. It is desirable to extend the applicability of this method as it would provide access to quenches where time evolution is governed by a non-integrable Hamiltonian; however, this requires a resolution of the singularities appearing for states in the multi-particle continuum.

\subsection*{Acknowledgements}

The authors are grateful to Dirk Schuricht and Bruno Bertini for useful discussions. This research
was partially supported by the National Research Development and Innovation
Office (NKFIH) under a K-2016 grant no. 119204, by the BME-Nanotechnology FIKP grant of EMMI (BME
FIKP-NAT), and an OTKA grant no. SNN118028. G.T.  acknowledges partial support by the Quantum
Technology National Excellence Program (Project No. 2017-1.2.1-NKP-2017-
00001), while M.K. was also supported by a ``Pr\'emium'' postdoctoral grant of the Hungarian Academy of Sciences.

\bibliography{pertolaps_paper}

\appendix
\section{Some form factor computations}
\subsection{Derivation of the bound state recursive relation}
\label{app:FF1brec}

To construct solutions to the form factor equations, one takes the following Ansatz for the $n$-particle form factors containing only the lightest particle:
\begin{equation}
\label{eq:FFAnsatz}
\begin{split}
F_{n}^{\phi}(\vt_1,\vt_2,\dots \vt_n)&\equiv F_{\underbrace{\scriptstyle 1\dots1}_n}^{\phi}(\vt_1,\vt_2,\dots \vt_n)\\
&=H_n\frac{\Lambda_n(x_1,\dots ,x_n)}{(\omega_n(x_1,\dots ,x_n))^n} \prod_{i<j}^{n}\frac{F_{11}^{\text{min}}(\vt_i-\vt_j)}{D_{11}(\vt_i-\vt_j)(x_i+x_j)}\,,
\end{split}
\end{equation}
where $x\equiv\exp(\vt)$ and $\omega_n$ denotes the elementary symmetric polynomials generated by
\begin{equation}
\label{eq:sympol}
\prod_{k=1}^{n}(x+x_k)=\sum_{j=0}^{n}x^{n-j}\omega_j(x_1,\dots ,x_n)\,,
\end{equation}
and $H_n$ is a constant factor. The operator-dependence is carried by $\Lambda_n(x_1,\dots ,x_n)$ that is an $n$-variable symmetric polynomial that can be expressed in terms of the elementary symmetric polynomials $\omega$. 

Using the above Ansatz, Eqs. \eqref{eq:FFkpol} and \eqref{eq:FFboundpol} yield recurrence relations for the $\Lambda_n$. Here we derive the recursion corresponding to the bound state pole equation  \eqref{eq:FFboundpol} which can be written as
\begin{align}
&-\imath \lim_{\vt_{ab}\rightarrow2\imath\pi/3}(\vt_{ab}-2\imath\pi/3) H_{n+2}\frac{\Lambda_{n+2}(x_a,x_b,x_1,\dots ,x_n)}{(\omega_{n+2})^{n+2}}\times\nonumber\\ &\times\prod_{i=1}^{n}\frac{F_{11}^{\text{min}}(\vt_a-\vt_j)F_{11}^{\text{min}}(\vt_b-\vt_j)}{D_{11}(\vt_a-\vt_j)D_{11}(\vt_b-\vt_j)(x_a+x_j)(x_b+x_j)}
\frac{F_{11}^{\text{min}}(\vt_a-\vt_b)}{D_{11}(\vt_a-\vt_b)(x_a+x_b)}=\\
&=\Gamma_{11}^1 H_{n+1}\frac{\Lambda_{n+1}(x_c,x_1,\dots ,x_n)}{(\omega_{n+1})^{n+1}} \prod_{i=1}^{n}\frac{F_{11}^{\text{min}}(\vt_c-\vt_j)}{D_{11}(\vt_c-\vt_j)(x_c+x_j)}\,.\nonumber
\end{align}
We can parameterise $\vt_a=\vt+\imath\pi/3$, $\vt_b=\vt-\imath\pi/3$ and $\vt_c=\vt$. The bound state poles are encoded in the factors $D_{11}$ which can be expressed as
\begin{equation}
D_{11}(\vt)=P_{2/3}(\vt)P_{2/5}(\vt)P_{1/15}(\vt)\,,
\end{equation}
where
\begin{equation}
\label{eq:Pgam}
P_{\gamma}(\vt)=\frac{\cos(\pi  \gamma)-\cosh (\vt )}{2 \cos ^2\left(\frac{\pi  \gamma }{2}\right)}\,.
\end{equation}
The residue can be calculated using l'Hospital's rule,
\begin{equation}
\label{eq:res}
-\imath \lim_{\vt_{ab}\rightarrow2\imath\pi/3}(\vt_{ab}-2\imath\pi/3)P_{2/3}(\vt_{ab})=\frac{2\cos[2](\pi/3)}{\sin(2\pi/3)}\,,
\end{equation}
resulting in
\begin{eqnarray}
&&\frac{2\cos[2](\pi/3)}{\sin(2\pi/3)}\frac{H_{n+2}}{H_{n+1}}\frac{\Lambda_{n+2}(x e^{\imath\pi/3},x e^{-\imath\pi/3},x_1,\dots ,x_n)}{(\omega_{n+2})^{n+2}}\times\nonumber\\ 
&&\times\prod_{i=1}^{n}\frac{F_{11}^{\text{min}}(\vt+\imath\pi/3-\vt_j)F_{11}^{\text{min}}(\vt-\imath\pi/3-\vt_j)}{D_{11}(\vt+\imath\pi/3-\vt_j)D_{11}(\vt-\imath\pi/3-\vt_j)(x e^{\imath\pi/3}+x_j)(x e^{-\imath\pi/3}+x_j)}\times\nonumber\\
&&\frac{F_{11}^{\text{min}}(2\pi\imath/3)}{P_{2/5}(2\pi\imath/3)P_{1/15}(2\pi\imath/3)(x e^{\imath\pi/3}+x e^{-\imath\pi/3})}=\nonumber\\
&&=\Gamma_{11}^1 \frac{\Lambda_{n+1}(x,x_1,\dots ,x_n)}{(\omega_{n+1})^{n+1}} \prod_{i=1}^{n}\frac{F_{11}^{\text{min}}(\vt-\vt_j)}{D_{11}(\vt-\vt_j)(x+x_j)}\,.	
\end{eqnarray}
The minimal form factor can be written in the form
\begin{equation}
F_{11}^\text{min}(\vt)=-\imath\sinh(\vt/2)G_{2/3}(\vt)G_{2/5}(\vt)G_{1/15}(\vt)\,.
\end{equation}
where the functions 
\begin{equation}
G_\lambda(\vartheta)=\exp\left\{
2\int\limits_0^\infty\frac{dt}{t}
\frac{\cosh\left(\lambda-1/2\right)t}{\cosh\frac{t}{2}\sinh t}\sin^2
\frac{(\imath\pi-\vartheta)t}{2\pi}
\right\}
\end{equation}
satisfy several functional identities \cite{1996NuPhB.473..469D}, of which the following two are especially important:
\begin{gather}
G_\lambda(\vt+\imath\pi\alpha)G_\lambda(\vt-\imath\pi\alpha)=\frac{G_\lambda(\imath\pi\alpha)G_\lambda(-\imath\pi\alpha)}{G_{\lambda+\alpha}(0)G_{\lambda-\alpha}(0)}G_{\lambda+\alpha}(\vt)G_{\lambda-\alpha}(\vt)\,,\\
G_{1-\lambda}(\vt)G_\lambda(\vt)=\frac{\sinh(1/2(\vt-\imath(\lambda-1)\pi))\sinh(1/2(\vt+\imath(\lambda+1)\pi))}{\sin[2](\pi\lambda/2)}\,.
\label{eq:Gids}
\end{gather}
We can then compute
\begin{multline}
\frac{F_{11}^{\text{min}}(\vt+\imath\pi/3-\vt_j)F_{11}^{\text{min}}(\vt-\imath\pi/3-\vt_j)}{F_{11}^{\text{min}}(\vt-\vt_j)}=\\
\frac{1}{\gamma}\prod_{\sigma_1,\sigma_2=\pm}\sinh((\vt-\vt_j)/2+\imath\pi\sigma_1/6)
\frac{\sinh(\frac{1}{2}(\vt-\vt_j+4\imath\pi\sigma_2/15))}{\sin[2](11\pi/30)}\,,
\label{eq:Fminsimp}
\end{multline}
where we used the identity $G_1(\vt)=-\imath\sinh(\vt/2)$, and denoted
\begin{equation}
\frac{1}{\gamma}\equiv\frac{\prod_{\sigma_1,\sigma_2=\pm}G_{1/15}(\sigma_1\imath\pi/3)G_{2/5}(\sigma_2\imath\pi/3)G_{2/3}(2\imath\pi/3)G_{2/3}(0)}{G_1(\imath\pi/3)G_{1/3}(\imath\pi/3)G_{11/15}(0)G_{1/15}(0)G_{2/5}(0)G_{-4/15}(0)}\,.
\end{equation}
Introducing further the function
\begin{equation}
G_{11}(\vartheta)=G_{1/15}(\vartheta)G_{2/5}(\vartheta)G_{2/3}(\vartheta)\,,
\end{equation}
one obtains
\begin{eqnarray}
&&\frac{H_{n+2}}{H_{n+1}}\frac{\Lambda_{n+2}(x e^{\imath\pi/3},x e^{-\imath\pi/3},x_1,\dots ,x_n)}{x^{n+4}\omega_{n}}	\frac{G_{11}(2\pi\imath/3)}{P_{2/5}(2\pi\imath/3)P_{1/15}(2\pi\imath/3)}\times\nonumber\\ 
&&\times\prod_{i=1}^{n}\frac{\prod_{\sigma_1,\sigma_2=\pm}\sinh((\vt-\vt_j)/2+\imath\pi\sigma_1/6)\sinh(\frac{1}{2}(\vt-\vt_j+4\imath\pi\sigma_2/15))}{D_{11}(\vt+\imath\pi/3-\vt_j)D_{11}(\vt-\imath\pi/3-\vt_j)(x e^{\imath\pi/3}+x_j)(x e^{-\imath\pi/3}+x_j)\sin[2](11\pi/30)\gamma}=\nonumber\\
&&=\frac{1}{2\cos[2](\pi/3)}\Gamma_{11}^1\Lambda_{n+1}(x,x_1,\dots ,x_n) \prod_{i=1}^{n}\frac{1}{D_{11}(\vt-\vt_j)(x+x_j)}\,.	
\end{eqnarray}
Considering now the $D$ factors
\begin{multline}
\label{eq:Dcomp}
\frac{D_{11}(\vt+\imath\pi/3-\vt_j)D_{11}(\vt-\imath\pi/3-\vt_j)}{D_{11}(\vt-\vt_j)}=\\
=\frac{\prod_{\sigma_1,\sigma_2,\sigma_3=\pm}P_{2/3}(\vt+\imath\pi\sigma_1/3-\vt_j)P_{2/5}(\vt+\imath\pi\sigma_2/3-\vt_j)P_{1/15}(\vt+\imath\pi\sigma_3/3-\vt_j)}{P_{2/3}(\vt-\vt_j)P_{2/5}(\vt-\vt_j)P_{1/15}(\vt-\vt_j)}
\end{multline}
and using the identity
\begin{equation}
\label{eq:keytrig}
\cos(\pi  \gamma)-\cosh(\vt )=2\sin(\left(\imath\vt-\pi\gamma\right)/2)\sin(\left(\imath\vt+\pi\gamma\right)/2) \equiv\prod_{\sigma=\pm}2\sin(\left(\imath\vt+\pi\gamma\sigma\right)/2)\,,
\end{equation}
Eq. \eqref{eq:Dcomp} can be simplified as
\begin{multline}
\label{eq:Dsimp}
\frac{D_{11}(\vt+\imath\pi/3-\vt_j)D_{11}(\vt-\imath\pi/3-\vt_j)}{D_{11}(\vt-\vt_j)}=\\
=\prod_{\sigma_1,\sigma_2=\pm}\sin(\frac{\imath}{2}(\vt-\vt_j)+\pi\sigma_1/2)\sin(\frac{\imath}{2}(\vt-\vt_j)+\pi\sigma_2/6)\times\\
\times\frac{\prod_{\sigma_3,\sigma_4=\pm}\sin(\frac{\imath}{2}(\vt-\vt_j)+ 2\pi\sigma_3/15)\sin(\frac{\imath}{2}(\vt-\vt_j)+11\pi\sigma_4/30)}{\prod_{\sigma=\pm}\cos[2](\pi/3)\cos[2](\pi/5)\cos[2](\pi/30)\sin(\frac{\imath}{2}(\vt-\vt_j)+\pi\sigma/3)}\,.
\end{multline}
Using 
\begin{equation}
P_{2/5}(2\pi\imath/3)P_{1/15}(2\pi\imath/3)=\frac{\sin\frac{8\pi}{15}\sin\frac{2\pi}{15}\sin\frac{3\pi}{10}\sin\frac{11\pi}{30}}{\cos[2](\pi/5)\cos[2](\pi/30)}
\end{equation}
and the identity
\begin{equation}
\prod_{\sigma=\pm}\sin(\frac{\imath}{2}(\vt-\vt_j)+\sigma\pi\gamma/2)=-\frac{1}{4 xx_j} (x-e^{-\imath\pi\gamma}x_j)(x-e^{\imath\pi\gamma}x_j)\,,
\end{equation}
the final form of the recurrence relation is:
\begin{equation}
\label{eq:bFFfinal}
\frac{\Lambda_{n+2}(x e^{\imath\pi/3},x e^{-\imath\pi/3},x_1,\dots ,x_n)}{x^4\prod_{i=1}^{n}(x-e^{-11\imath\pi/15}x_j)(x-e^{11\imath\pi/15}x_j)(x+x_j)}=(-1)^n\Lambda_{n+1}(x,x_1,\dots ,x_n)\,,
\end{equation}
provided the $H_n$ are chosen to satisfy
\begin{multline}
\frac{H_{n+2}}{H_{n+1}}=\frac{\Gamma_{11}^1}{2\cos[2](\pi/3)\cos[2](\pi/5)\cos[2](\pi/30)G_{11}(2\pi\imath/3)}\times\\
\times\left[\frac{\sin[2](11\pi/30)\gamma}{4\cos[2](\pi/3)\cos[2](\pi/5)\cos[2](\pi/30)}\right]^n\,.
\label{eq:Hrec}
\end{multline}

\subsection{Form factors involving higher species from bound state equations}
\label{app:FFbdown}
Form factors involving species other than $A_1$ can be obtained using the bound state singularity equation \eqref{eq:FFboundpol}. Consider a particle $A_c$ that is the bound state of two $A_1$ particles, then the relevant singularity takes the form
\begin{equation}
\label{eq:FFbound2}
-\imath \lim_{\vt_{ab}\rightarrow\imath u_{11}^c}(\vt_{ab}-\imath u_{11}^c) F_{n+2}^{\phi}(\vt_a,\vt_b,\vt_1,\vt_2,\dots \vt_n)=\Gamma_{11}^c F_{c,n}^{\phi}(\vt_c,\vt_1,\vt_2,\dots \vt_n)\,,
\end{equation}
where the index $n$ is a short-hand for  $n$ particles of type $A_1$. 

For form factors containing particles of different species, the Ansatz \eqref{eq:FFAnsatz} must be generalised to the form 
\begin{equation}
\label{eq:FFAnsatz2}
F_{a_1,a_2,\dots a_n}^{\phi}(\vt_1,\vt_2,\dots \vt_n)=Q_{a_1,a_2,\dots a_n}^\phi(x_1,\dots ,x_n) \prod_{i<j}^{n}\frac{F_{a_i,a_j}^{\text{min}}(\vt_i-\vt_j)}{D_{a_i,a_j}(\vt_i-\vt_j)(x_i+x_j)^{\delta_{a_i,a_j}}}\,,
\end{equation}
where the $D_{ab}$ factors ensure the correct position of the bound state poles, the functions $F^\text{min}_{ab}$ are minimal form factors that have no poles in the strip $\Im\vt\in[0,\pi]$ and satisfy 
\begin{equation}
 F_{ab}^{\text{min}}(\imath\pi-\vt)=F_{ab}^{\text{min}}(\imath\pi+\vt)\qquad
 F_{ab}^{\text{min}}(\vt)=S_{11}(\vt)=F_{ab}^{\text{min}}(-\vt)\,
\end{equation}
$\delta_{a,b}$ is the Kronecker delta, and 
\begin{equation}
\label{eq:QAns}
Q_{a_1,\dots ,a_n}(x_1,\dots ,x_n)=\sum_{\alpha_1=\alpha_1'}^{\alpha_1''}\dots \sum_{\alpha_n=\alpha_n'}^{\alpha_n''}d_{a_1,\dots a_n}^{\alpha_1,\dots ,\alpha_n}x_1^{\alpha_1}\dots x_n^{\alpha_n}
\end{equation}
is a polynomial symmetric under exchange of particles of the same species and its degree is constrained by the asymptotic behaviour of the form factors \cite{2006NuPhB.737..291D}. Using this Ansatz  Eq. \eqref{eq:FFbound2} can be written as
\begin{align}
&-\imath \lim_{\vt_{ab}\rightarrow\imath u_{11}^c}(\vt_{ab}-\imath u_{11}^c) Q_{n+2}^\phi(x_a,x_b,x_1,\dots ,x_n)\times\nonumber\\ 
&\times\prod_{i=1}^{n}\frac{F_{11}^{\text{min}}(\vt_a-\vt_j)F_{11}^{\text{min}}(\vt_b-\vt_j)}{D_{11}(\vt_a-\vt_j)D_{11}(\vt_b-\vt_j)(x_a+x_j)(x_b+x_j)}
\frac{F_{11}^{\text{min}}(\vt_a-\vt_b)}{D_{11}(\vt_a-\vt_b)(x_a+x_b)}=\\
&=\Gamma_{11}^c Q_{c,n}^\phi(x_c,x_1,\dots ,x_n) \prod_{i=1}^{n}\frac{F_{c1}^{\text{min}}(\vt_c-\vt_j)}{D_{c1}(\vt_c-\vt_j)}\,.\nonumber
\end{align}
For definiteness we choose the pole at $2\imath\pi/5$ which corresponds to the particle $A_2$ (the other pole at $\imath\pi/15$ corresponding to $A_3$ can be handled similarly). The residue can be computed using
\begin{equation}
\label{eq:res2}
-\imath \lim_{\vt_{ab}\rightarrow2\imath\pi/5}(\vt_{ab}-2\imath\pi/5)P_{2/5}(\vt_{ab})=\frac{2\cos[2](\pi/5)}{\sin(2\pi/5)}\,,
\end{equation}
which leads to
\begin{align}
&Q_{n+2}^\phi(x e^{\imath\pi/5},x  e^{-\imath\pi/5},x_1,\dots ,x_n) 
\frac{\sin(\pi/5)G_{11}(2\imath\pi/5)\cos(\pi/5)}{P_{2/3}(2\imath\pi/5)P_{1/15}(2\imath\pi/5)x\sin(2\pi/5)}
\times\nonumber\\
&\times \prod_{i=1}^{n}\frac{\prod_{\sigma=\pm}F_{11}^{\text{min}}(\vt-\vt_j+\imath\pi\sigma/5)D_{12}(\vt-\vt_j)}{\prod_{\sigma'=\pm}F_{12}^{\text{min}}(\vt-\vt_j)D_{11}(\vt-\vt_j+\imath\pi\sigma'/5)(xe^{\imath\pi/5}+x_j)(xe^{-\imath\pi/5}+x_j)}\nonumber \\
&=\Gamma_{11}^2 Q_{2,n}^\phi(x,x_1,\dots ,x_n)\,.
\end{align}
Using the identities \eqref{eq:Gids} the form factor product simplifies once again, leading to an analogous calculation to that detailed above. The final result reads
\begin{multline}
\label{bFF2final}
\frac{	Q_{n+2}^\phi(x e^{\imath\pi/5},x  e^{-\imath\pi/5},x_1,\dots ,x_n)\prod_{i=1}^{n}\prod_{\sigma_=\pm}(x-e^{4\imath\pi\sigma/5}x_j)}{x\prod_{i=1}^{n}\prod_{\sigma_1,\sigma_2=\pm}(x-e^{13\imath\pi\sigma_1/15}x_j)(x+x_j e^{\imath\pi\sigma_2/5})}=\\
=-(C_2)^{-n}\frac{\sin(\pi/5)\sin\frac{8\pi}{15}\sin\frac{7\pi}{30}\sin\frac{\pi}{6}\sin\frac{2\pi}{15}}{\sin(2\pi/5)\cos[2](\pi/3)\cos(\pi/5)\cos[2](\pi/30)G_{11}(2\imath\pi/5)}\Gamma_{11}^2 Q_{2,n}^\phi(x,x_1,\dots ,x_n)
\end{multline}
with
\begin{equation}
C_2=\frac{(\cos[4](\pi/3)\cos[4](\pi/5)\cos[4](\pi/30))\gamma_2}{\cos[2](2\pi/15)\cos[2](7\pi/30)\cos[2](3\pi/10)\cos[2](2\pi/5)\sin[2](13\pi/30)}
\end{equation}
and
\begin{equation}
\gamma_2=\frac{\prod_{\sigma_1,\sigma_2,\sigma_3=\pm}G_{1/15}(\sigma_1\imath\pi/5)G_{2/5}(\sigma_2\imath\pi/5)G_{2/3}(\sigma_3\imath\pi/5)}{G_{13/15}(0)G_{7/15}(0)G_{3/5}(0)G_{1/5}(0)G_{-2/15}(0)G_{4/15}(0)}\,.
\end{equation}


\section{Perturbative calculations}
\label{app:pert}

\subsection{Rayleigh--Schrödinger expansion}
\label{app:RS}
Taking a Hamiltonian $H=H_0+\lambda V$, its spectrum and the eigenstates  can be found using Rayleigh--Schrödinger perturbation theory based on the Hamiltonian $H_0$ with eigenstates $\ket*{n^{(0)}}$,
\begin{equation}
\label{eq:basspec}
H_0\ket*{n^{(0)}}=E^{(0)}_n\ket*{n^{(0)}}\,,
\end{equation}
expressed as a power series in $\lambda$:
\begin{align}
\label{eq:pert}
E_n= & E_n^{(0)}+\lambda E_n^{(1)}+ \lambda^2 E_n^{(2)}+\dots \,, \\
\ket*{n}= & \ket*{n^{(0)}}+\lambda \ket*{n^{(1)}}+ \lambda^2 \ket*{n^{(2)}}+\dots \,,
\end{align}
where the ellipses denote higher order contributions in $\lambda$. Here we only need the expansion for the eigenstates which to second order takes the form
\begin{eqnarray}
\ket*{n}&=&\ket*{n^{(0)}}+\lambda\sum_{k\neq n}\frac{\bra*{k^{(0)}}V\ket*{n^{(0)}}}{E^{(0)}_n-E^{(0)}_k}\ket*{k^{(0)}}+ 
\lambda^2\Bigg[\sum_{k\neq n}\sum_{l\neq n}\frac{\bra*{l^{(0)}}V\ket*{k^{(0)}}\bra*{k^{(0)}}V\ket*{n^{(0)}}}{(E^{(0)}_n-E^{(0)}_k)(E^{(0)}_n-E^{(0)}_l)}\ket*{l^{(0)}}
\nonumber\\
&&-\sum_{k\neq n}\frac{\bra*{n^{(0)}}V\ket*{n^{(0)}}\bra*{k^{(0)}}V\ket*{n^{(0)}}}{(E^{(0)}_n-E^{(0)}_k)^2}\ket*{k^{(0)}}\Bigg]+O(\lambda^3)\,.
\label{eq:pert_es}
\end{eqnarray}
Note that the resulting expression for the states is not normalised. Quantities expressed on this basis must be normalised by dividing with the norm of the ground state, which is $\mathcal{N}=1+\mathcal{O}(\lambda^2)$. However, for our calculations up to $\mathcal{O}(\lambda^2)$ this is irrelevant since the leading order of the overlaps is always $\mathcal{O}(\lambda)$.

\subsection{Dealing with disconnected pieces}
\label{app:twopcalcs}
\subsubsection{The case $K_{aa}(\vt)$}
Disconnected contributions appear in the second order of perturbation theory, the relevant contributions can be extracted from Eq. \eqref{eq:twop01}. Restoring particle labels the relevant contribution to $K_{aa}(\vt,-\vt)$ is given as
\begin{equation}
\label{eq:infdisc01}
D_{ab}(\vt,-\vt)\equiv\frac{\tilde{\rho}_a(\vt)}{\sqrt{\rho_{aa}(\vt,-\vt)}}\lambda^2L^2 \sum_{\vt'}\frac{\mathbin{_{ab}\braket*{\{\vt',\vt'_{ab}\}}{\phi|0}_L} \,\mathbin{_{aa}\braket*{\{\vt,-\vt\}}{\phi|\{\vt',\vt'_{ab}\}}_{ab,L}}}{2m_a\cosh\vt(m_a\cosh\vt'+m_b\cosh\vt'_{ab})}
\end{equation}
which must be summed over the intermediate species label $b$, and where $\vt'_{ab}$ is defined similarly to \eqref{eq:vtbc}:
\begin{equation}
\label{eq:vtab}
\vt'_{ab}=-\text{arcsinh}\left(\frac{m_a\sinh\vt'}{m_b}\right)\,,
\end{equation}
and the summation over $\vt'$ runs over the solutions of the following Bethe--Yang equation:
\begin{equation}
\label{eq:BY0}
\tilde{Q}(\vt')=m_a\sinh\vt'+\delta_{ab}(\vt'-\vt'_{ab})=2\pi I\,,
\end{equation}
indexed by the quantum number $ I\in\mathbb{Z}$. Expressing the finite volume matrix elements with the form factors yields
\begin{equation}
\label{eq:infdisc02}
D_{ab}(\vt,-\vt)=\frac{\lambda^2}{2m_a^2\cosh[2](\vt)}\sum_{\vt'}\frac{F_{ab}^{\phi*}(\vt',\vt'_{ab})F_{aaab}^\phi(\imath\pi+\vt,\imath\pi-\vt,\vt',\vt'_{ab})}{(m_a\cosh\vt'+m_b\cosh\vt'_{ab})m_a\cosh\vt'\tilde{\rho}_a(\vt')}\,,
\end{equation}
where $\tilde{\rho}_a({\vt'})$ is obtained by differentiating the function  $\tilde{Q}$ defined in Eq. \eqref{eq:BY0}.  Since the four-particle form factor has poles for $\vt'=\pm\vt$ it is not possible to replace the sum with a simple integral. This problem can avoided by using contour integrals to express the sum. Denoting
\begin{equation}
\label{eq:Dabnot}
d_{ab}\equiv\sum_{\vt'} \frac{f(\vt')}{\tilde{\rho}_a(\vt')}\equiv
\sum_{\vt'}\frac{F_{ab}^{\phi*}(\vt',\vt'_{ab})F_{aaab}^\phi(\imath\pi+\vt,\imath\pi-\vt,\vt',\vt'_{ab})}{(m_a\cosh\vt'+m_b\cosh\vt'_{ab})m_a\cosh\vt'\tilde{\rho}_a(\vt')} \,.
\end{equation}
one can write
\begin{equation}
d_{ab}=-\sum_{\vt'}\oint_{\vt'}\frac{\dd \theta}{2\pi}\frac{f(\theta)}{1-e^{\imath \tilde{Q}(\theta)}}\,,
\end{equation}
where the contours go around each root $\vt'$ counterclockwise. The contours can be deformed to give
\begin{equation}
d_{ab}=\left(\int_{-\infty+\imath\epsilon}^{\infty+\imath\epsilon}-\int_{-\infty-\imath\epsilon}^{\infty-\imath\epsilon}+\oint_\vt+\oint_{-\vt}\right)\frac{\dd \theta}{2\pi}\frac{f(\theta)}{1-e^{\imath \tilde{Q}(\theta)}}\,,
\end{equation}
where $\epsilon$ is a small shift. The second integral vanishes in the infinite volume limit due to
\begin{equation}
\label{eq:Qinf}
\lim_{L\rightarrow\infty}\frac{1}{1-e^{\imath \tilde{Q}(\theta+\imath\epsilon)}}=\lim_{L\rightarrow\infty}\frac{1}{1-e^{\imath m_aL(\sinh\theta\cos\epsilon+\imath\cosh\theta\sin\epsilon)+O\left(L^0\right)}}=\begin{cases}
1\,, & \epsilon>0 \\
0\,, & \epsilon<0
\end{cases}
\end{equation}
and in the first one only the numerator remains. Moreover, the integral contour can be pulled back to the real axis using
\begin{equation}
\int_{-\infty}^{\infty}\dd{\vt'}\coth(\vt'\pm\vt)=0\,,
\end{equation}
which can be proved by shifting the integration contour to $\Im\vt'=\pi/2$: 
\begin{equation}
\int_{-\infty+\imath\epsilon}^{\infty+\imath\epsilon}\frac{\dd \theta}{2\pi}f(\theta)=\int_{-\infty}^{\infty}\frac{\dd \vt'}{2\pi}\left[f(\vt')-R_1(\vt)\coth(\vt'-\vt)-R_2(-\vt)\coth(\vt'+\vt)\right]\,,
\end{equation}
where $R_1$ and $R_2$ are the residues of $f(\vt')$ at $\vt'=\vt$ and $\vt'=-\vt$, respectively, which also appear in the contributions of the two isolated poles:
\begin{align}
\label{eq:K11res}
\oint_\vt\frac{\dd \theta}{2\pi}\frac{f(\theta)}{1-e^{\imath \tilde{Q}(\theta)}}=&\imath\frac{R_1(\vt)}{1-e^{\imath \tilde{Q}(\vt)}}\,,\nonumber\\
\oint_{-\vt}\frac{\dd \theta}{2\pi}\frac{f(\theta)}{1-e^{\imath \tilde{Q}(\theta)}}=&\imath\frac{R_2(-\vt)}{1-e^{\imath \tilde{Q}(-\vt)}}\,.
\end{align}
The residues can be calculated using Eqs. \eqref{eq:FFkpol} and \eqref{eq:ordSmat}:
\begin{align}
\label{eq:discRes}
R_1(\vt)=&\frac{-\imath F_{ab}^\phi(\imath\pi-\vt,\vt_{ab})F_{ab}^{\phi*}(\vt,\vt_{ab})S_{aa}(2\vt)(1-S_{aa}^*(2\vt)S_{ab}(\vt-\vt_{ab}))}{(m_a\cosh\vt+m_b\cosh\vt_{ab})m_a\cosh\vt}\,\nonumber\\
R_2(-\vt)=&\frac{R_1(-\vt)}{S_{aa}(-2\vt)}\,.
\end{align}
The denominators in \eqref{eq:K11res} can be simplified using that $\pm\vt$ are solutions to another Bethe--Yang equation:
\begin{equation}
\label{eq:BY02}
\tilde{Q}'(\vt)=m_aL\sinh\vt+\delta_{aa}(2\vt)=2\pi J\qquad J\in\mathbb{Z}+\frac{1}{2}\,,
\end{equation}
with
\begin{equation}
S_{aa}(\vt)=-e^{\imath\delta_{aa}(\vt)}\,.
\end{equation}
Comparing with Eq. \eqref{eq:BY0} yields
\begin{equation}
\label{eq:etadef}
1-e^{\imath \tilde{Q}(\vt)}=1-S_{aa}^*(2\vt)S_{ab}(\vt-\vt_{ab})\equiv\eta^{-1}(\vt)\,.
\end{equation}
Putting everything together one obtains
\begin{align}
\label{eq:infdisc03}
D_{ab}(\vt,-\vt)=\frac{\lambda^2}{2m_a^2\cosh[2](\vt)}
\Big[ & \int_{-\infty}^{\infty}\frac{\dd \vt'}{2\pi}\left(f(\vt')-R_1(\vt)\coth(\vt'-\vt)-R_2(-\vt)\coth(\vt'+\vt)\right)+\nonumber\\
&+\imath R_1(\vt)\eta(\vt)+\imath R_2(-\vt)\eta(-\vt)\Big]\,,
\end{align}
with
\begin{equation}
\label{eq:fdef}
f(\vt')=\frac{F_{ab}^{\phi*}(\vt',\vt'_{ab})F_{aaab}^\phi(\imath\pi+\vt,\imath\pi-\vt,\vt',\vt'_{ab})}{(m_a\cosh\vt'+m_b\cosh\vt'_{ab})m_a\cosh\vt'}\,.
\end{equation}

Using the above results, one can easily write down the generalisation of Eq. \eqref{eq:twop03} to multiple particle species:
\begin{align}
\label{eq:Kaa_app}
&K_{aa}(-\vt,\vt)=
-\lambda \frac{F_{aa}^{\phi*}(-\vt,\vt)}{2m_a^2\cosh^2\vt}+
\lambda^2\Bigg(\sum_{b=1}^{N_{\text{spec}}}\frac{F_b^\phi F_{baa}^{\phi*}(\imath\pi,-\vt,\vt)}{2m_a^2\cosh^2\vt m_b^2}+
\frac{F_{aa}^{\phi*}(-\vt,\vt)F_{aa}^\phi(\imath\pi, 0)}{2m_a^4\cosh^4\vt}+
\nonumber\\
&+\sum_{(c,d)\neq(a,b)}\frac{1}{(2\delta_{cd})!}\int_{-\infty}^{\infty}\frac{\dd{\vt'}}{2\pi} 
\frac{F_{aacd}^{\phi,s}(\imath\pi+\vt,\imath\pi-\vt,\vt',\vt'_{cd})F_{cd}^{\phi*}(\vt',\vt'_{cd})}{2m_a^2\cosh^2\vt(m_c\cosh\vt'+m_d\cosh\vt'_{cd})m_d{\cosh\vt'_{cd}}} + ...\Bigg)+\nonumber\\
&+\sum_{b=1}^{N_{\text{spec}}}D_{ab}(\vt,-\vt)+ O\left(\lambda^3\right)\,,
\end{align}
where $(c,d)\neq(a,b)$ means that the sum excludes those pairs in which exactly one particle is of species $a$.

\subsubsection{The case $K_{ab}(\vt)$}
The above results can be extended to a pair with different particles. The overlap function is then $K_{ab}(\vt,\vt_{ab})$, with $\vt_{ab}$ as in Eq. \eqref{eq:vtab}. The terms involving to rapidity integrals are given by
\begin{align}
\label{eq:Kab01}
K_{ab}(\vt,&\vt_{ab})=
-\lambda \frac{F_{ab}^{\phi*}(\vt,\vt_{ab})}{(m_a\cosh\vt+m_b\cosh\vt_{ab})m_b\cosh\vt_{ab}}+
\nonumber\\
&+\lambda^2\left(\sum_{c=1}^{N_{\text{spec}}}\frac{F_c^\phi F_{cab}^{\phi*}(\imath\pi,\vt,\vt_{ab})}{m_c^2(m_a\cosh\vt+m_b\cosh\vt_{ab})m_b\cosh\vt_{ab}}+\right.\\\left.
+\right.&\left.\frac{F_{ab}^{\phi*}(\vt,\vt_{ab})}{(m_a\cosh\vt+m_b\cosh\vt_{ab})^2 m_b\cosh\vt_{ab}}\left(\frac{F_{aa}^\phi(\imath\pi, 0)}{m_b\cosh\vt_{ab}}+ \frac{F_{bb}^\phi(\imath\pi, 0)}{m_a\cosh\vt}\right)+ ...\right)+ O\left(\lambda^3\right)\,.\nonumber
\end{align}
Terms containing integrals are similar to the $D_{ab}$ contribution of Eq. \eqref{eq:infdisc01}, and the ones with disconnected contributions have the form
\begin{equation}
\label{eq:infdiscb01}
G_{ab}^c(\vt,\vt_{ab})\equiv\frac{\tilde{\rho}_a(\vt)}{\sqrt{\rho_{ab}(\vt,\vt_{ab})}}\lambda^2L^2 \sum_{\vt'}\frac{\mathbin{_{cc}\braket*{\{\vt',-\vt'\}}{\phi|0}_L} \mathbin{_{ab}\braket*{\{\vt,\vt_{ab}\}}{\phi|\{-\vt',\vt'\}}_{cc,L}}}{2m_c\cosh\vt'(m_a\cosh\vt+m_b\cosh\vt_{ab})}\,,
\end{equation}
with $c=a$ or $b$. The computation required here is similar to the one above, so we only present the results. The $c=a$ term can be expressed as
\begin{align}
\label{eq:Gaba}
G_{ab}^a(\vt,\vt_{ab})=&\frac{\lambda^2}{(m_a\cosh\vt+m_b\cosh\vt_{ab})m_b\cosh\vt_{ab}}\times\\
&\times\Big[
\int_{-\infty}^{\infty}\frac{\dd \vt'}{2\pi}\left(g(\vt')-R_3(\vt)\coth(\vt'-\vt)-R_4(-\vt)\coth(\vt'+\vt)\right)\nonumber\\
&+\imath R_3(\vt)\eta_2(\vt)+\imath R_4(-\vt)\eta_2(-\vt)
\Big]\nonumber\,,
\end{align}
with
\begin{equation}
\label{eq:gdef}
g(\vt')=\frac{F_{aa}^{\phi*}(-\vt',\vt')F_{abaa}^\phi(\imath\pi+\vt,\imath\pi+\vt_{ab},-\vt',\vt')}{2m_a^2\cosh[2](\vt')}\,,
\end{equation}
and
\begin{align}
\label{eq:discbRes}
&R_3(\vt)=\frac{-\imath F_{ba}^\phi(\imath\pi+\vt_{ab},-\vt)F_{aa}^{\phi*}(-\vt,\vt)S_{aa}^*(2\vt)S_{ab}(\vt-\vt_{ab})(1-S_{aa}(2\vt)S_{ab}(\vt_{ab}-\vt))}{2m_a^2\cosh[2](\vt)}\,,\nonumber\\
&R_4(-\vt)=R_3(\vt)\,,\nonumber\\
&\eta_2(\vt)=\frac{1}{1-S_{aa}(2\vt)S_{ab}^*(\vt-\vt_{ab})}\,,
\end{align}
while $G_{ab}^b$ is slightly different:
\begin{align}
\label{eq:Eabb}
G_{ab}^a(\vt,\vt_{ab})&=\frac{\lambda^2}{(m_a\cosh\vt+m_b\cosh\vt_{ab})m_b\cosh\vt_{ab}}\times\\
&\Big[\int_{-\infty}^{\infty}\frac{\dd \vt'}{2\pi}\left(h(\vt')-R_5(\vt)\coth(\vt'-\vt)-R_6(\vt)\coth(\vt'+\vt)\right)+\nonumber\\
&\imath R_5(\vt)\eta_3(\vt)+\imath R_6(\vt)\eta_3(-\vt)\Big]\nonumber
\end{align}
with
\begin{equation}
\label{eq:hdef}
h(\vt')=\frac{F_{bb}^{\phi*}(-\vt',\vt')F_{abbb}^\phi(\imath\pi+\vt,\imath\pi+\vt_{ab},-\vt',\vt')}{2m_b^2\cosh[2](\vt')}
\end{equation}
and
\begin{align}
\label{eq:discbRes2}
&R_5(\vt)=\frac{-\imath F_{ab}^\phi(\imath\pi+\vt,-\vt_{ab})F_{bb}^{\phi*}(-\vt_{ab},\vt_{ab})S_{bb}^*(2\vt_{ab})(1-S_{ab}(\vt-\vt_{ab})S_{bb}(2\vt_{ab}))}{2m_b^2\cosh[2](\vt_{ab})}\,,\nonumber\\
&R_6(\vt)=R_5(\vt)S_{ab}(\vt-\vt_{ab})S_{ab}(-\vt-\vt_{ab})\,,\qquad\nonumber\\
&\eta_2(\vt)=\frac{1}{1-S_{aa}(2\vt)S_{ab}^*(\vt-\vt_{ab})}\,.
\end{align}
There are additional terms corresponding to inserting a state with only one $a$ or $b$ particle; however, terms including such multi-particle form factors are expected to give very small contributions so we neglect them. 

The final result for the $K_{ab}$ function is
\begin{align}
\label{eq:Kab_app}
&K_{ab}(\vt,\vt_{ab})=
-\lambda \frac{F_{ab}^{\phi*}(\vt,\vt_{ab})}{C_{ab}(\vt,\vt_{ab})}
+\lambda^2
\Bigg[
\sum_{c=1}^{N_{\text{spec}}}\frac{F_c^\phi F_{cab}^{\phi*}(\imath\pi,\vt,\vt_{ab})}{m_c^2C_{ab}(\vt,\vt_{ab})}+\nonumber\\
&+\frac{F_{ab}^{\phi*}(\vt,\vt_{ab})}{C_{ab}(\vt,\vt_{ab})(m_a\cosh\vt+m_b\cosh\vt_{ab})}\left(\frac{F_{aa}^\phi(\imath\pi, 0)}{m_b\cosh\vt_{ab}}+\frac{F_{bb}^\phi(\imath\pi, 0)}{m_a\cosh\vt}\right)+\nonumber\\[1em]
&+\sum_{\substack{c\neq a,b \\ d\neq a,b}} \frac{1}{(2\delta_{cd})!}\int_{-\infty}^{\infty}\frac{\dd{\vt'}}{2\pi}  \frac{F_{abcd}^{\phi,s}(\imath\pi+\vt,\imath\pi+\vt_{ab},\vt',\vt'_{cd})F_{cd}^{\phi*}(\vt',\vt'_{cd})}{C_{ab}(\vt,\vt_{ab})(m_c\cosh\vt'+m_d\cosh\vt'_{cd})m_d\cosh\vt'_{cd}}
+\nonumber\\[0.5em]
&
+G_{ab}^a(\vt,\vt_{ab})+G_{ab}^b(\vt,\vt_{ab})+ ...
\Bigg]
+ O\left(\lambda^3\right)
\end{align}
with
\begin{equation}
C_{ab}(\vt,\vt_{ab})=(m_a\cosh\vt+m_b\cosh\vt_{ab}) m_b\cosh\vt_{ab}\,.
\end{equation}

\subsection{Numerical evaluation of the perturbative expressions}
\label{app:tables}

Although our final results Eqs. \eqref{eq:onepmultspec}, \eqref{eq:Kaafin} and \eqref{eq:Kabfin} look quite complicated, numerical evaluation reveals that not all contributions are equally important.  First, let us examine the second order contributions to the one-particle overlap \eqref{eq:onepmultspec}. Table \ref{tab:onep}. contains the eight largest coefficients multiplying $\lambda^2$  and shows that some of them are suppressed by orders of magnitude,  which reflects the fast convergence of the form factor expansions. 
\begin{table}[t!]
\centering
\begin{tabular}{|c|c|c|c|c|c|c|c|c|}
\hline
\multirow{2}{*}{$g_1$} & $\text{A}_1$ & $\text{A}_2$ & $\text{A}_1\text{-A}_1$ & $\text{A}_3$ & $\text{A}_4$ & $\text{A}_1\text{-A}_2$ & $\text{A}_1\text{-A}_3$ & $\text{A}_5 $ \\\cline{2-9}
& 5.84879 & -0.94431 & -0.38934 & 0.28140 & 0.03816 & 0.02552 & 0.01286 & -0.01184 \\\hline
\hline
\multirow{2}{*}{$g_2$} & $\text{A}_1$ & $\text{A}_2$ & $\text{A}_1\text{-A}_1$ & $\text{A}_3$ & $\text{A}_4$ & $\text{A}_6$ & $\text{A}_1\text{-A}_2$ & $\text{A}_5$ \\\cline{2-9}
& 1.78700 & -1.18054 & 0.25376 & 0.14533 & -0.05807 & -0.00571 & -0.00539 & 0.00449 \\\hline
\hline
\multirow{2}{*}{$g_3$} & $\text{A}_1$ & $\text{A}_3$ & $\text{A}_2$ & $\text{A}_4$ & $\text{A}_1\text{-A}_1$ & $\text{A}_1\text{-A}_2$ & $\text{A}_5$ & $\text{A}_2-\text{A}_2$ \\\cline{2-9}
& 0.96636 & 0.43049 & -0.26373 & -0.09194 & -0.06197 & 0.01725 & -0.00891 & -0.00161 \\\hline
\end{tabular}
\caption{Contributions to $g_a$ at order $\lambda^2$ sorted by magnitude, with the particle content of the inserted state shown in the top rows.}
\label{tab:onep}
\end{table}

For the pair overlap functions, the second order contributions are collected in Table \ref{tab:twop}. Again, it is the lowest-lying states that contribute the most, but the coefficients decrease less drastically with the energy of the state, which is the reason why it was important to construct form factors beyond the ones available previously. 
\begin{table}[t!]
\centering
\begin{tabular}{|c|c|c|c|c|c|c|c|}
\hline
\multirow{3}{*}{$K_{11}$} & $\text{A}_1$ & $\text{A}_1-\text{A}_1^\text{disc}$ & $\text{A}_2$ & $\text{A}_3$ & $\text{A}_1-\text{A}_1^\text{conn}$ & $\text{A}_5$ & $\text{A}_4$ \\\cline{2-8}
& -0.2931 & 0.2232 & -0.1541& -0.0154 & -0.0106 & -0.0079 & -0.0072\\
&+2.3520$\imath$&-1.7915$\imath$&+1.2366$\imath$&+0.1235$\imath$& +0.0850$\imath$&+0.0636$\imath$&+0.0575$\imath$\\\hline
\hline
\multirow{3}{*}{$K_{12}$}
& $\text{A}_1$ & $\text{A}_2$ & $\text{A}_1-\text{A}_1$ & $\text{A}_1-\text{A}_2^{\text{disc}}$ & $\text{A}_1-\text{A}_2^{\text{conn}}$ & $\text{A}_3$ & $\text{A}_5$ \\\cline{2-8}
&-0.2135&0.0968&-0.1366&-0.1396&0.0130&-0.0365&0.0041\\
&+0.0161$\imath$&-0.1739$\imath$&-0.0705$\imath$&+0.0105$\imath$&+0.0523$\imath$&-0.0085$\imath$&-0.0116$\imath$\\\hline
\end{tabular}
\begin{tabular}{|c|c|c|c|c|c|}
\hline
\multirow{3}{*}{$K_{22}$} &
$\text{A}_1$ & $\text{A}_2-\text{A}_2^{\text{disc}}$ & $\text{A}_2$ & $\text{A}_1-\text{A}_1$ & $\text{A}_3$ \\\cline{2-6}
& 0.3187 & -0.2242 & 0.1211 & -0.0380 & 0.0200 \\
&-0.0497 $\imath$ &+0.0350 $\imath$ &-0.0189 $\imath$ &-0.0897 $\imath$ &-0.0031 $\imath$ \\\hline
\end{tabular}
\caption{Most sizeable contributions in second order to $K_{ab}(\vt)$ at $\vt=0.45$. Upper row for each particle indicates the inserted state. The superscript ``disc'', where present, indicates whether it is the disconnected or the connected part of the diagonal form factor.}
\label{tab:twop}
\end{table}

\section{TCSA extrapolation}
\label{app:epol}

\begin{figure}[t!]
\begin{tabular}{cc}
\begin{subfloat}
{\includegraphics[width=0.45\textwidth]{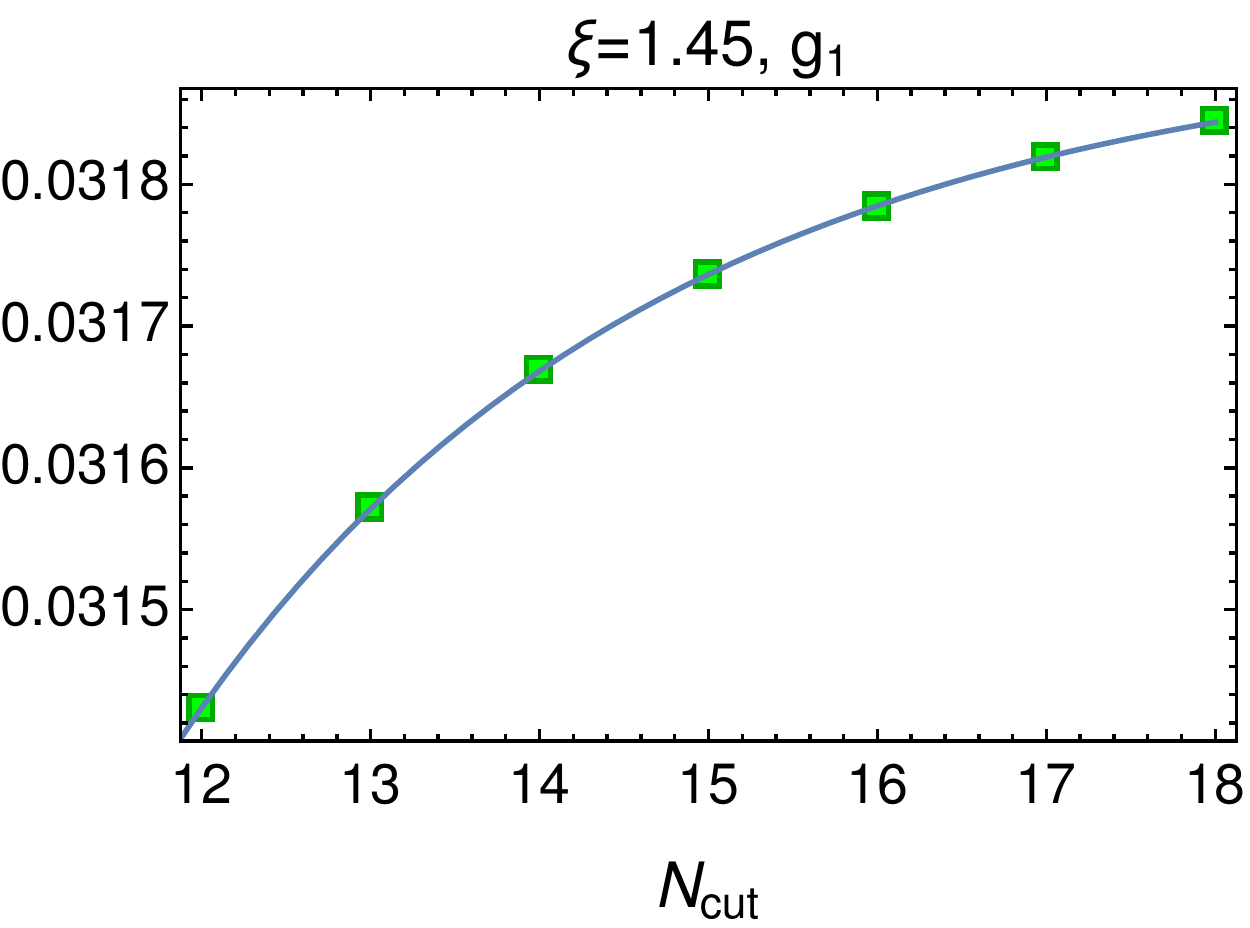}
\label{subfig:epolg1}}
\end{subfloat} &
\begin{subfloat}
{\includegraphics[width=0.45\textwidth]{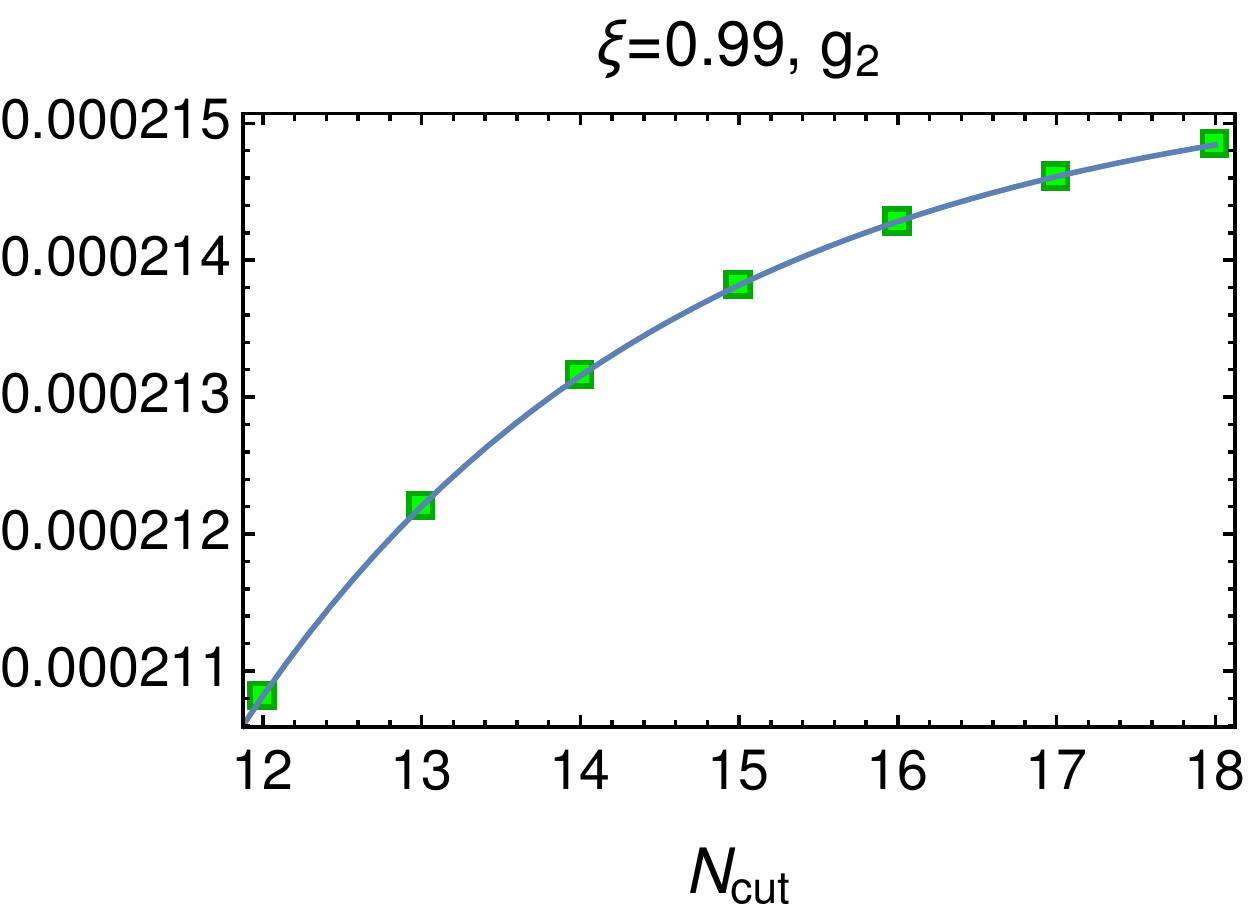}
\label{subfig:epolg2}}
\end{subfloat} \\
\begin{subfloat}
{\includegraphics[width=0.45\textwidth]{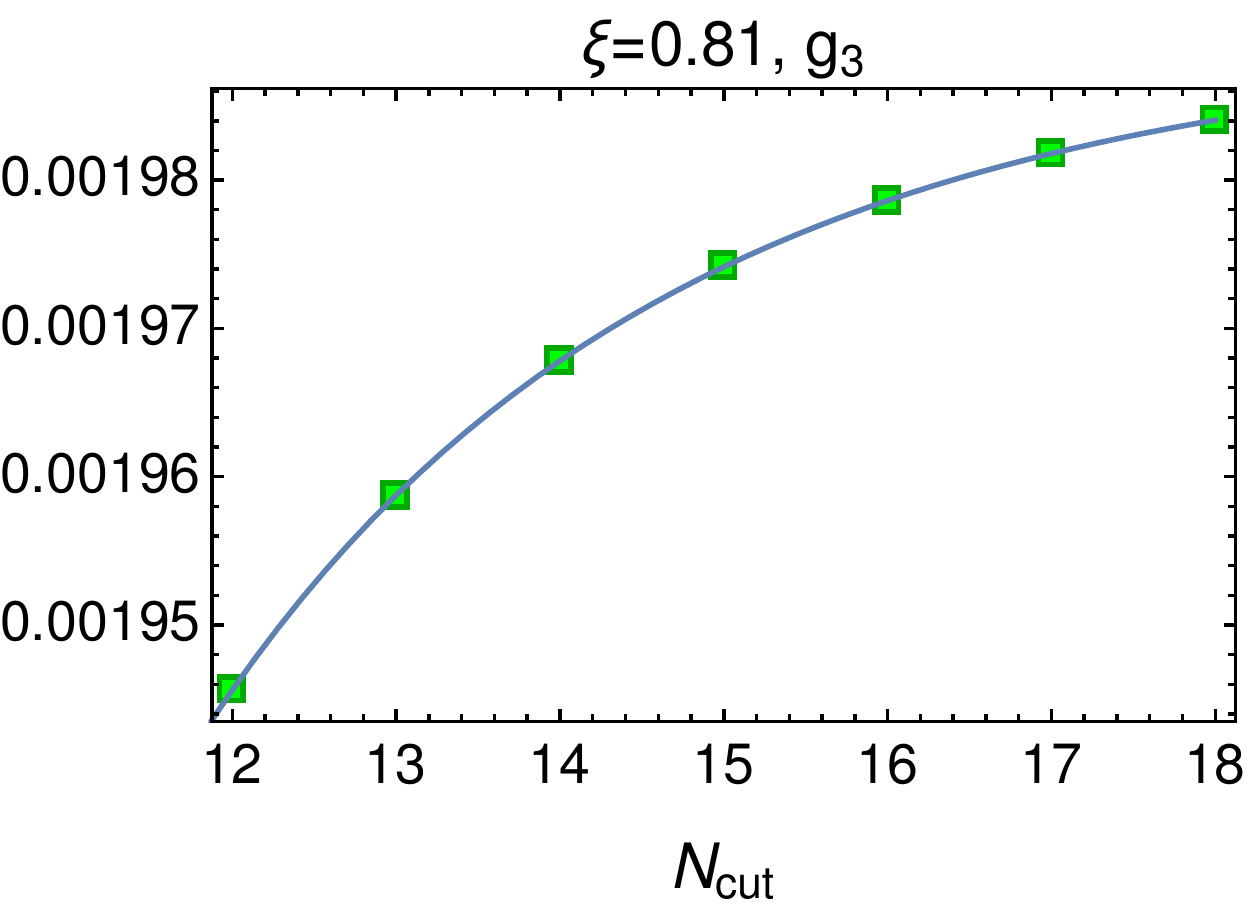}
\label{subfig:epolg3}}
\end{subfloat} &
\begin{subfloat}
{\includegraphics[width=0.45\textwidth]{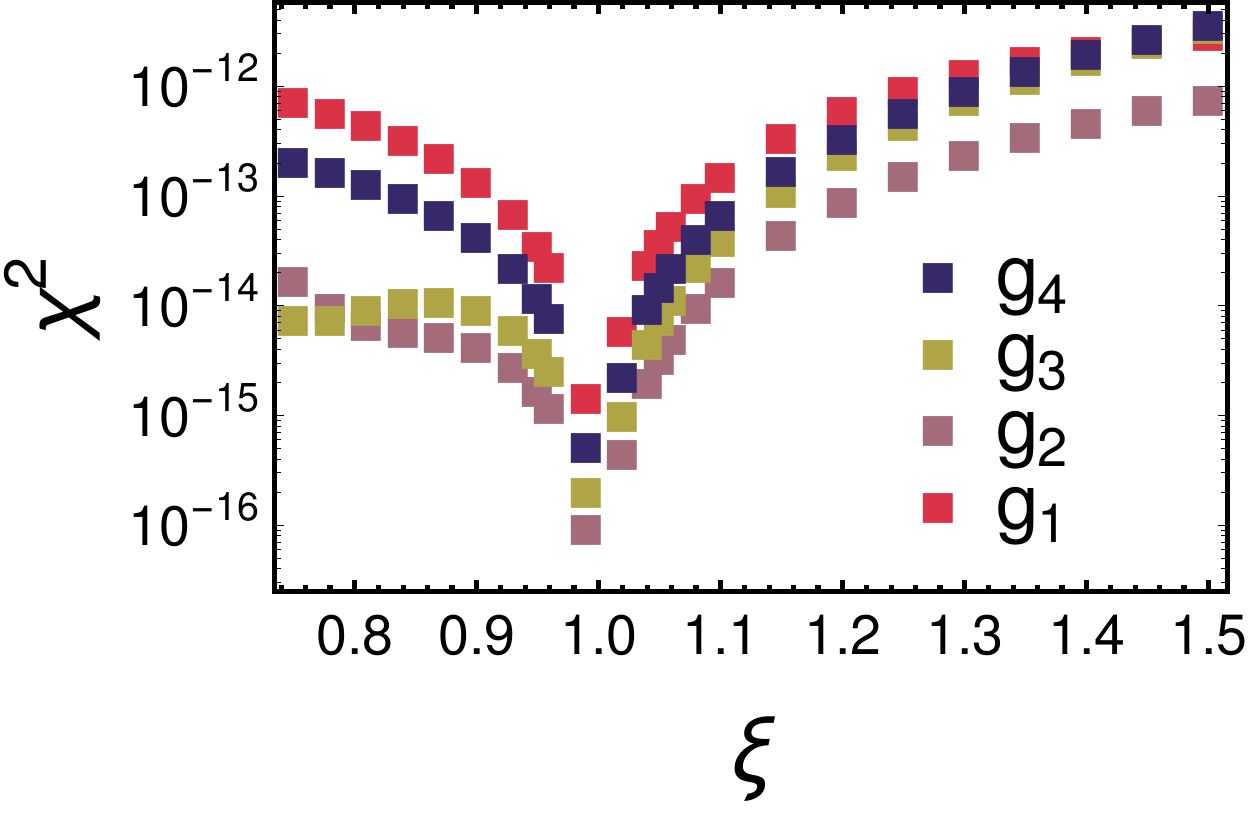}
\label{subfig:epol_frs}}
\end{subfloat}
\end{tabular}
\caption{Illustration of the extrapolation scheme in the case of one-particle overlaps of the Ising Field Theory. The first three subfigures show that the data can be fitted well with the function \eqref{eq:epol}. The fourth panel quantifies this observation by displaying the sum of the squared fit residuals denoted by $\chi^2$ as functions of the quench parameter $\xi$.}
\label{fig:epol_gfacs}
\end{figure}

\begin{figure}[t!]
\begin{tabular}{cc}
\begin{subfloat}
{\includegraphics[width=0.45\textwidth]{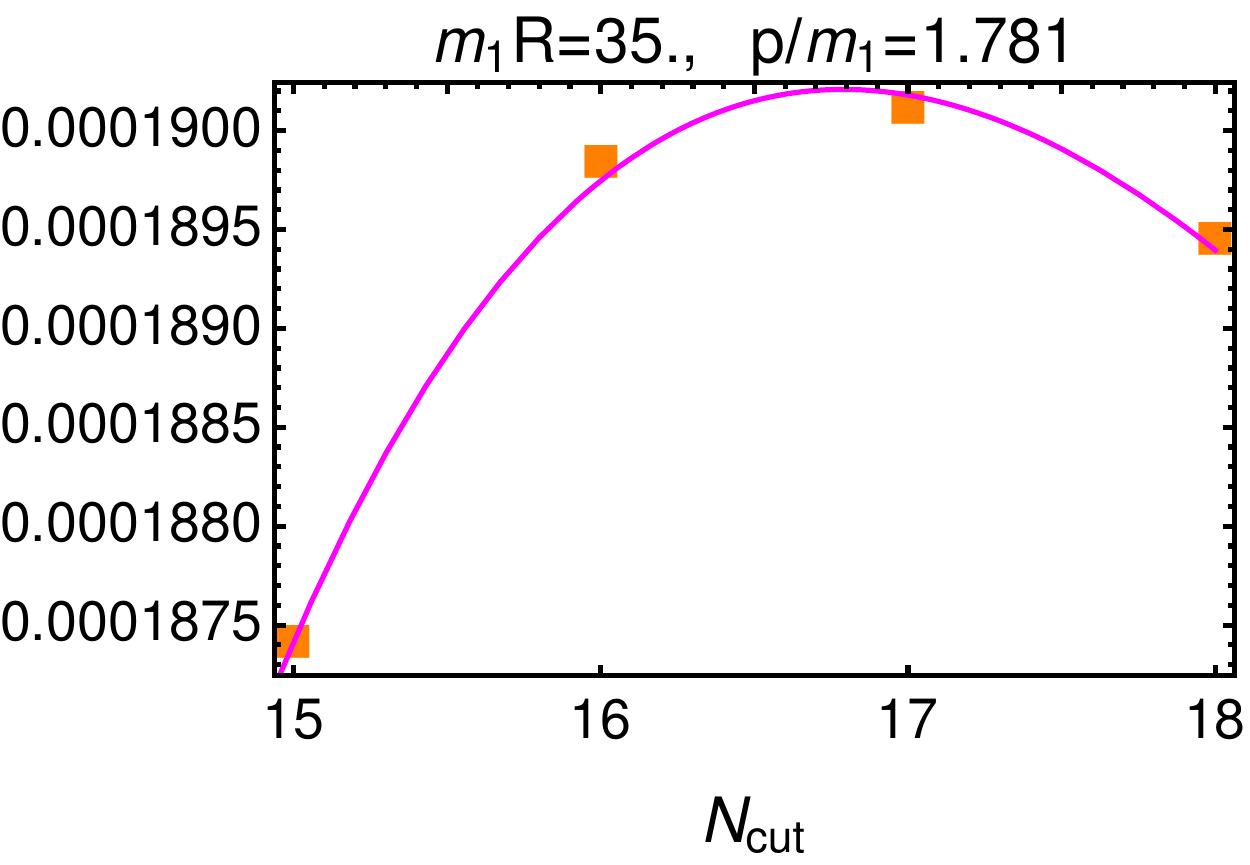}
\label{subfig:epolK11r35}}
\end{subfloat} &
\begin{subfloat}
{\includegraphics[width=0.45\textwidth]{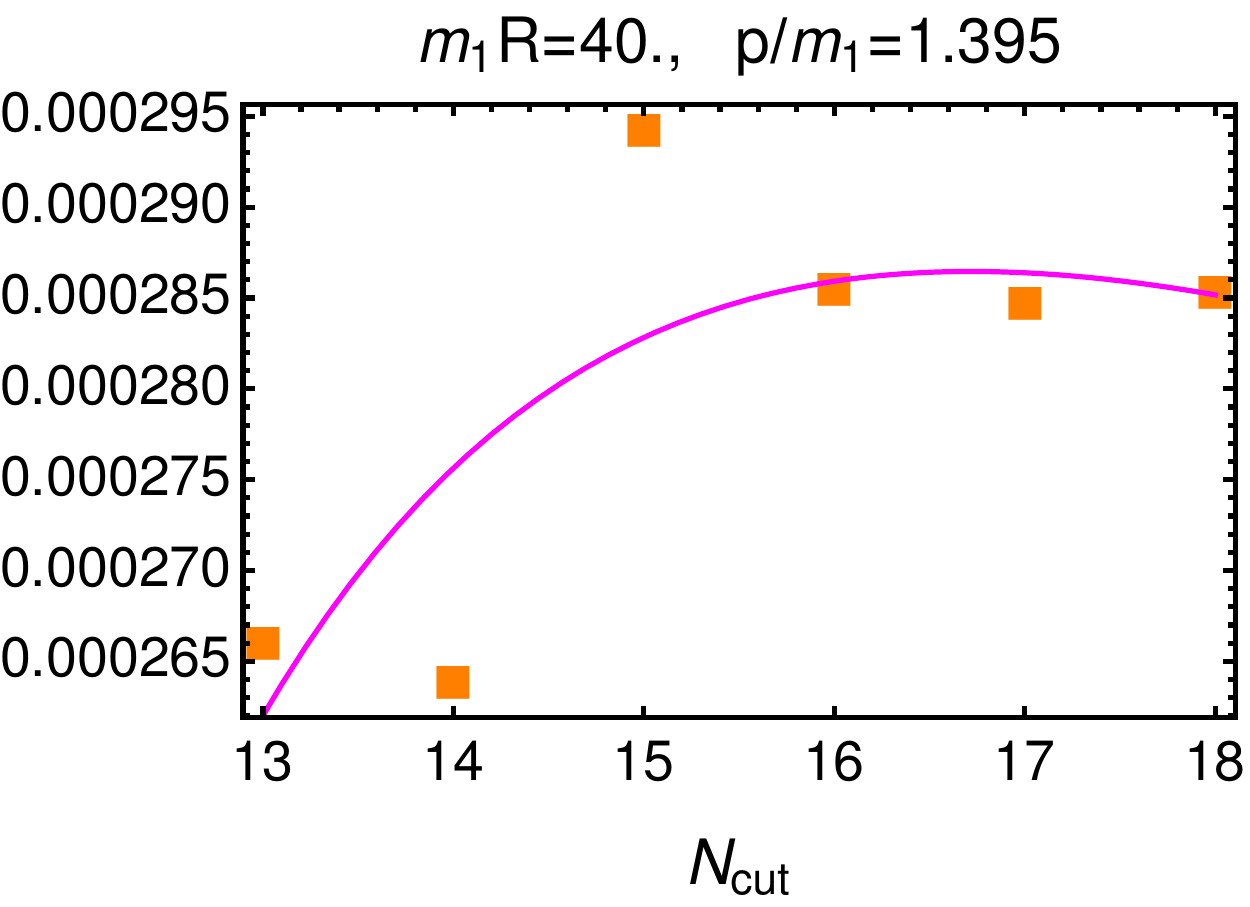}
\label{subfig:epolK11r40}}
\end{subfloat} \\
\begin{subfloat}
{\includegraphics[width=0.45\textwidth]{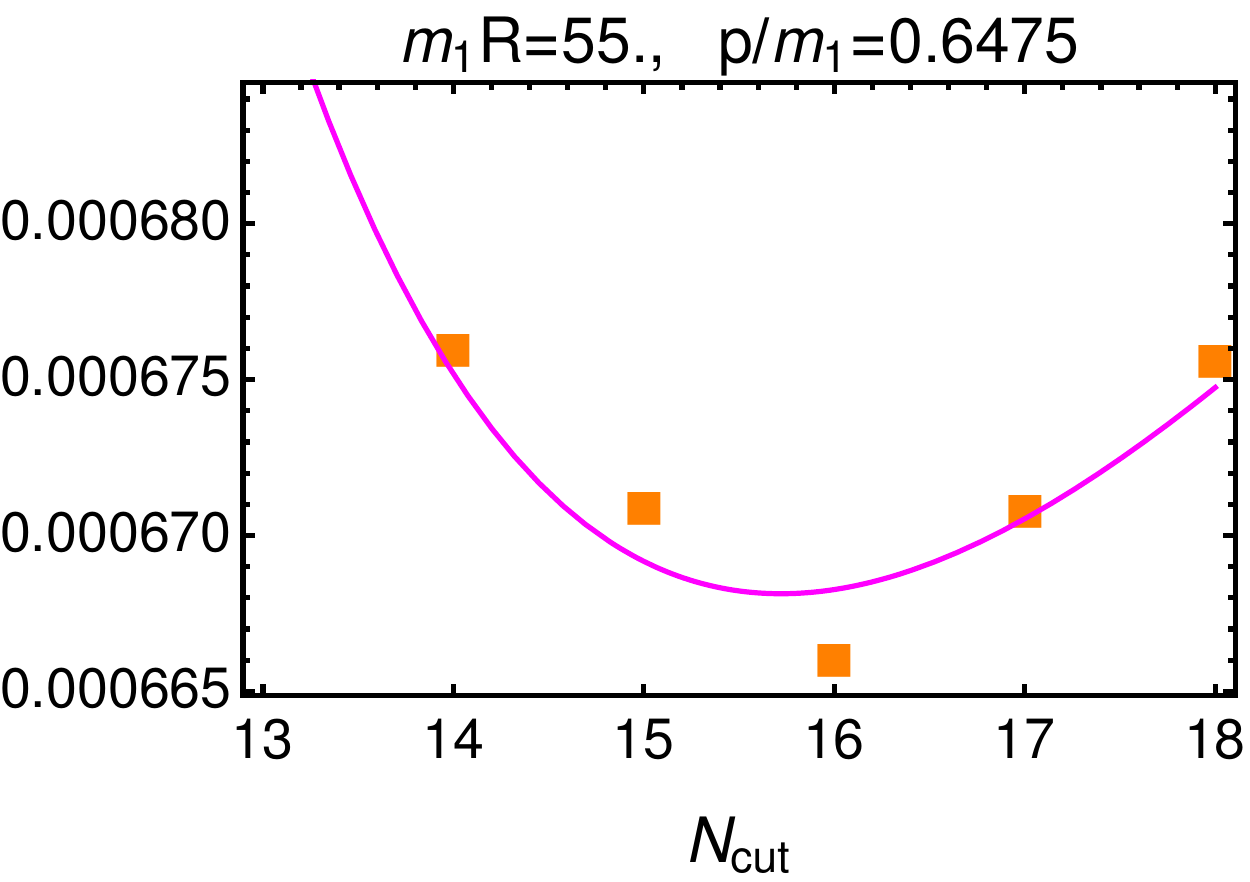}
\label{subfig:epolK11r55}}
\end{subfloat} &
\begin{subfloat}
{\includegraphics[width=0.45\textwidth]{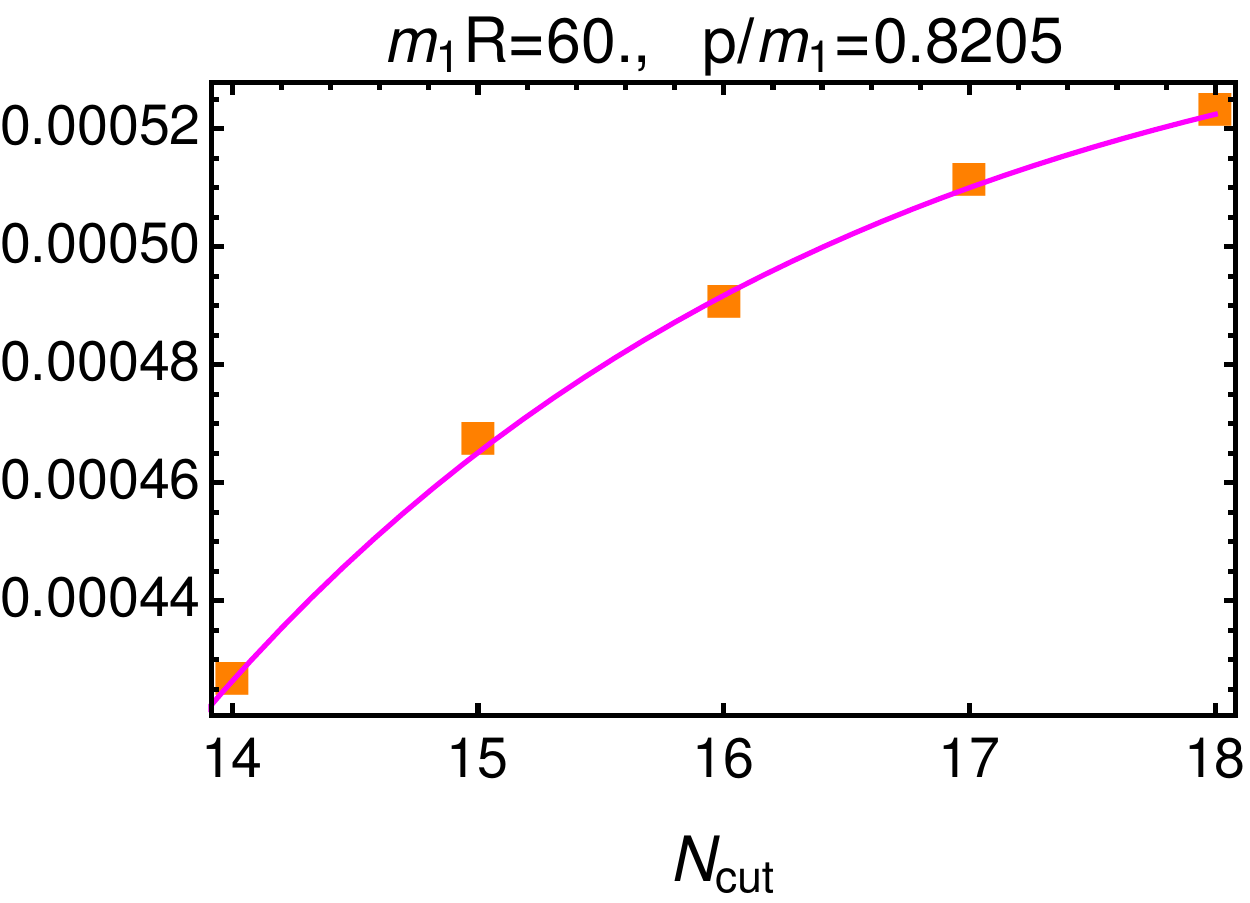}
\label{subfig:epolK11r60}}
\end{subfloat}
\end{tabular}
\caption{Illustration of cut-off extrapolation in the case of two-particle overlaps $K_{11}(p)$ of the Ising Field Theory. The first three panels show that problems with the fit can occur at various volumes and energies. The fourth panel illustrates that a precise fit can also be obtained at large volume.}
\label{fig:epol_Kfun}
\end{figure}

In this Appendix we illustrate the efficiency of Eq. \eqref{eq:epol} in accounting for the cut-off dependence of the numerical overlaps obtained from TCSA. For one-particle overlap functions it gave very good fits to the cut-off dependence as shown in Fig. \ref{fig:epol_gfacs}. The fourth figure shows that the sum of the squared residuals remains consistently small throughout the explored parameter region.
To evaluate the $K(p)$ function numerically at various values of the momenta we performed quenches varying the dimensionless volume of the system between $m_1R=30\dots 65$. The cut-off dependence fitted to the data is the result of a second-order perturbative calculation which breaks down for larger volumes or in the case of computing the overlaps for states with higher energy. Unfortunately, as shown in Fig. \ref{fig:epol_Kfun}, the quality of the fits varies very much with the volume and the state under consideration, which prevented a systematic extrapolation. As a result, for a comparison with the analytic predictions in the main text we used the data from the highest available cut-off $N_\text{cut}=18$.
\end{document}